\documentclass[onecolumn,letterpaper]{IEEEtran}
\usepackage[utf8]{inputenc}
\usepackage[OT1]{fontenc}
\usepackage{indentfirst}
\usepackage{cite}
\usepackage{hyperref}
\usepackage{amssymb}
\usepackage{float}
\usepackage{amsmath}
\DeclareMathOperator*{\argmax}{argmax} 
\DeclareMathOperator*{\argmin}{argmin}
\usepackage{mathtools}
\usepackage{CJK}
\newtheorem{theorem}{Theorem}

\newtheorem{lemma}{Lemma}
\newtheorem{proposition}{Proposition}
\newtheorem{corollary}{Corollary}
\newtheorem{fact}{Fact}

\newtheorem{remark}{Remark}
\newtheorem{conjecture}{Conjecture}
\newtheorem{example}{Example}
\newtheorem{algo}{Algorithm}
\usepackage{graphicx}
\usepackage{color}
\usepackage{enumitem}
\usepackage{tikz}
\usepackage{mathtools}
\usepackage{subcaption}
\usepackage{pgfplots}
\usepackage{rotating}
\newtheorem{innercustomdef}{Definition}
\newenvironment{customdef}[1]
  {\renewcommand\theinnercustomdef{#1}\innercustomdef}
  {\endinnercustomdef}

\usepackage{amsmath}
\usepackage{algorithm}
\usepackage{algpseudocode}

\newcommand{\ts}{{\ell}}

\newcommand{\cN}{{\cal N}}
\newcommand{\cD}{{\cal D}}
\newcommand{\cM}{{\boldsymbol{M}}}
\newcommand{\cO}{{\cal O}}
\newcommand{\cTheta}{{\Theta}}

\newcommand{\nT}{{T}}
\newcommand{\cp}{{q}}

\newcommand{\nt}{{m}}
\newcommand{\nti}{{\nt_{i}}}
\newcommand{\ntip}{{\nt_{i}^+}}
\newcommand{\ntin}{{\nt_{i}^-}}

\newcommand{\mo}{{s}}
\newcommand{\moi}{{\mo_{i}}}
\newcommand{\moip}{{\mo_{i}^+}}
\newcommand{\moin}{{\mo_{i}^-}}

\newcommand{\pn}{{P(-,\cp)}}
\newcommand{\pp}{{P(+,\cp)}}
\newcommand{\pmin}{{P_{\min}(\cp)}}
\newcommand{\pgap}{{\Delta(\cp)}}

\newcommand{\cpn}{{\cp^*}}
\newcommand{\pminn}{{P_{min}(\cpn)}}
\newcommand{\pgapn}{{\Delta(\cpn)}}

\newcommand{\fd}{{P_{\text{FA}}}}
\newcommand{\fn}{{P_{\text{MD}}}}
\newcommand{\spn}{{\rho}}
\newcommand{\spp}{{\nu}}
\newcommand{\spmin}{{P_{\min}}}
\newcommand{\spgap}{{\Delta}}

\newcommand{\cEl}{{\hat{\mathcal{E}}}}
\newcommand{\fdl}{{\hat{P}_{\text{FA}}}}
\newcommand{\fnl}{{\hat{P}_{\text{MD}}}}
\newcommand{\spnl}{{\hat{\spn}}}
\newcommand{\sppl}{{\hat{\spp}}}
\newcommand{\spgapl}{{\nabla}}

\newcommand{\up}{{U}}
\newcommand{\lw}{{L}}
\newcommand*{\dif}{{\,\mathrm{d}}}

\newcommand{\ap}{{\alpha}}

\newcommand{\gm}{{\varphi}}
\newcommand{\qp}{{t}}

\newcommand{\pnm}{{Q(-,\cp)}}
\newcommand{\ppm}{{Q(+,\cp)}}

\newcommand{\pgapm}{{\nabla(\cp)}}

\newcommand{\cx}{{X}}
\newcommand{\cy}{{Y}}
\newcommand{\cxh}{{\hat{X}}}
\newcommand{\etp}{{H}}
\newcommand{\mi}{{I}}
\newcommand{\cA}{{Z}}

\newcommand{\na}{{a}}
\newcommand{\plsize}{{\chi}}
\newcommand{\mean}{{\mu}}
\newcommand{\var}{{\sigma^{2}}}
\newcommand{\vara}{{b}}
\newcommand{\varb}{{c}}
\newcommand{\LHS}{{\text{L.H.S.}}}
\newcommand{\RHS}{{\text{R.H.S.}}}

\newcommand{\nan}{{\kappa}}
\newcommand{\apn}{{\gamma}}
\newcommand{\cE}{{\mathbb{E}}}
\newcommand{\mnp}{{\vartheta}}
\newcommand{\idex}{{i}}
\newcommand{\arb}{{\eta}}

\newcommand{\lwm}{{\hat{L}}}
\newcommand{\upm}{{\hat{U}}}

\usepackage{accents}

\newcommand{\fgap}{{\delta}}

\newcommand{\fh}{{h(f)}}
\newcommand{\fH}{{H(f)}}

\newcommand{\bet}{{\beta}}
\newcommand{\exifinal}{{ 376017 \times \frac{1}{\bet^2} \left( \frac{\up-\lw}{f(\up)-f(\lw)} \right)^2  d\log \left( \frac{2n}{\varepsilon } \right)}}

\newcommand{\exi}{{i}}
\newcommand{\exind}{{\ell}}
\newcommand{\sig}{{\lambda}}
\newcommand{\sigi}{{\sig_{\exi}}}
\newcommand{\sigio}{{\sig_{\exi+1}}}
\newcommand{\sigt}{{\sig_{\ta}}}
\newcommand{\sigto}{{\sig_{\ta-1}}}
\newcommand{\sigind}{{\sig_{\exind}}}
\newcommand{\sigindo}{{\sig_{\exind+1}}}
\newcommand{\ta}{{\tau}}
\newcommand{\alp}{{\ap}}
\newcommand{\alpi}{{\alp_{\exi}}}
\newcommand{\alpind}{{\alp_{\exind}}}

\newcommand{\nod}{{\cA_i}}

\newcommand{\suppress}[1]{}

\newcommand{\ary}{\boldsymbol{X}}
\newcommand{\simdoc}{\boldsymbol{\hat{X}}}
\newcommand{\mtx}{\boldsymbol{M}}

\newcommand{\rs}{\boldsymbol{Z}}
\newcommand{\rse}{Z}

\newcommand{\tmin}{T_{min}}
\newcommand{\tmax}{T_{max}}
\newcommand{\tgap}{\Delta_T}
\newcommand{\vari}{j}

\newcommand{\ftau}{{\Gamma}}
\newcommand{\ftaup}{{\hat{\Gamma}}}
\newcommand{\fT}{{\ftau(\cp)}}
\newcommand{\fTo}{\ftau_1(\cp)}
\newcommand{\fTt}{\ftau_2(\cp)}
\newcommand{\fTp}{\ftaup (\cp)}
\newcommand{\fTn}{\ftau (\cpn)}
\newcommand{\fTpn}{\ftaup (\cpn)}
\newcommand{\cpns}{\cp_0}

\newcommand{\fTpns}{\ftaup (\cpns)}
\newcommand{\pgapns}{{\Delta(\cpns)}}

\newcommand{\cph}{{\hat{\cp}^*}}
\newcommand{\cphglobal}{{\hat{\cp}}}
\newcommand{\pgaph}{{\Delta(\cph)}}
\newcommand{\pph}{{P(+,\cph)}}
\newcommand{\ppmh}{{Q(+,\cph)}}
\newcommand{\ppn}{{P(+,\cpn)}}
\newcommand{\ppmn}{{Q(+,\cpn)}}
\newcommand{\pminh}{{P_{min}(\cph)}}
\newcommand{\fTh}{{\Gamma(\cph)}}
\newcommand{\fThglobal}{{\Gamma(\cphglobal)}}

\newcommand{\seq}{{S}}
\newcommand{\ups}{{\upsilon}}
\newcommand{\logu}{{k}}
\newcommand{\ind}{{i}}
\newcommand{\exiind}{{\ell}}
\newcommand{\fbar}{{\Tilde{f}}}

\newcommand{\xw}[1]{\textcolor{cyan}{#1}}
\definecolor{applegreen}{rgb}{0.55, 0.71, 0.0}

\newcommand{\lom}{{\texttt{LoM}}}
\newcommand{\lomerr}{{\epsilon}}
\newcommand{\ud}{{\hat{d}}}
\newcommand{\lomp}{{P}}
\newcommand{\lomcp}{{\zeta}}
\newcommand{\lomrs}{{t(\ud,\lomcp,\lomerr)}}
\newcommand{\lompo}{{t^{+}(\ud,\lomcp,\lomerr)}}
\newcommand{\lomppd}{{\lomp(d,\lomcp)}}
\newcommand{\lompp}{{\lomp(\ud,\lomcp)}}
\newcommand{\lomppo}{{\lomp(\ud-1,\lomcp)}}
\newcommand{\lompgap}{{\Delta(\ud,\lomcp)}}

\newcommand{\esterr}{{\varepsilon}}
\newcommand{\dup}{{d_u}}
\newcommand{\dlw}{{d_l}}
\newcommand{\dmd}{{d_m}}
\newcommand{\lomcps}{{\lomcp^*}}
\newcommand{\lomrss}{{t(\ud,\lomcps,\lomerr)}}
\newcommand{\lompgapup}{{\Delta(\dup,\lomcp)}}
\newcommand{\lompgapmd}{{\Delta(\dmd,\lomcp)}}
\newcommand{\lompgaps}{{\Delta(\ud,\lomcps)}}

\title{{Generalized Group Testing}}
\author{Xiwei Cheng, Sidharth Jaggi, Qiaoqiao Zhou
\thanks{X.\ Cheng (e-mail: xwcheng@link.cuhk.edu.hk) is with the Department of Computer Science and Engineering, The Chinese University of Hong Kong, Hong Kong.}
\thanks{S.\ Jaggi (e-mail: sid.jaggi@bristol.ac.uk) is with the School of Mathematics, University of Bristol, U.K.}
 \thanks{Q.\ Zhou (email: zhouqq@comp.nus.edu.sg) is with the Department of Computer Science, National University of Singapore.}
\thanks{A preliminary version of this work appeared in the {\it Proceedings of the $25^{th}$ International Conference on Artificial Intelligence and Statistics (AISTATS)}, Valencia,
Spain, 2022.}}

\begin{document}

\maketitle

\begin{abstract}
In the problem of classical group testing one aims to identify a small subset (of size $d$) diseased individuals/defective items in a large population (of size $n$). This process is based on a minimal number of suitably-designed group tests on subsets of items, where the test outcome is positive iff the given test contains at least one defective item.

Motivated by physical considerations, such as scenarios with imperfect test apparatus, we consider a generalized setting that includes as special cases multiple other group-testing-like models in the literature. In our setting the test outcome is governed by an arbitrary {\it monotonically increasing} (stochastic) test function $f(\cdot)$, with the test outcome being positive with probability $f(x)$, where $x$ is the number of defectives tested in that pool. This formulation subsumes as special cases a variety of noiseless and noisy group-testing models in the literature.

Our main contributions are as follows.

Firstly, for any monotone test function $f(\cdot)$ we present a non-adaptive scheme  that with probability $1-\varepsilon$ identifies all defective items. Our scheme requires at most ${\cal O}\left( \fH d\log\left(\frac{n}{\varepsilon}\right)\right)$ tests, where $\fH$ is a suitably defined ``sensitivity parameter" of $f(\cdot)$, and is never larger than ${\cal O}(d^{1+o(1)})$, but indeed can be substantially smaller for a variety of $f(\cdot)$.

Secondly, we argue that any non-adaptive group testing scheme needs at least $\Omega \left((1-\varepsilon)\fh d\log\left(\frac n d\right)\right)$ tests to ensure high reliability recovery. Here $\fh$ is a suitably defined ``concentration parameter" of $f(\cdot)$, and $\fh \in  \Omega{(1)}$.

Thirdly, we prove that our sample-complexity bounds for generalized group testing are information-theoretically near-optimal for a variety of sparse-recovery group-testing models in the literature.
That is, for {\it any} ``noisy" test function $f(\cdot)$
(i.e., $0< f(0) < f(d) <1$), and for a variety of ``(one-sided) noiseless" test functions $f(\cdot)$ (i.e., either $f(0)=0$, or $f(d)=1$, or both) studied in the literature we show that $\frac {\fH} {\fh} \in \Theta(1)$.
As a by-product we tightly characterize the heretofore open information-theoretic order-wise sample-complexity for the well-studied model of threshold group-testing. For general (near)-noiseless test functions $f(\cdot)$ we show that $\frac {\fH} {\fh} \in {\cal O}(d^{1+o(1)})$.  We also demonstrate a ``natural" test-function $f(\cdot)$ whose sample complexity scales ``extremally" as $\Theta(d^2\log n )$, rather than $\Theta(d\log n)$ as in the case of classical group-testing.

Some of our techniques may be of independent interest -- in particular our achievability requires a delicate saddle-point approximation, our impossibility proof relies on a novel bound relating the mutual information of pair of random variables with the mean and variance of a specific function, and as a by-product of our proof showing that our sample-complexity upper and lower bounds are close we derive novel structural results about monotone functions.
\end{abstract}

\section{Introduction}
\label{sec:introduction}
Group testing~\cite{dorfman1943detection} is the non-linear sparse recovery process of identifying a small subset of defective items from a larger set of items based on a series of judiciously designed tests. Each test is carried out on a subset of items, and each binary outcome indicates whether or not the test includes at least one defective item. In other words, the test outcome is specified by the ``OR'' function. In designing the testing scheme, it is desirable to minimize the number of tests while still enabling high probability of correct identification of the subset of defective items.\footnote{A significant part of the group-testing literature focuses on zero-error recovery -- see for instance the survey in~\cite{du2000combinatorial}. However, even in the context of classical group-testing more stringent recovery criterion comes at the cost of requiring a number of tests than high-probability recovery requires. Further, in the context of this work, where test outcomes are probabilistic in nature, zero-error recovery is impossible, hence we focus on high-probability recovery.} 
The group testing paradigm has found applications in a wide variety of contexts, including biology~\cite{ngo2000survey}, pattern finding~\cite{macula2004}, wireless communications~\cite{berger1984,wolf1985}, and testing for diseases recently COVID-19 testing~\cite{gollier2020group}.

Many variants of the classical group testing paradigm have already been considered in the literature. For example, Damaschke~\cite{damaschke2006threshold} considered {\it threshold test functions}: the test outcome is negative if the number of defectives in a test is no larger than the lower threshold $\ell$; positive if no smaller than the upper threshold $u$; and arbitrary (negative or positive) otherwise. Let $n$ and $d$ be the number of all items and the number of defective items, respectively. For $u=\ell+1$,~\cite{damaschke2006threshold} proposed an adaptive algorithm with the number of tests scaling as ${\cal O}\left( (d+u^2) \log n\right)$ to exactly identify the defectives. However, for $u>\ell+1$, they proved that the defectives cannot be exactly identified, but ${\cal O}\left( (dn^b+d^u) \log n\right)$ adaptive tests suffice to identify the defectives if up to $(u-1)(1+\frac{1}{b})-\ell$ misidentifications are allowed (here $b>0$ is an arbitrary constant). Chen and Fu \cite{chen2009nonadaptive} proposed a non-adaptive algorithm for which the number of tests scales as ${\cal O}\left( \sigma d^{u+1} \log (\frac n d)\right)$ if up to $u-\ell-1$ misidentifications and $\sigma$ erroneous tests are allowed. Subsequently, Cheraghchi \cite{cheraghchi2010improved} showed that it can be reduced to ${\cal O}\left( d^{u-\ell+1} \log d \log (\frac n d)\right)$. More recently, for the special case $u=\ell+1$, \cite{de2020} reduced it further to ${\cal O}\left( d^{\frac 3 2}  \log (\frac n d)\right)$ when $u$ is asymptotically close to $\frac d 2$. The works of~\cite{bui2019,bui2020} sought to find schemes that admit low decoding complexity. Chan {\it et al.} \cite{sid13itw} studied stochastic threshold group testing. They introduced two stochastic variants of the threshold test function: Bernoulli gap stochasticity and linear gap stochasticity. For Bernoulli gap stochasticity, the test outcome is equally likely to be negative or positive whenever the number of defectives in a test is in the interval $(\ell, u)$. For linear gap stochasticity, the probability of having positive outcome increases linearly as the number of defectives ranges from $\ell$ to $u$. By allowing a small error probability $\varepsilon$, they proposed a two-stage adaptive algorithm with $11.09e^2d \log n +{\cal O}\left(d\log( \frac{1} {\varepsilon})\right)$ number of tests and a non-adaptive algorithm with ${\cal O}\left(\log(\frac {1} {\varepsilon})d\sqrt{\ell}\log n\right)$ number of tests for Bernoulli gap stochasticity, and a non-adaptive algorithm with ${\cal O}\left((u-\ell-1)^2d\log n\right)+{\cal O}\left(d\log(\frac {1} {\varepsilon})\right)$ number of tests for linear gap stochasticity. Recently,  for Bernoulli gap stochasticity,  Reisizadeh {\it et al.} \cite{reisizadeh2018} improved the number of tests required to ${\cal O}\left(\sqrt{u}d\log ^3 n \right)$.

\subsection{Our contributions}
In this paper, motivated by physical considerations such as the effect of dilution on the chemistry of group tests, we formulate and analyze group testing with a general monotonically increasing stochastic test function $f(\cdot)$ (i.e., $x \geq y \Rightarrow f(x) \geq f(y)$) that takes as input the number of defective items in a test and outputs the probability of the given test having a positive outcome. This formulation subsumes as special cases a variety of noiseless and noisy group-testing models in the literature. Our contributions are as follows. 
\begin{itemize}
\item For any monotone test function $f(\cdot)$ we present a non-adaptive generalized group-testing scheme that identifies all defective items with probability at least $1-\varepsilon$. Our scheme requires at most ${\cal O}\left( \fH d\log\left(\frac {n} {\varepsilon}\right)\right)$ tests, where $\fH$ is a suitably defined ``sensitivity parameter" of $f(\cdot)$, and is never larger than ${\cal O}(d^{1+o(1)})$, but indeed can be substantially smaller for a variety of $f(\cdot)$.  The computational complexity of decoding is $\cO \left( n\fH d  \log \left(\frac {n} {\varepsilon}\right) \right)$.

\item We argue that any non-adaptive group testing scheme that has a probability of error of at most $\varepsilon$ requires at least $\Omega \left((1-\varepsilon)\fh d\log\left(\frac n d\right)\right)$ tests, where $\fh$ is a ``concentration parameter" of $f(\cdot)$, and $\fh \in \Omega{(1)}$. 

\item We prove that our sample-complexity bounds for generalized group testing are information-theoretically near-optimal for a variety of sparse-recovery group-testing models in the literature. That is, 
\begin{itemize}
    \item For {\it any} ``noisy" test function $f(\cdot)$
(i.e., $0< f(0) < f(d) <1$) we show that
$\frac{\fH}{\fh} \in \Theta(1)$. This implies that for noisy test functions our non-adaptive scheme has order-wise optimal sample complexity.

\item For a variety of ``noiseless" test functions $f(\cdot)$ (i.e. $f(0)=0$, $f(d)=1$) studied in the literature we also show that $\frac{\fH}{\fh} \in \Theta(1)$. Hence for these test-functions as well the sample complexity of our non-adaptive scheme is order-wise optimal. 

\item As a by-product we tightly characterize the heretofore open information-theoretic order-wise sample-complexity for the well-studied model of threshold group-testing.

\item  Perhaps surprisingly, for technical reason our results are somewhat weaker for general ``(one-sided) near-noiseless" test functions $f(\cdot)$ (i.e., either $f(0) \xrightarrow{d\rightarrow \infty} 0$, or $f(d)  \xrightarrow{d\rightarrow \infty} 1$, or both) than for general noisy test functions. For general near-noiseless test functions $f(\cdot)$ we show that $\frac{\fH}{\fh} \in {\cal O}(d^{1+o(1)})$. 

\item   We also demonstrate a ``natural" test-function $f(\cdot)$ whose sample complexity scales ``extremally" as $\Theta \left( d^2\log n\right)$ -- our results above exclude the possibility of any monotone test function having optimal sample-complexity higher than this. Note that the optimal sample-complexity scales as $\Theta ( d\log n)$ in the case of classical group-testing.
\end{itemize}

\item Some of our techniques may be of independent interest -- in particular: 
\begin{itemize} 
\item Our achievability requires a delicate saddle-point approximation,
\item Our impossibility proof relies on a novel bound relating the mutual information of pair of random variables with the mean and variance of a specific function, and
\item As a by-product of our proof showing that our sample-complexity upper and lower bounds are close we derive novel structural results about monotone functions.
\end{itemize}
\end{itemize}

The rest of this paper is organized as follows. We formulate the generalized (non-adaptive) group testing problem in Section \ref{sec:problem}. Section \ref{sec:test_design} presents our proposed non-adaptive algorithm. In Section \ref{sec:main_results} we state the main results of this work. In Section  \ref{sec:intuition} we describe the intuition behind our results, and proof sketches, with full proofs deferred to Section \ref{sec:proof:thm_main}-\ref{sec:proof:match} and corresponding Appendices. Finally, Section \ref{sec:simulation} contains simulation results of our proposed algorithm.


\section{Problem Formulation}
\label{sec:problem}
A set $\mathcal{N}:=\{1,\dots, n\}$ of $n$ items contains a subset $\mathcal{D}\subsetneq \mathcal{N}$ of  \textit{defective} items -- elements in $\mathcal{N}\setminus \mathcal{D}$ are called \textit{non-defective}. We follow the ``combinatorial group testing model'', which assumes that the size of defective set $\cD$ is fixed as $d$, and each such subset has the same probability. 
The identity of $\cD$ is unknown {\it a priori} -- the goal of group testing is to correctly identify $\cD$ through a minimal series of {\it group} tests on subsets of items. In ``classical'' group testing a test outcome is negative if every item in the pool is non-defective, and is positive if at least one item is defective. As such this may be viewed as a {\it disjunctive} measurement, i.e., viewing each item as a $0$ or a $1$ depending on whether it is non-defective or defective, each test performs an OR of the items in its pool. A canonical setting in which this measurement model is pertinent is when a small number of individuals in a large population are diseased but only a small number of testing kits are available; in this case samples from different individuals may be ``pooled'' together in different combinations and the set of test outcomes analyzed jointly to infer $\cD$. In this work we assume that the {\it number} $d = |{\cD}|$ of defectives is known {\it a priori}.\footnote{\label{fn:est-def} Another branch of the group-testing literature (see for instance~\cite{damaschke2010competitive,falahatgar2016estimating,bshouty2018adaptive}) concerns itself with the problem of reliably approximating the number $d$ itself with a minimal number of adaptive or non-adaptive tests. In classical group-testing, most algorithms are reasonably robust to small perturbations in the value of $d$, and the task of roughly estimating $d$ is an ``easier" task (requiring asymptotically fewer tests than the task of estimating the set $\cD$). In this generalized group-testing setting, our algorithms and bounds are sensitive to small perturbations in the value of $d$, and require the exact value of $d$. In Appendix~\ref{app:esti_d}, we present an algorithm for exactly estimating $d$. It turns out to be a ``harder" task (requiring asymptotically  more tests than the task of estimating the set $\cD$).}

Instantiating disjunctive tests which are sensitive to even a single defective in a testing pool may be tricky, for instance due to the impact of dilution on the chemistry used in pooled tests~\cite{zenios1998pooled}. Our primary contribution in this work is  to consider a very general class of (probabilistic) measurement functions  $f(\cdot): \mathbb{Z}_{\geq 0} \rightarrow [0, 1]$. The input, say $x$, to the measurement function $f(x)$ is the number of defective items $x$ in a given pool, and the value of $f(x)$ is the probability that the given test results in a positive test outcome. 

A slight notational subtlety here -- since we will be interested in asymptotic results (when $d$ and $n$ are ``large"), in the interest of generality we allow the function $f(\cdot)$ to also depend on the value of $d$, the overall number of defectives in the population of size $n$. Hence our notation $f(\cdot)$ actually indexes a set $\{f_d(\cdot)\}_{d}$ of measurement functions. Since we assume the number $d$ to be known in advance (see the discussion above, and Footnote~\ref{fn:est-def}) thus the actual function $f_d(\cdot)$ in the set $f(\cdot)$ is well-specified. Thus for notational convenience we suppress the dependence of $f(\cdot)$ on $d$ in the remainder of this paper. Hence a statement like $f(0) \xrightarrow{d\rightarrow \infty}0$ should be interpreted as meaning that $\lim_{d \rightarrow \infty} f_d(0) = 0$.

In this work we restrict ourselves to the natural class of measurement, {\it monotone} measurement functions, i.e., $x\geq y \Rightarrow f(x)\geq f(y)$. Monotone measurement functions subsume many existing models of group-testing as special cases. For instance, when 
\begin{equation} 
\label{f(x)_classical}
    f(x)= 
    \left\{
    \begin{array}{cc}
    0& \; x = 0, \\
    1     &  \; x \geq 1,
    \end{array} 
    \right.
\end{equation}
this reduces to the problem of classical group testing. 
Observe that when 
\begin{equation} 
    f(x)= 
    \left\{
    \begin{array}{cc}
    0     & \; x \leq \ell, \\
    \frac{x-\ell}{u-\ell}     &  \; \ell < x< u,\\
    1   & \; x \geq u, 
    \end{array} \label{f(x)_linear_gap}
    \right.
\end{equation}
for some integers $0 < \ell< u < d$, this reduces to the ``linear gap'' stochastic group testing examined by~\cite{sid13itw}. To avoid triviality, we further assume that $f(0)<f(d)$. (If $f(0)=f(d)$, no sequence of tests can ever reliably recover the defective set $\cD$.)

{
It will be helpful to distinguish between two types of test functions $f(\cdot)$.
\begin{itemize}
    \item If $0 < f(0) < f(d) < 1$,  we say that $f(\cdot)$ is {\it noisy}. 
    For such $f(\cdot)$, even pools with no defective items have a probability $f(0)$ (some positive constant independent of $d$) of resulting in a positive test outcome, and pools with one or more defective items  have a probability of at least $1-f(d)$ (again, a constant bounded away from $0$, independent of $d$) of resulting in a negative test outcome. The corresponding notion of noisy test outcomes in the classical group-testing literature (see for instance~\cite{scarlett2018near}) often focuses on test functions of the form \begin{equation}
        \label{f(x)_classical_noisy}
    f(x)= 
    \left\{
    \begin{array}{cc}
    a& \; x = 0, \\
    b     &  \; x \geq 1,
    \end{array} 
    \right.
    \end{equation}
    for some $0 < a < b < 1$ (with the {\it symmetric noise setting}, i.e. $b = 1-a$, receiving the most attention). 
    \item In contrast, \begin{itemize}
        \item If either $ f(0) \xrightarrow{d\rightarrow \infty} 0$ or $f(d) \xrightarrow{d\rightarrow \infty} 1$  we say that $f(\cdot)$ is {\it one-sided near-noiseless}, and
        \item If both $ f(0) \xrightarrow{d\rightarrow \infty} 0$ and $f(d) \xrightarrow{d\rightarrow \infty} 1$ hold we say that $f(\cdot)$ is {\it near-noiseless}.
        \item Analogously, if either $ f(0) = 0$ or $f(d) = 1$ we say that $f(\cdot)$ is {\it one-sided noiseless}, and
        \item If both $ f(0) = 0$ and $f(d) = 1$ hold we say that $f(\cdot)$ is {\it noiseless}. 
    \end{itemize}
 Note that (one-sided) near-noiselessness is a significantly weaker requirement on $f(\cdot)$ than in much of the noiseless group-testing literature, where it is often assumed that $f(0) = 0$ and $f(1) = 1$ (the corresponding one-sided noiseless version was studied in~\cite{scarlett2020noisy}).
\end{itemize}
}

Group testing schemes can be adaptive (where each test may be designed based on the outcomes of all preceding tests) or non-adaptive (where all tests must be chosen prior to observing any test outcomes). Here, we focus on non-adaptive group testing.\footnote{The adaptive version of group-testing has also been extensively studied -- see for instance the survey in~\cite{du2000combinatorial}. However, since non-adaptive tests allow for test-parallelization, and also make it easier to design hardware to perform the tests (unlike adaptive test designs, where the composition of (at least some) tests may depend on prior test outcomes), we restrict our attention in this work to designing non-adaptive schemes.} 

{
Non-adaptive generalized group-testing test designs are specified by a (possibly randomly chosen) binary matrix $\boldsymbol{M}\in\{0,1\}^{T\times n}$, where $M_{ji} = 1$ if test $j$ includes item $i$ and $M_{ji} = 0$ otherwise. The rows of $\boldsymbol{M}$ correspond to tests, and the columns correspond to items.  The probability of error of any non-adaptive algorithm (with a specified test matrix $\boldsymbol{M}$) is defined as the probability 
that the estimated defective set $\hat{\mathcal{D}}$ differs from the true $\mathcal{D}$. 
This probability is over the elements comprising $\cD$ (which is assumed to be distributed uniformly at random from all $d$-sized subsets of $\{1,\ldots,n\}$), over the randomness in test outcomes (since each test with $x$ defectives may result in a positive test outcome with probability $f(x)$ and a negative test outcome with probability $1-f(x)$), and randomness if any in the decoding rule.}
We require that the probability of error is bounded from above by some $\varepsilon$. That is,  $\Pr(\hat{\mathcal{D}}\neq \mathcal{D})\leq \varepsilon$ -- such test designs will be called $(1-\varepsilon)$-reliable.




\section{Test Design and Decoding}
\label{sec:test_design}
We now present our non-adaptive test designs, and the corresponding decoding rules. We emphasize here that the algorithm below depends critically on {prior} knowledge of the size $d$ of the defective set -- the setting where $d$ is not known {\it a priori } is the context of Lemma~\ref{lem:estimation:d} in  Section~\ref{subsec:est-d}. 

{\bf Test design:} We consider Bernoulli designs -- see, for example, \cite{AtiaS12,ScarlettC16,ScarlettC17,Aldridge17,Monograph19}. That is, the test matrix $\cM$ is a $\nT \times n$ binary matrix in which each entry is independently chosen to equal $1$ with probability $\cp$ and $0$ otherwise, for some design parameters $T$ and $\cp\in (0,1)$ to be specified later.

{\bf Parameters for the decoding rule:}
Given these tests and their outcomes we now specify two decoding rules we use to produce an estimate $\hat{{\cal D}}$ of the defective set ${\cal D}$. Before presenting the algorithm, let us first introduce some definitions and notation.

\textit{Definitions:}
\begin{enumerate}
\item {\it Item-included test-positivity probability:} For any item $i$ in a test, the quantity $\pn$ denotes the probability that  the test has a positive outcome conditioned on the event that item $i$ is non-defective.\footnote{\label{fn:no-i} Due to the symmetry of randomness in the defective set ${\cD}$ this value is independent of the index value $i$, hence in our notation we do not index the notation for these probabilities with $i$.} Analogously $\pp$ denotes$^{\ref{fn:no-i}}$ the probability that the test has a positive outcome conditioned on the event that item $i$ is defective. Mathematically,
\begin{equation}
\label{eq:def:p-p+}
\begin{aligned}
\pn &:= \sum\limits_{j=0}^{d} \binom{d}{j} \cp^j (1-\cp)^{d-j} f(j),\text{ and} \\
\pp &:= \sum\limits_{j=0}^{d-1} \binom{d-1}{j} \cp^j (1-\cp)^{d-1-j} f(j+1).
\end{aligned}
\end{equation}
\item {\it Item-not-included test-positivity probability:} For any item $i$ not in a test, the quantity $\pnm$ denotes$^{\ref{fn:no-i}}$ the probability (over the randomness in the defective set ${\cD}$) that the test has a positive outcome conditioned on the event that item $i$ is non-defective. Analogously, $\ppm$, denotes$^{\ref{fn:no-i}}$ the probability that the test has a positive outcome conditioned on the event that item $i$ is defective. Mathematically,
\begin{equation}
\label{eq:def:p-pp+p}
\begin{aligned}
\pnm &:= \sum\limits_{j=0}^{d} \binom{d}{j} \cp^j (1-\cp)^{d-j} f(j),\text{ and} \\
\ppm &:= \sum\limits_{j=0}^{d-1} \binom{d-1}{j} \cp^j (1-\cp)^{d-1-j} f(j).
\end{aligned}
\end{equation}
\item \label{item:def:Delta}{\it Item test sensitivity:} For any item $i$ in a test, its test sensitivity $\pgap$ (respectively $\pgapm$) is defined as the difference between the probabilities of positive test outcomes conditioned on item $i$ being defective or not.$^{\ref{fn:no-i}}$ Mathematically,
\begin{equation}\label{eq:def:delta}
\begin{aligned}
    &\pgap := \pp - \pn, \text{ and} \\
    &\pgapm := \pnm - \ppm .
\end{aligned}
\end{equation}

\item {\it Minimal test sensitivity:} A parameter that will be useful in our code design and analysis is the minimal test sensitivity $\pmin$, defined as
\begin{equation} \label{eq:def:Pmin}
\begin{aligned}
    &\pmin := \min\left\{ \pp,1-\ppm \right\}.
\end{aligned}
\end{equation}
Lemma~\ref{lem:relation:quantities} below, whose proof is given in Appendix \ref{app:convert_delta_nabla},
 provides (in)equalities relating the quantities defined thus far.
 \begin{lemma}
\label{lem:relation:quantities}
For all $\cp\in(0,1)$, we have
\begin{align}
\label{ineq:relation:P:Q}
    f(d)\ge  \pp > \pn = \pnm > \ppm \ge f(0),
\end{align}
\begin{align}
\label{eq:delta:convert}
    \pgapm = \frac{\cp}{1-\cp} \pgap,
\end{align}
\begin{align}
\label{eq:1>=pmin>max:Delta}
    1\geq \pmin \geq \max \left\{\pgap, \pgapm\right\}.
\end{align}
\end{lemma}

\item {\it Test participation parameter:} It also helps to define the test participation parameter $\nt$ as in~\eqref{eq:def:m} below.
\begin{equation}\label{eq:def:m}
    \nt := \frac{8.32\pmin}{\left(\pgap\right)^2} \log \left( \frac{2n}{\varepsilon } \right) .
\end{equation}
As shown in Lemma~\ref{lem:num_tests_lb} in Section~\ref{sec:proof_thm_1_1}, with high probability each item in $\cN$ participates in at least $\nt$ tests.
Correspondingly, we define the test participation parameter $\mo$ as below.
\begin{equation}
\label{eq:def:mo}
    \mo := \frac{8.32\pmin}{\left(\pgapm\right)^2} \log \left( \frac{2n}{\varepsilon } \right) .
\end{equation}
Also, as shown in Lemma \ref{lem:num_tests_lb_l} in Section \ref{sec:proof_thm_1_1}, with high probability each item in $\cN$ participates in at most $T-\mo$ tests.





\item {\it Number of tests:}  
Finally, we set the number of tests as any integer $T$ such that
\begin{equation}
\label{eq:def:T}
\begin{aligned}
 T  \geq \fT:=  \frac{13(1-\cp)}{3\cp} \nt ,
    \end{aligned}
\end{equation}
e.g., we can set $T = \lceil \fT \rceil$.

\item {\it Choice of $\cp$:} One wishes to choose $\cp$ such that the defective set $\cD$ can be reliably recovered while minimizing the required number of tests. As shown in Section \ref{sec:proof_thm_1_1}, for all $\cp\in(0,1)$, $\fT$ tests of the above design can reliably recover the defective set $\cD$. Therefore we should choose $\cp$ as the value in $ (0,1)$ that minimizes $\fT$.
However, such a minimizing $\cp$ is hard to analyze and analytically characterize. Instead, we consider the quantity $\fTp$ defined as
\begin{equation}
\label{eq:def:fTp}
    \fTp := \frac{\fT}{\pmin}  = \frac{36.06(1-\cp)}{\cp (\pgap)^2} \log \left( \frac{2n}{\varepsilon} \right).
\end{equation}
The equality in \eqref{eq:def:fTp} follows by using \eqref{eq:def:m} and \eqref{eq:def:T}.
Note that $\fTp \ge \fT$ for all $\cp \in (0,1)$ since $\pmin \le 1$ by \eqref{eq:1>=pmin>max:Delta}, and hence an upper bound on $\fTp$ is also an upper bound for $\fT$. We will choose $\cp$ as
\begin{equation} \label{eq:def:cpn}
    \cpn := \argmin_{\cp\in(0,1)} \frac{36.06(1-\cp)}{\cp (\pgap)^2} \log \left( \frac{2n}{\varepsilon} \right).
\end{equation}
It turns out such a $\cpn$ can be efficiently characterized -- see Theorem \ref{thm:main}-\ref{thm1:item4} for details.
\end{enumerate}	

{\bf Decoding rules:} We are now ready to describe our two decoding rules, each of which proceeds by separately estimating whether or not each item $i\in\cN$ is defective or not (instead of jointly estimating the (non)-defective status of all $i$ simultaneously).
Note that both decoding rules work for all $\cp \in (0,1)$. However, as we will elaborate in Section \ref{sec:proof_thm_1_1} that, the first rule requires fewer tests for $(1-\varepsilon)$-reliable recovery when $\cp \in (0,1/2]$, whereas the second rule requires fewer tests for $(1-\varepsilon)$-reliable recovery when $\cp \in (1/2,1)$. Hence we use decoding rule $1$ if $\cp\leq \frac 1 2$, and  decoding rule $2$ otherwise.

{\underline{\it Decoding Rule 1}:} The first decoding rule makes use of the tests that each item participates in. More precisely, denote by $\nti$ the number of tests that item $i$ participates in, and denote by $\ntip$ (respectively $\ntin$) the number of tests with positive (respectively negative) outcome within these $\nti$ tests. We then classify $i$ as follows:
\begin{equation} \label{eq:classification_small_q}
    i =
    \left\{
    \begin{array}{ll}
    \text{non-defective} & \text{if} \; \frac{\ntip}{\nti} \le \frac{\pn+\pp}{2}, \\
    \text{defective}     & \text{if} \; \frac{\ntip}{\nti} > \frac{\pn+\pp}{2}.
    \end{array}
    \right.
\end{equation}

{\underline{\it Decoding Rule 2}:} In contrast, the second decoding rule makes use of the tests that do not include the item. More precisely, let $\moi$ denote the number of tests that item $i$ is excluded from, and let $\moip$ (respectively $\moin$) denote the number of tests with positive (respectively negative) outcome within these $\moi$ tests. We then classify $i$ as follows:
\begin{equation} \label{eq:classification_large_q}
    i =
    \left\{
    \begin{array}{ll}
    \text{non-defective} & \text{if} \; \frac{\moip}{\moi} > \frac{\pnm+\ppm}{2}, \\
    \text{defective}     & \text{if} \; \frac{\moip}{\moi} \le \frac{\pnm+\ppm}{2}.
    \end{array}
    \right.
\end{equation}



\section{Main Results}
\label{sec:main_results}
As described in this Section, we have four main results, corresponding respectively to an an upper bound on the number of tests required for $(1-\varepsilon)$-reliable recovery via the test-design/decoder proposed in Section~\ref{sec:test_design}, an information-theoretic lower bound on the number of tests required by {\it any} non-adaptive test-design/decoder guaranteeing $(1-\varepsilon)$-reliable recovery, a comparison between our algorithmic upper bound and our information-theoretic lower bound, and an algorithm for estimating the exact number of defectives.

\subsection{Achievability/Upper bound} \label{subsec:ach}
Before stating our achievability, it is useful to define the ``sensitivity parameter" $\fH$ of a given monotone test function $f(\cdot)$. This sensitivity parameter, in a certain manner, measures the ``fastest rate of change" of $f(\cdot)$, maximized over all intervals $[\lw,\up] \subseteq [0,d]$. 


\begin{customdef}{8}\label{def:sensi:para}
{\it Sensitivity parameter:}  Given a monotone test function $f(\cdot)$, its sensitivity parameter $\fH$ is defined as 
\begin{align}
\label{eq:def:sensi:para}
     \fH:=\min_{0\leq \lw<\up \leq d} \left(\frac{1}{\min\left\{\up-\lw , \sqrt{\lw+1} , \sqrt{d-\up+1}\right\}}\times \frac{\up-\lw}{f(\up)-f(\lw)} \right)^2. 
\end{align} 
\end{customdef}

Here the $\frac{\up-\lw}{f(\up)-f(\lw)}$ term bounds the inverse slope of the test function $f(\cdot)$ in the region $[\lw,\up]$, and the $$\min\left\{\up-\lw , \sqrt{\lw+1} , \sqrt{d-\up+1}\right\}$$ term is an amortization factor that, at a high level, relates to the standard deviation of $f(\cdot)$ w.r.t. a certain hypergeometric distribution in that interval. 

Further, for any monotone test function $f(\cdot)$, Lemma~\ref{lem:sensi:para:bound} below (whose proof may be found in Appendix \ref{app:sensi:para:bound}) bounds the sensitivity parameter $\fH$ of $f(\cdot)$, and asserts $\fH\in  {\cal O}\left(d^{1+o(1)}\right)$. 

\begin{lemma} 
\label{lem:sensi:para:bound}
For any monotone test function $f(\cdot)$ and $d \geq 2$, we have\footnote{For $d=1$, we directly have $\fH=\frac{1}{\left(f(d)-f(0)\right)^2}$ by definition.}
\begin{align}
\label{eq:sensi-para:lb:up}
 \frac{1}{\left(f(d)-f(0)\right)^2} \leq \fH\leq \frac {16}{\left(f(d)-f(0)\right)^2}\left(\log\log d +2\right)^2 d.  
\end{align}
\end{lemma}


While the bounds in Lemma~\ref{lem:sensi:para:bound} hold universally for any monotone $f(\cdot)$, the upper bound in~\eqref{eq:sensi-para:lb:up} may be unduly pessimistic. 
For instance, for any $w\in [0,1]$, consider the natural class of test functions for which the slope is $\frac {1}{d^w}$. Namely:

\begin{example}\label{ex:dwf} Let the test function $f(\cdot)$ be defined as 
\begin{equation} 
\label{eq:linear-alpha:function}
    f(x) = 
    \left\{
    \begin{array}{ll}
    \frac{x}{d^{w}} & x \in \left[0, d^{w}\right]\cap \mathbb{Z}^+, \\
    1 & \text{otherwise} ,
    \end{array} 
    \right.
\end{equation}
For such $f(\cdot)$, one can see that $\fH\in {\cal O} \left(d^w\right)$, for instance by choosing $\lw = \left \lfloor \frac{d^w}{3} \right \rfloor$ and $\up = \left \lceil \frac{2d^w}{3} \right \rceil$. 
\end{example}

With the definition of $\fH$ and the corresponding bounds at hand, we can now state our main achievability result, including the computation of the test design parameter $\cpn$ in \eqref{eq:def:cpn}.
\begin{theorem} 
\label{thm:main}
The non-adaptive test design and decoding outlined in Section \ref{sec:test_design} has the following performance:
\begin{enumerate}[label=\alph*)]
  \item  
  The probability of error is at most $\varepsilon$; 
  \label{thm1:item1}
  \item  \label{thm1:item3}
  The number of tests $T$ satisfies
  \begin{align}
  \label{eq:thm:T:ub}
    T \leq 376017  \pminn  \fH d\log \left( \frac{2n}{\varepsilon } \right)+1 \leq 376017 \fH d\log\left( \frac{2n}{\varepsilon} \right)+1;
  \end{align}
  \item 
  The test design parameter $\cpn$ in \eqref{eq:def:cpn} can be efficiently approximated in $\cO(d^7 \log^2(d))$ time; \label{thm1:item4}
  \item  
  The computation complexity of decoding is $\cO \left( n\fH d  \log \left( \frac{n}{\varepsilon} \right) \right)$. \label{thm1:item2}
\end{enumerate}
\end{theorem}

Proofs of each part of Theorem~\ref{thm:main} may be found in consecutive sub-sections in Section~\ref{sec:proof:thm_main}. 

\begin{remark}
Our current method of finding an appropriate $\cpn$ is via a brute-force method, hence the somewhat high computational complexity of $\cO(d^7 \log^2(d))$ (though it is the one-time cost of {\it designing} our algorithm parameters, rather than the many-time cost of {\it decoding}). We are currently exploring methods to choose $\cpn$ with lower computational complexity.
\end{remark}

\begin{remark}
\label{remark:ub}
Due to Lemma~\ref{lem:sensi:para:bound}, Theorem~\ref{thm:main} guarantees that our scheme requires at most ${\cal O}\left(d^{2+o(1)}\log \left(\frac n \varepsilon\right) \right)$ tests for $(1-\varepsilon)$-reliable recovery. However, as noted in Example~\ref{ex:dwf}, the universal bound on $\fH$ presented in Lemma~\ref{lem:sensi:para:bound} may be loose -- for the class of test functions in Example~\ref{ex:dwf}, the number of tests required by our scheme actually scales as ${\cal O}\left(d^{1+w}\log \left(\frac {n}{\varepsilon}\right) \right)$.
\end{remark}

\subsection{Converse/Lower bound}\label{subsec:conv}
To complement our achievability result in Theorem~\ref{thm:main}, we also present an information-theoretic lower bound on the number
of tests required by any non-adaptive group testing algorithm  that has a probability of error of at most $\varepsilon$. To this end, it is useful to define the ``concentration parameter" $\fh$ of a given monotone function $f(\cdot)$. This definition parallels (but is distinct from) the definition of $\fH$ in the previous Section~\ref{subsec:ach}) -- it may be thought of as a measure of concentration of $f(\cdot)$ under hypergeometric sampling. 

Note that for a pool size of $\plsize$, the quantities $\mean(\plsize)$ and $\var(\plsize)$, defined respectively as
\begin{equation}
\label{eq:def:u_v_x}
\begin{aligned}
        \mean(\plsize) &\coloneqq \sum\limits_{\na=0}^{d} \frac{\binom{d}{\na}\binom{n-d}{\plsize-\na}}{\binom{n}{\plsize}} f(\na), \\ 
    \var(\plsize) &\coloneqq \sum\limits_{\na=0}^{d} \frac{\binom{d}{\na}\binom{n-d}{\plsize-\na}}{\binom{n}{\plsize}} (f(\na)-\mean(\plsize))^2, 
\end{aligned}
\end{equation}
\noindent correspond respectively to (the hypergeometrically weighted) mean and variance of the measurement function $f(\cdot)$, given that the pool-size is $\plsize$. That is, conditioned on choosing a random pool of size $\plsize$, these are the mean and variance of the test positivity probability.
{\begin{remark}\label{eq:mu:lb:ub:mono}
We have $f(0) \leq \mean(\plsize)\leq f(d) $ by the monotonicity of $f(\cdot)$.
\end{remark}}


\begin{customdef}{9}
For test function $f(\cdot)$ and $\plsize\in \{0,1,\dots, n\}$, we define the concentration parameter $\fh$ of $f(\cdot)$ as 
\begin{align}
\label{eq:def:concen:para}
    \fh:=\min_{\plsize\in \{1,\dots,n-1\}}\frac{\mean(\plsize)\left(1-\mean(\plsize)\right)}{\var(\plsize)},
\end{align}
\end{customdef}


We are now in a position to state our main converse/impossibility result (whose proof can be found in Section~\ref{sec:proof:converse}
) as follows:
\begin{theorem} \label{thm:converse}
For any non-adaptive group testing algorithm that ensures a reconstruction error of at most $\varepsilon$, the number of tests $T$ must satisfy  
\begin{equation}
\label{eq:T:lower:bound}
    T \geq \frac{1}{\log e} \fh \left( (1-\varepsilon)\log\binom{n}{d} - 1 \right).
\end{equation}
\end{theorem}
\begin{remark}
\label{remark:lb}
Note also that for any test function $\fh\geq 1$. This is because $f(a)\leq 1$ for all $a$, so using the identity  $(f(\na) - \mean(\plsize))^2=f^2(\na) +\mean^2(\plsize)-2f(\na)\mean(\plsize)$, we have that for all $\plsize$,
\begin{align*}
\var(\plsize) =\sum\limits_{\na=0}^{d} \frac{\binom{d}{\na}\binom{n-d}{\plsize-\na}}{\binom{n}{\plsize}} f^2(\na)-\mean^2(\plsize)\leq \sum\limits_{\na=0}^{d} \frac{\binom{d}{\na}\binom{n-d}{\plsize-\na}}{\binom{n}{\plsize}} f(\na)-\mean^2(\plsize)=\mean(\plsize)\left(1-\mean(\plsize)\right).
\end{align*}
Hence the lower bound in \eqref{eq:T:lower:bound} scales as $ \Omega \left(\log\binom{n}{d}\right)$, which in turns scales as $ \Omega \left(d\log\left(\frac {n} {d}\right)\right)$. However, as noted in Section~\ref{subsec:comp} (see Corollary~\ref{coro:special:cases}.\ref{coro_linear} below), the bound $\fh \geq 1$ is in general loose -- there exist test functions  for which $\fh$ can be as large as $\Omega(d)$. Hence, due to the $\fh$ term in~\eqref{eq:T:lower:bound}, this impossibility result may scale as $\Omega \left(d^2\log\left(\frac {n} {d}\right)\right)$, which is a strict tightening of the information-theoretic lower bounds $\Omega \left(d\log\left(\frac {n} {d}\right)\right)$ extant in the literature (see for instance~\cite{sid2014}).
\end{remark}

\subsection{Comparison between upper and lower bounds}\label{subsec:comp}
Comparing the lower bound in \eqref{eq:T:lower:bound} with the upper bound in \eqref{eq:thm:T:ub}, we note that their ratio scales order-wise as $\frac{\fH}{\fh}$. Corollary~\ref{coro:special:cases} below (whose proof may be found in Appendix~\ref{app:coro:cases}) demonstrates that for a variety of test functions in the literature $\frac{\fH}{\fh}\in \Theta\left(1\right)$. Thus, our bounds are order-wise tight for those test functions. 
\begin{corollary}
\label{coro:special:cases}
Consider the sparse regime $d=n^\theta, 0\leq\theta <1$.
\begin{enumerate}[label=\alph*)]
    \item \label{coro_classical}For the classical group testing measurement function $f(\cdot)$ given in \eqref{f(x)_classical}, both the $\fH$ and $\fh$ functionals equal $1$, enabling us to recover the well-known fact (see for instance~\cite{sid2014}) that the sample-complexity of classical group-testing is $\Theta(d\log n)$ 
    \item \label{coro_threshold}For the threshold test function, i.e., for some $\ell\in\{0,\dots, d-1\}$, 
\begin{equation} 
\label{eq:threshold:function}
    f(x) = 
    \left\{
    \begin{array}{ll}
    0 & \text{if} \; x \le \ts, \\
    1 & \text{if} \; x > \ts,
    \end{array} 
    \right.
\end{equation}
both the upper bound on the number of tests required for $(1-\varepsilon)$-reliable recovery in Theorem \ref{thm:main} and the corresponding lower bound in Theorem \ref{thm:converse} scale as $\Theta \left(d\log n\right)$. {To the best of our knowledge this is the first order-wise tight characterization of the optimal sample-complexity of threshold group testing.}
\item \label{coro_linear}For the ``linear" test function, i.e., 
\begin{equation}
\label{eq:linear:function}
    f(x) = \frac{x}{d}, \quad x\in\{0,\dots,d\},
\end{equation} 
the upper bound on the number of tests required for $(1-\varepsilon)$-reliable recovery in Theorem~\ref{thm:main} matches (order-wise) the lower bound in Theorem \ref{thm:converse},  both scaling as $\Theta \left(d^2\log n\right)$. Hence, by Lemma~\ref{lem:sensi:para:bound}, this test function is essentially extremal in its sample complexity. 
\end{enumerate}
\end{corollary}



For general (near-noiseless) monotone test functions $f(\cdot)$, while we are not able to show that the sample-complexities in Theorems~\ref{thm:main} and~\ref{thm:converse} match up to constant factors, we nonetheless show in Theorem~\ref{thm:match} below (for which a proof may be found in Section~\ref{sec:proof:match}) that they match up to a $\cO\left( \frac{\pminn}{\mean(\plsize^*) \left( 1-\mean(\plsize^*) \right)} \right)$ factor. 
{Here $\plsize^*$ denotes an optimal solution to \eqref{eq:def:concen:para}, i.e., 
\begin{equation} 
\label{eq:def:optimal-size}
\begin{aligned} 
    \plsize^*&\coloneqq\argmin_{\plsize\in \{1,\dots,n-1\}} \frac{\mean(\plsize)(1-\mean(\plsize))}{\var(\plsize)}
\end{aligned}
\end{equation}
is the pool-size minimizing $\frac{\mean(\plsize)\left(1-\mean(\plsize)\right)}{\var(\plsize)}$.} 

\begin{theorem} \label{thm:match}
In the sparse regime $d=n^\theta, 0\leq\theta <1$, the number of tests required in Theorem \ref{thm:main} is no more than a $\cO\left( \frac{\pminn}{\mean(\plsize^*) \left( 1-\mean(\plsize^*) \right)} \right)$ factor larger than the lower bound presented in Theorem \ref{thm:converse}. In particular, the number of tests required in Theorem \ref{thm:main} is never more than a $\cO\left(d^{1+o(1)}\right)$ factor larger than the information-theoretic lower bound in Theorem~\ref{thm:converse}.
\end{theorem}

Using Theorem \ref{thm:match}, we show in Corollary \ref{coro:noisy:function} below (whose proof is in Appendix \ref{app:coro:noisy}) that for {\it any} noisy test function our non-adaptive scheme is indeed order-wise optimal.
\begin{corollary}
\label{coro:noisy:function}
In the sparse regime $d=n^\theta, 0\leq\theta <1$, the upper and lower bounds are order-wise tight for all noisy test functions.
\end{corollary}

Indeed, we present the following conjecture, whose positive resolution would tightly characterize the order-wise sample-complexity for {\it any} monotone test function.

\begin{conjecture} \label{conj:optimal}
We conjecture that for any monotone test function $f(\cdot)$, $\frac{\min_{\cp \in (0,1)} \fT}{\fh d \log n} \in \Theta (1)$. 
\end{conjecture}

\subsection{Estimating the Exact Number of Defectives}\label{subsec:est-d}
As highlighted in Section \ref{sec:test_design}, our algorithms depend critically on the assumption that the number of defectives $d$ is known {\it a priori}. Moreover, different from the classical group-testing in which most algorithms are robust to small perturbations in the value of $d$, our algorithms require the exact value of $d$. If the value of $d$ is not available, which is likely to be the case in practice, it will be useful to have an algorithm for exactly estimating this. In Appendix \ref{app:esti_d}, we develop such an algorithm, whose performance is summarized in the following result. 

\begin{lemma}
\label{lem:estimation:d}
For any fixed $\esterr\in (0,1)$, there exists an adaptive algorithm that outputs the exact value of $d$ with probability at least $1-\esterr$, using $\cO \left(\left(\fH d\right)^2 \log d \log\left(\frac{\log n}{\esterr }\right)\right)$ tests, where $\fH$ is the sensitivity parameter \eqref{eq:def:sensi:para}. Moreover, it uses at most $2\log (2d)$ stages of adaptivity.
\end{lemma}

The idea of the algorithm may be found in Section \ref{sec:intuition}. The details are given in Appendix \ref{app:esti_d}.

Comparing Lemma \ref{lem:estimation:d} with Theorem \ref{thm:main}, we see the number of tests required by our algorithm for estimating $d$ is about a  $\cO (\fH d)$ factor larger than the number of tests required by our algorithm for estimating $\cD$, because we need an exact estimate of $d$.   Therefore, as future work we are exploring algorithms which can estimate the set $\cD$ even when $d$ is only known up to a (small) multiplicative factor.

\section{Intuition and Proof Sketches}
\label{sec:intuition}
We give here some high-level intuition behind our main results and provide corresponding proof sketches. For readers' convenience, we also provide a road-map of the intermediate results leading to our main results in Figs.~\ref{fig:workflow_1} and \ref{fig:workflow_2_3}.

\def\pgfsysdriver{pgfsys -dvipdfmx.def}
\tikzset{
thm/.style ={
rectangle, 
rounded corners = 5pt, 
minimum width = 100pt, 
minimum height = 20pt,
inner sep = 5pt, 
draw=blue }
}
\tikzset{
lem/.style ={
rectangle, 
rounded corners = 5pt, 
minimum width = 100pt, 
minimum height = 20pt,
inner sep = 5pt, 
draw=black }
}
\tikzset{
prop/.style ={
rectangle, 
rounded corners = 5pt, 
minimum width = 100pt, 
minimum height = 20pt,
inner sep = 5pt, 
draw=green }
}
\tikzset{
mr/.style ={
rectangle, 
rounded corners = 5pt, 
minimum width = 100pt, 
minimum height = 20pt,
inner sep = 5pt, 
draw=red }
}

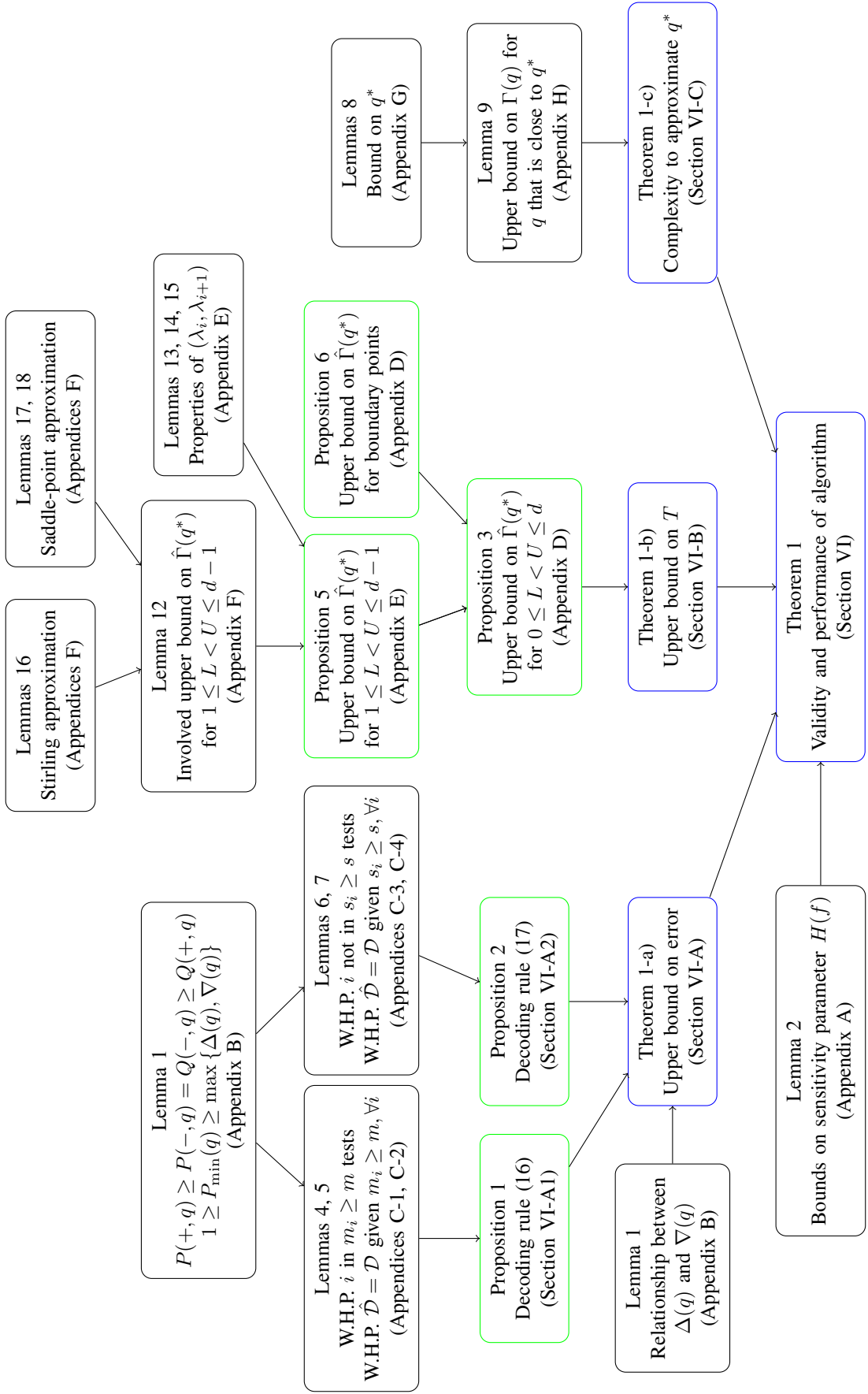
\begin{figure}
\centering
\rotatebox{90}{\begin{tikzpicture}

\node[lem,align=center] (lem2new) at(-8,0) 
{Lemma \ref{lem:sensi:para:bound} \\ Bounds on sensitivity parameter $\fH$ \\ (Appendix \ref{app:sensi:para:bound})};

\node[thm,align=center] (thm1) at(0,0) 
{Theorem \ref{thm:main} \\Validity and performance of algorithm \\ (Section \ref{sec:proof:thm_main})};

\node[thm,align=center] (thm1a) at(-7,2.5) 
{Theorem \ref{thm:main}-\ref{thm1:item1} \\Upper bound on error \\ (Section \ref{sec:proof_thm_1_1})};

\node[thm,align=center] (thm1b) at(0,2.5) 
{Theorem \ref{thm:main}-\ref{thm1:item3} \\Upper bound on $T$ \\ (Section \ref{sec:proof_thm_1_3})};

\node[thm,align=center] (thm1c) at(7.5,2.5) 
{Theorem \ref{thm:main}-\ref{thm1:item4} \\Complexity to approximate $\cpn$ \\ (Section \ref{sec:proof_thm_1_4})};

\node[prop,align=center] (prop1) at(-11,5) 
{Proposition \ref{prop:error_small_q} \\Decoding rule \eqref{eq:classification_small_q} \\ (Section \ref{sec:prop1})};

\node[prop,align=center] (prop2) at(-7,5) 
{Proposition \ref{prop:error_large_q} \\Decoding rule \eqref{eq:classification_large_q} \\ (Section \ref{sec:prop2})};

\node[prop,align=center] (prop3) at(0,5) 
{Proposition \ref{prop:existence_final} \\Upper bound on $\fTpn$ \\ for $0 \le \lw < \up \le d$ \\ (Appendix \ref{sec:existence_final})};

\node[prop,align=center] (prop5) at(-1,7.75) 
{Proposition \ref{prop:existence} \\ Upper bound on $\fTpn$ \\ for $1 \le \lw < \up \le d-1$ \\ (Appendix \ref{sec:existence}) };

\node[prop,align=center] (prop6) at(3,7.75) 
{Proposition \ref{prop:existence_margin} \\ Upper bound on $\fTpn$ \\ for boundary points \\ (Appendix \ref{sec:existence_final}) };

\node[lem,align=center] (lem12) at(-11,7.75) 
{Lemmas \ref{lem:num_tests_lb}, \ref{lem:num_tests_correctness} \\ W.H.P. $i$ in $\nti \ge \nt$ tests \\ W.H.P. $\hat{\mathcal{D}} = \mathcal{D}$ given $\nti \ge \nt,\forall i$ \\ (Appendices \ref{sec:num_tests_lb}, \ref{sec:num_tests_correctness})};

\node[lem,align=center] (lem34) at(-5.8,7.75) 
{Lemmas \ref{lem:num_tests_lb_l}, \ref{lem:num_tests_correctness_l} \\ W.H.P. $i$ not in $\moi \ge \mo$ tests \\ W.H.P. $\hat{\mathcal{D}} = \mathcal{D}$ given $\moi \ge \mo,\forall i$ \\ (Appendices~\ref{sec:num_tests_lb_l}, \ref{sec:num_tests_correctness_l})};

\node[lem,align=center] (lem5) at(-11.5,2.5) 
{Lemma \ref{lem:relation:quantities} \\Relationship between \\ $\pgap$ and $\pgapm$ \\ (Appendix \ref{app:convert_delta_nabla})};

\node[lem,align=center] (lem8_new) at(-8.5,10.5) 
{Lemma \ref{lem:relation:quantities} \\ $\pp \geq \pn = \pnm \geq \ppm$ \\ $1\geq \pmin \geq \max \left\{\pgap, \pgapm\right\}$ \\ (Appendix \ref{app:convert_delta_nabla}) };

\node[lem,align=center] (lem8) at(-1,10.5) 
{Lemma \ref{lem:existence_pre} \\ Involved upper bound on $\fTpn$ \\ for $1 \le \lw < \up \le d-1$ \\ (Appendix \ref{proof:lem:existence_pre}) };

\node[lem,align=center] (lem91011) at(3.75,10.5) 
{Lemmas \ref{lem:exi:alp}, \ref{lem:exi:alp_dif}, \ref{lem:exi:slp} \\ Properties of $(\sigi , \sigio)$ \\ (Appendix \ref{sec:existence}) };

\node[lem,align=center] (lem121314) at(2.5,13) 
{Lemmas \ref{lem:int_j_alpha}, \ref{lem:int_U_L} \\ Saddle-point approximation \\ (Appendices \ref{proof:lem:existence_pre})};

\node[lem,align=center] (lemstirling) at(-2,13) 
{Lemmas \ref{lem:stirling} \\ Stirling approximation \\ (Appendices \ref{proof:lem:existence_pre})};

\node[lem,align=center] (lemc7) at(7.5,7.5) 
{Lemmas \ref{lem:q_range} \\ Bound on $\cpn$ \\ (Appendix \ref{app:lem_q_range}) };

\node[lem,align=center] (lemc8) at(7.5,5) 
{Lemma \ref{lem:fT_error} \\ Upper bound on $\fT$ for \\ $\cp$ that is close to $\cpn$ \\ (Appendix \ref{app:lem_fT_error}) };

\draw[->] (thm1a) --(thm1);
\draw[->] (thm1b) --(thm1);
\draw[->] (thm1c) --(thm1);
\draw[->] (lem2new) --(thm1);
\draw[->] (prop1) --(thm1a);
\draw[->] (prop2) --(thm1a);
\draw[->] (lem5) --(thm1a);
\draw[->] (lem12) --(prop1);
\draw[->] (lem34) --(prop2);
\draw[->] (prop3) --(thm1b);
\draw[->] (prop5) --(prop3);
\draw[->] (prop6) --(prop3);
\draw[->] (prop5) --(prop3);
\draw[->] (lem8) --(prop5);
\draw[->] (lem91011) --(prop5);
\draw[->] (lem121314) --(lem8);
\draw[->] (lem8_new) --(lem12);
\draw[->] (lem8_new) --(lem34);
\draw[->] (lemc7) --(lemc8);
\draw[->] (lemc8) --(thm1c);

\draw[->] (lemstirling) --(lem8);

\end{tikzpicture}}
\caption{Organization of Propositions, Lemmas, and Theorems for our proof of achievability.}
\label{fig:workflow_1}
\end{figure}


\begin{figure}
\centering
\begin{tikzpicture}

\node[thm,align=center] (thm1b) at(11,0) 
{Theorem \ref{thm:main}-\ref{thm1:item3} \\Upper bound on $T$ \\ (Section \ref{sec:proof_thm_1_3}) };

\node[thm,align=center] (thm2) at(0,0) 
{Theorem \ref{thm:converse} \\Lower bound on $T$ \\ (Section \ref{sec:proof:converse}) };

\node[thm,align=center] (thm3a) at(6,0) 
{Theorem \ref{thm:match} \\Near-optimality \\ (Section \ref{sec:proof:match}) };

\node[prop,align=center] (prop4) at(6,2.3) 
{Proposition \ref{prop:match} \\ Existence of proper $(\lwm,\upm)$ \\ (Section \ref{sec:proof:match}) };

\node[lem,align=center] (lem6) at(0,2.3) 
{Lemma \ref{lem:cv_H} \\ Bound on the mutual \\ information of each test \\ (Appendix \ref{app:lem_cv}) };

\node[lem,align=center] (lem7) at(6,4.6) 
{Lemma \ref{lem:match} \\ Existence of a ``large-increment-interval" \\ (Appendix \ref{app:lem_match}) };

\node[lem,align=center] (lem15) at(0,4.6) 
{Lemma \ref{lem:cv:series} \\ Perturbation inequality \\
on $x\ln(x)$ \\ (Appendix \ref{app:lem_cv}) };

\draw[->] (thm2) --(thm3a);
\draw[->] (thm1b) --(thm3a);
\draw[->] (lem6) --(thm2);
\draw[->] (lem15) --(lem6);
\draw[->] (prop4) --(thm3a);
\draw[->] (lem7) --(prop4);

\end{tikzpicture}
\caption{Organization of Propositions, Lemmas, and Theorems for our proof of converse and its tightness.}
\label{fig:workflow_2_3}
\end{figure}
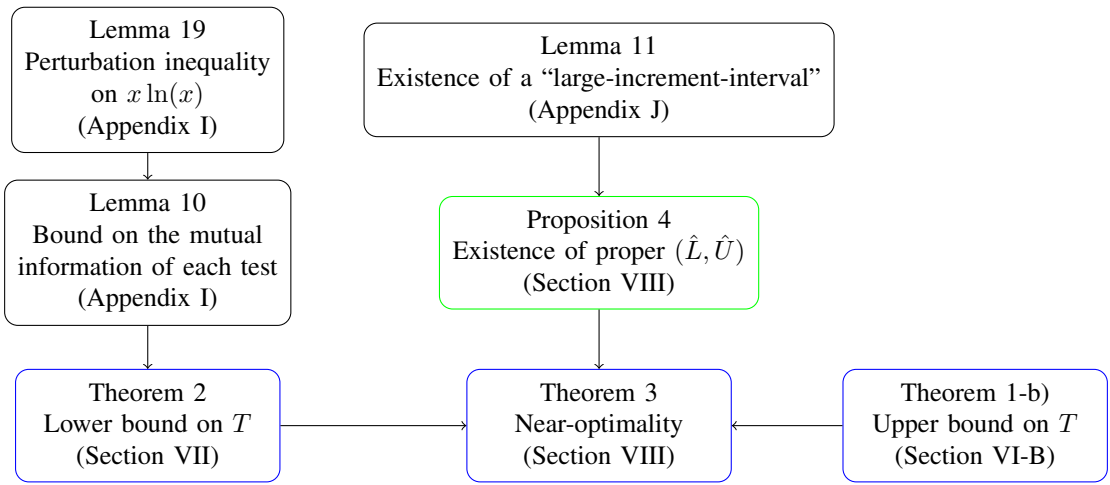
\begin{table*}
    \centering
    \begin{tabular}{|c|l|}
        \hline
        \multicolumn{2}{|c|}{{\bf Parameters in the Problem Formulation}}
        \\\hline
        $\cN$ & The set of all items.
        \\\hline
        $n$ & The total number of items, and $n = |\cN|$.
        \\\hline
        $\cD$ & The unknown subset of defective items, distributed uniformly at random over all $\binom{n}{d}$ sets of size $d$.
        \\\hline
        $d$ & The number of defective items, with $d = |\cD|$.
        \\\hline
        $f(\cdot)$ & The {\it test function}, a monotone function indicating the probability $f(x)$ that a test pool with $x$ defectives has a positive test outcome.
        \\\hline
        $\varepsilon$ & The pre-specified upper bound on the probability of incorrect reconstruction of $\cD$.
        \\\hline
        $T$ & The number of tests.
        \\\hline
        
        \multicolumn{2}{c}{}
        \\\hline
     \multicolumn{2}{|c|}{{\bf Test Design and Decoding Parameters}}
        \\\hline
        $\cM$ & The $T \times n$ binary test matrix: $M_{ji} = 1$ if item $i$ is in test $j$; $M_{ji} = 0$ otherwise.
        \\\hline
        $\cp$ & Probability with which each element in $\cM$ is chosen as $1$ in an i.i.d. manner.
        \\\hline
        $\pn$ & The probability that a test containing $i$ has a positive outcome when $i$ is non-defective, as defined in \eqref{eq:def:p-p+}.
        \\\hline
        $\pp$ &  The probability that a test containing $i$ has a positive outcome when $i$ is defective, as defined in \eqref{eq:def:p-p+}.
        \\\hline
        $\pnm$ & The probability that a test excluding $i$ has a positive outcome when $i$ is non-defective, as defined in \eqref{eq:def:p-pp+p}.
        \\\hline
        $\ppm$ & The probability that a test excluding $i$ has a positive outcome when $i$ is defective, as defined in \eqref{eq:def:p-pp+p}.
        \\\hline
        $\pgap$,$\pgapm$ & The difference of test-positivity probability conditioned on item $i$ being defective or not, as defined in \eqref{eq:def:delta}.
        \\\hline
        $\pmin$ & Minimal test sensitivity, as defined in \eqref{eq:def:Pmin}.
        \\\hline
        $\nt$, $\mo$ & Test participation parameter such that one item is included in $[\nt,T-\mo]$ tests with high probability, as defined in \eqref{eq:def:m} and \eqref{eq:def:mo}.
        \\\hline
        $\fT$ & The number of tests required in our algorithm, as defined in \eqref{eq:def:T}.
        \\\hline
        $\fTp$ & As defined in \eqref{eq:def:fTp}, an upper bound on $\fT$ that is easier to analyze and optimize than $\fT$.
        \\\hline
        $\cpn$ & The parameter that minimizes $\fTp$, as defined in \eqref{eq:def:cpn} -- it can be efficiently approximated by Theorem \ref{thm:main}-\ref{thm1:item4}.
        \\\hline
        
        \multicolumn{2}{c}{}
        \\\hline
        
     \multicolumn{2}{|c|}{{\bf Parameters in the Achievability/Theorem \ref{thm:main}}}
        \\\hline
        $\fH$ & The sensitivity parameter (as defined in \eqref{eq:def:sensi:para}) which helps bound $T$ from above in Theorem \ref{thm:main}-\ref{thm1:item3} and is bounded in Lemma \ref{lem:sensi:para:bound}.
        \\\hline
        $\fTo$& The number of tests required by Decoding Rule 1 in \eqref{eq:classification_small_q}, as defined in \eqref{eq:T_small_q}. 
        \\\hline
        $\fTt$ & The number of tests required by Decoding Rule 2 in \eqref{eq:classification_large_q}, as defined in \eqref{eq:T_large_q}. 
        \\\hline
        $\cph$ & Any $\cp$ that is ``close enough" to $\cpn$ -- the corresponding $\fT$ can be bounded by $\Theta(\fTn)$ in Lemma \ref{lem:fT_error}.
        \\\hline
        $\cphglobal$ & An approximation to $\cpn$. 
        \\\hline
        $\ap$ & A useful parameter w.r.t. $\lw'$ and $\up'$ in the saddle-point approximation in Lemma \ref{lem:int_U_L}, defined as $\frac{1}{2} \min\left\{\up'-\lw',\sqrt{\lw'},\sqrt{d-\up'}\right\}$. 
        \\\hline
        $\bet$ & The minimum term in $\fH$ w.r.t. $\lw$ and $\up$, as defined in \eqref{eq:def:beta}. 
        \\\hline

        \multicolumn{2}{c}{}
        \\\hline
     \multicolumn{2}{|c|}{{\bf Parameters in the Converse/Theorem \ref{thm:converse}}}
        \\\hline
        $\plsize$ & The pool size. 
        \\\hline
        $\mean(\plsize)$,$\var(\plsize)$ & The mean and variance of the test positivity probability, given that the pool size is $\plsize$.
        \\\hline
        $\fh$ & The concentration parameter (as defined in \eqref{eq:def:concen:para}) which helps bound $T$ from below in Theorem \ref{thm:converse}.
        \\\hline
        $\plsize^*$ & The pool size that minimizes $(\mean(1-\mean))/\var$, as defined in \eqref{eq:def:optimal-size}. 
        \\\hline
        $\boldsymbol{\cx}$ & The length-$n$ binary vector that is a weight $d$ vector representing the locations of the defective set $\cD$.
        \\\hline
        $\boldsymbol{\cy}$ &The length-$T$ binary vector representing the outcome of each test. 
        \\\hline
        $\boldsymbol{\cA}$ &The length-$T$ vector representing the number of defectives in each test. 
        \\\hline
        $\boldsymbol{\cxh}$ & The length-$n$ binary vector representing the estimated locations of the defective set $\cD$. 
        \\\hline

        \multicolumn{2}{c}{}
        \\\hline
     \multicolumn{2}{|c|}{{\bf Tightness Parameters/Theorem \ref{thm:match}}}
        \\\hline
        $\mnp$ & The expected number of defectives in a test of pool size $\plsize^*$, as defined in \eqref{eq:def:vartheta}.
        \\\hline
        $(\lwm,\upm)$ & A pair of parameters for $\fH$ such that $\fH$ is upper bounded by $\Theta(1/\var(\plsize^*))$.        
        \\\hline
        $\arb$ & The closest integer to $\mnp$, as defined in \eqref{eq:match:def:eta}, which is one of the pair $(\lwm,\upm)$.
        \\\hline
        $\nan$ & The other one of the pair $(\lwm,\upm)$, whose existence is proved in Lemma \ref{lem:match}.
        \\\hline
        $\apn(\cdot)$ & A parameter similar to $\bet$ defined as $\apn(\nan):=\min\left\{ |\nan-\mnp|, \sqrt{\nan+1}, \sqrt{d-\nan+1}, \sqrt{\mnp+1}, \sqrt{d-\mnp+1} \right\}$.
        \\\hline
        
    \end{tabular} 
    \caption{Table of frequently used notation.} 
\end{table*}

\noindent {\bf A. Achievability/Theorem \ref{thm:main}:} 
As noted in Section~\ref{sec:test_design}  we use a Bernoulli test design, where each item participates in a test in an i.i.d. manner with probability $\cp$. Both of our two decoding rules, specified in~\eqref{eq:classification_small_q} and~\eqref{eq:classification_large_q}, proceed by classifying each item $i \in \{1,\ldots,n\}$ as defective or non-defective independently of any other item.

In particular, Decoding Rule $1$, specified in~\eqref{eq:classification_small_q}, proceeds as follows. For any test including item $i$, we denote by $\pn$ the probability of having a positive outcome if item $i$ is non-defective and by $\pp$ if item $i$ is defective. Due to the monotonicity of our test function $f(\cdot)$, $\pp > \pn$.  By the law of large numbers, when item $i$ participates in a large enough number of tests, the fraction of positive-outcome tests converges to either $\pn$ or $\pp$ depending on whether the item is non-defective or defective respectively. Therefore, decoding rule \eqref{eq:classification_small_q} proceeds by classifying $i$ as defective or not by checking that the fraction of positive test outcomes is closer to $\pp$ or $\pn$. 

Analogously, Decoding Rule 2 specified in~\eqref{eq:classification_large_q} is similar, but now makes use of the tests {\it not} including item $i$, with $\pnm$ denoting the probability of having a positive outcome if item $i$ is non-defective, and $\ppm$ denoting the corresponding probability if item $i$ is defective. As above, due to the monotoncity of $f(\cdot)$, $\pnm > \ppm$. The fraction of positive-outcome tests converges to $\pnm$ and $\ppm$, respectively, and $i$ is classified according to which fraction is closer. 

The reason we have two different decoding rules is since, as shown in Fig.~\ref{fig:T} when $\cp \in (0,1/2]$,
 the first rule requires fewer tests than the second does for $(1-\varepsilon)$-reliable recovery, with the situation reversed in $\cp \in (1/2,1)$.

$\bullet$ \noindent {\it Lemma \ref{lem:sensi:para:bound} -- $\fH$:} Lemma~\ref{lem:sensi:para:bound}, whose proof can be found in Appendix~\ref{app:sensi:para:bound}, helps provide a universal bound on the the sensitivity parameter for any monotone function $f(\cdot)$.

In order to simplify the $\min\left\{\up-\lw , \sqrt{\lw+1} , \sqrt{d-\up+1}\right\}$ term, we first split our analysis into two cases, corresponding to the scenarios $f\left(\left\lceil d/2\right \rceil\right)\geq (f(0)+f(d))/2$ and $f\left(\left\lceil d/2\right \rceil\right) < (f(0)+f(d))/2$ -- we choose $\lw$ and $\up$ from $[0,d/2]$ or $(d/2,d]$ accordingly. This enables us to argue that only one of $\sqrt{\lw+1}$ and $\sqrt{d-\up+1}$ is active in the $\min\left\{\up-\lw , \sqrt{\lw+1} , \sqrt{d-\up+1}\right\}$ term, simplifying our argumentation. We focus on the first case -- the proof of the second case is analogous. Considering any $\ups \in (0,1]$ and $\logu = \lceil \log (1/\ups) + 1 \rceil$, we construct a sequence  $\{\seq_{\ind}\}$ with $\logu+1$ elements such that $\frac{\seq_{\ind+1}^2} {  \seq_\ind+1  } \le 2d^{1+\ups}$ for all $\ind$. We then argue that there exists two adjacent elements $\seq_\exiind$ and $\seq_{\exiind+1}$ such that $f(\seq_{\exiind+1}) - f(\seq_{\exiind}) \ge (f(d)-f(0)) / 2\logu$. Finally, using the definition of $\fH$ in \eqref{eq:def:sensi:para} with $\lw=\seq_{\exiind}$ and $\up=\seq_{\exiind+1}$ and letting $\ups = 1/\log d$, one can prove Lemma \ref{lem:sensi:para:bound}.

$\bullet$ \noindent {\it Theorem~\ref{thm:main}.\ref{thm1:item1} -- Probability of error:
}
We first collect various inequalities relating quantities such as $\pp$, $\pn$, $\ppm$, $\pnm$, etc, in Lemma~\ref{lem:relation:quantities}, whose proof may be found in Appendix~\ref{app:convert_delta_nabla}. With these relations at hand, the probability of error of these two decoding rules can be analyzed via standard concentration inequalities, collected in Appendix~\ref{app:tail-bound}. 

In particular Proposition~\ref{prop:error_small_q}, posits that for Decoding Rule 1, $T \geq \frac{13}{6\cp}\nt$ tests allow for $(1-\varepsilon)$-reliable recovery of the set of $\cD$ of defectives. This is proved in two steps. 
First, in Lemma~\ref{lem:num_tests_lb}, we compute the probability that an arbitrary item participates in less than $\nt$ tests, which, by using the Chernoff bound \cite{chernoff}, is no larger than $\frac{\varepsilon}{2n}$. Second, conditioning on the event that each item participates in at least $\nt$ tests, in Lemma~\ref{lem:num_tests_correctness} we compute the probability of misidentification (either false alarm or missed detection). Using the Chernoff bound, this probability is again no larger than $\frac{\varepsilon}{2n}$. Combining these two parts with a union bound yields Proposition~\ref{prop:error_small_q}. 

Proposition~\ref{prop:error_large_q} similarly analyzes Decoding Rule 2 -- here we show that $T \geq \frac{13}{6(1-\cp)} \mo$ tests allow for $(1-\varepsilon)$-reliable recovery of the set of $\cD$ of defectives. This is proved in a similar manner to Proposition~\ref{prop:error_small_q}. In Lemma \ref{lem:num_tests_lb_l}, we compute the probability that an arbitrary item participates in more than $T-\mo$ tests is no larger than $\frac{\varepsilon}{2n}$. Then, conditioning on the event that each item participates in at most $T-\mo$ tests, in Lemma \ref{lem:num_tests_correctness_l} we compute the probability of misidentification (either false alarm or missed detection) is again no larger than $\frac{\varepsilon}{2n}$. Taking the union bound over these two parts gives us Proposition \ref{prop:error_large_q}. 

Next, we compare the number of tests required by the two decoding rules. This is accomplished by using the identity $\pgapm = \frac{\cp}{1-\cp} \pgap$ shown in Lemma \ref{lem:relation:quantities}. With this identity, we argue that $T$ tests in \eqref{eq:def:T} suffice. 

$\bullet$ \noindent {\it Theorem~\ref{thm:main}.\ref{thm1:item3} -- Bound on $T$:
}
The number of tests chosen in~\eqref{eq:def:T} as $\lceil \fT \rceil$ suffices to ensure $(1-\varepsilon)$-reliable recovery, but it is not immediately apparent how this quantity relates to the bound claimed in~\eqref{eq:thm:T:ub}. As a simplifying first step, as noted in the text surrounding~\eqref{eq:def:fTp}, 
instead of bounding $\fT$ directly, we bound instead $\fTp = \frac{\fT}{\pmin}$ -- since $\pmin \leq 1$ (see Lemma~\ref{lem:relation:quantities}) this suffices to give an upper bound on $T$ allowing $(1-\varepsilon)$-reliable recovery.

Perhaps the most technically involved part of our work focuses on providing a reasonably tight upper bound -- as accomplished in Proposition~\ref{prop:existence_final} -- on $\fTpn$ in terms of the sensitivity parameter $\fH$ defined in Definition~\ref{def:sensi:para}.

Before discussing Proposition~\ref{prop:existence_final} in the context of general monotone functions, consider first the example of the threshold group-testing function described in~\eqref{eq:threshold:function} (corresponding to negative test outcomes if there are at most $\ts$ defectives in a pool).
Intuitively speaking, for accurately classifying each item, we should choose some $q$ such that the gap $\pgap =\pp- \pn$ is ``large" (bounded away from $0$). This can also be seen from \eqref{eq:def:fTp}, wherein to minimize $\fTp$, we would like $\pgap$ to be as large as possible. 
For the threshold group testing scenario, 
if we choose $\cp=\ts/d$, then on the one hand if an item $i$ is non-defective the expected number defectives in a pool is $\ts$; and on the other hand if item $i$ is defective the expected number of  defectives is $ \frac{\ts}{d} (d-1) + 1 \approx \ts+1$, and hence the gap $\pgap$ in test positivity is about as large as can be hoped for -- one can see that choices of $q$ signficantly larger or smaller than this would result in a smaller gap $\pgap$. 

For a general test function $f(\cdot)$, the gap $\pgap$ can be regarded as the ``binomially-weighted mean of the increment'' of $f(\cdot)$ (see~\eqref{eq:convert_delta} for the precise expression). To make $\pgap$ larger, we should attempt to assign a larger weight to a carefully chosen region ``large-increment-interval'' $[\lw,\up]$ where $f(\cdot)$ exhibits a large increment. 

More precisely, for general test functions $f(\cdot)$ our pathway to proving  Proposition \ref{prop:existence_final} relies on bounding the integral of $\pgap$ from $\lw/d$ to $ \up/d$, which, by the mean value theorem, gives a bound on $\pgapns$ for some $\cpns\in [\lw/d,\up/d]$. To this end in Lemmas~\ref{lem:int_j_alpha}~and~\ref{lem:int_U_L} we provide a delicate saddle-point-approximation style bound for $\int_{\lw/d}^{\up/d} \cp^j (1-\cp)^{d-j} \dif \cp$. This together with Stirling's approximation summarized in Lemma~\ref{lem:stirling} results in Lemma~\ref{lem:existence_pre}, which almost gets us to Proposition \ref{prop:existence_final}, except for two issues.

Firstly, there is a $\frac{ (d-\lw') \up' }{(d-\up')\lw'}$ multiplicative term that appears in Lemma~\ref{lem:existence_pre} but not in Proposition \ref{prop:existence_final}. Towards this end, we divide $[\lw',\up')$ into suitably small intervals $[\sigi,\sigio),\; \exi\in \{0,\dots, \ta-1\}$. The intervals are chosen in this manner to satisfy the following two constraints, which are in tension with each other:  
\begin{enumerate}[label=(\roman*)]
    \item In Lemma \ref{lem:exi:alp} -- each $\sigio - \sigi$ should be larger than $\min\{ \up'-\lw' , \sqrt{\lw'} , \sqrt{\up'-\lw'} \}$ that appears in the denominator of Lemma~\ref{lem:existence_pre},
    \item In Lemma \ref{lem:exi:alp_dif} -- each $\sigio - \sigi$ should  also be small enough so that $\frac{ (d-\sigi) \sigio }{(d-\sigio)\sigi}$ can be bounded from above by a constant.
\end{enumerate}
 By substituting $\lw' = \sigind$ and $\up' = \sigindo$ for a specific $\exind$ identified by the mean value theorem in Lemma \ref{lem:exi:slp} allows us to obtain Proposition~\ref{prop:existence}.





Secondly, the ``boundary points", i.e., $\lw=0$ or $\up=d$, are handled separately in Proposition \ref{prop:existence_margin}. 

Proposition \ref{prop:existence_final} is then proved by unifying Propositions \ref{prop:existence} and \ref{prop:existence_margin}. 

$\bullet$ \noindent {\it Theorem~\ref{thm:main}.\ref{thm1:item4} -- Complexity of approximating $\cpn$:
}
A critical part of our test-design and decoding algorithms is an appropriate choice of $\cp \in (0,1)$. We proceed as follows: we uniformly quantize the interval $(0,1)$ into $\Theta(d^4)$ intervals and calculate the corresponding $\fT$ for each $\cp$, and then set $\cp$ to equal the value $\cphglobal$ that minimizes $\fT$. We then prove that choosing $\cp = \cphglobal$ results in our scheme having similar performance to using the value $\cp = \cpn$, i.e., $\fThglobal = \Theta(\fTn)$. 

To this end, in Lemma~\ref{lem:q_range} we first show that $\cpn$ can never be either ``too small" or ``too large" -- i.e., $\cpn \in \left( \frac{1}{376017d^3} , 1-\frac{1}{376017d^3}\right)$. Next, we prove in Lemma~\ref{lem:fT_error} that for all $\cph$ that is close enough to $\cpn$, i.e., $\left|\cph -\cpn\right|\leq \frac{1}{376017d^4}$, we have $\fTh\leq 64e^2\fTn$. 


$\bullet$ \noindent {\it Theorem~\ref{thm:main}.\ref{thm1:item2} -- Complexity of decoding:
}
Since our decoder only needs to count the number of tests and tests with positive outcomes that one item is included in (respectively not included in) and check the ratio via Decoding Rule 1 in \eqref{eq:classification_small_q} (respectively Decoding Rule 2 in \eqref{eq:classification_large_q}), the computational complexity of decoding is $\cO(n T) = \cO \left( n\fH d \log \left( \frac{n}{\varepsilon} \right) \right)$.

\noindent {\bf B. Converse/Theorem \ref{thm:converse}:} The proof of our converse argument, which may be found in Section~\ref{sec:proof:converse}, proceeds as follows.
Let $\boldsymbol{\cx}$ be the input vector and $\boldsymbol{\cy}$ be the outcome vector. We decompose the entropy $H(\boldsymbol{\cx})$ into the conditional entropy term $H(\boldsymbol{\cx}|\boldsymbol{\cy})$ and the mutual information $I(\boldsymbol{\cx};\boldsymbol{\cy})$. By the assumption that $\cD$ is uniformly distributed over all $\binom{n}{d}$ size-$d$ subsets of $\{1,\ldots,n\}$, we have $H(\boldsymbol{\cx})=\log \binom{n}{d}$. Using Fano's inequality, $H(\boldsymbol{\cx}|\boldsymbol{\cy})$ can be bounded in terms of the error probability $\varepsilon$. Following techniques similar to the channel coding literature  (see for instance \cite[Sec. 7.3]{yeung08}), one can upper bound $\mi(\boldsymbol{\cx};\boldsymbol{\cy}) \le \sum_{i=1}^{T} [\etp(\cy_i) - \etp(\cy_i|\cA_i)]$, where $\nod$ denotes the number of defectives in the $i$-th pool. One salient feature of generalized group testing is that the test outcome is no longer deterministic when given the number of defectives in the test pool. That is, $\etp(\cy_i|\cA_i)>0$ and is not negligible. 
Next (by resorting to an inequality on the $\ln(\cdot)$ function presented in Lemma~\ref{lem:cv:series}) in Lemma \ref{lem:cv_H} we bound each $\etp(\cy_i) - \etp(\cy_i|\cA_i) \le \frac{\var(\plsize_i)}{\mean(\plsize_i)(1-\mean(\plsize_i))}$, where $\plsize_i$ is the size of the $i$-th pool, $\mean(\plsize_i)$ and $\var(\plsize_i)$ are, respectively, the mean and variance of being positive. Finally, by optimizing the pool-size parameter $\plsize$, we prove the lower bound.

\noindent {\bf C. Tightness/Theorem \ref{thm:match}:} 
To prove the tightness of our achievability and converse in Theorem~\ref{thm:match} (whose proof may be found in Section~\ref{sec:proof:match}), we show there exists a pair of $(\lwm,\upm)$ such that $\frac{1}{\min\left\{\upm-\lwm , \sqrt{\lwm+1} , \sqrt{d-\upm+1}\right\}}\times \frac{\upm-\lwm}{f(\upm)-f(\lwm)} = \cO \left( \frac{1}{\sigma(\plsize^{*})} \right)$. From the definition of $\plsize^*$ in \eqref{eq:def:optimal-size}, we know that $\var(\plsize^*)$ is ``relatively large". This implies that $f(\cdot)$ increases rapidly in the region adjacent to $\plsize^* \frac{d}{n}$ (which quantity equals the expected number of defectives in the test pool). Therefore it is reasonable to choose $\plsize^* \frac{d}{n}$ as one of the pair $(\lwm,\upm)$. The existence of the other one is shown in a proof by contradiction in Lemma~\ref{lem:match},  making use of the mean and variance formulae for hypergeometric distributions. Based on Lemma \ref{lem:match}, in Proposition \ref{prop:match} we deal with integrality issues and specify $(\lwm,\upm)$. Finally, invoking  Proposition \ref{prop:existence_final} with this pair of $(\lwm,\upm)$, one can prove Theorem \ref{thm:match}.

\noindent {\bf D. Estimation of $d$/Lemma \ref{lem:estimation:d}:} Let $\ud \geq 2$ be a putative number of defective items. We devise a subroutine which decides whether $d\leq \ud-1$ or $d\geq \ud$. The idea of the subroutine is similar to that of the algorithm for deciding whether an item is defective or not in Section \ref{sec:test_design}. More precisely, in a Bernoulli test design where items are included with probability $\lomcp$, denote by $\lompp$ (respectively $\lomppo$) 
the probability of having a positive outcome if there are $\ud$ (respectively $\ud-1$) defectives. By the law of large numbers, when taking a large
enough number of tests, the fraction of positive-outcome tests should be either at least $\lompp$ or at most $\lomppo$, depending on whether
there are at least $\ud$ defectives or at most $\ud-1$ defectives respectively. Therefore, we can reliably estimate whether or not $d$ exceeds $\ud$ by checking the fraction of positive test outcomes. By an analogous argument to that of Lemma~\ref{lem:num_tests_correctness}, one can bound the probability of error of the subroutine. 
By a similar argument to the one in Theorem~\ref{thm:main}.\ref{thm1:item3}, one can find an upper bound on the number of tests used by the subroutine. Using this subroutine, we can obtain an exact estimate of $d$ via the following two steps.
\begin{enumerate}[label=(\roman*)]
    \item To obtain an upper bound on $d$, we perform a sequence of subroutines to check whether $d\leq 2^i$ for $i\in \{1,2, \dots\}$, until the answer is affirmative. 
    \item To find the exact value of $d$, we perform another sequence of subroutines via a binary search. 
\end{enumerate}

\section{Proof of Theorem \ref{thm:main}}
\label{sec:proof:thm_main}
\subsection{Proof of Theorem \ref{thm:main}-\ref{thm1:item1}} \label{sec:proof_thm_1_1}
First, we separately discuss the performance of the decoding rules proposed in \eqref{eq:classification_small_q} and \eqref{eq:classification_large_q}. 

\subsubsection{Performance of Decoding Rule 1 in (\ref{eq:classification_small_q})} \label{sec:prop1}
\begin{proposition} 
\label{prop:error_small_q}
Suppose that the decoding rule used is \eqref{eq:classification_small_q}. Then we show that the probability of error is at most $ \varepsilon$ 
if 
\begin{equation} 
\label{eq:T_small_q}
\nT \ge \fTo := \frac{13}{6\cp} \nt.
\end{equation}
\end{proposition}

This is proved in two steps. 
First, we compute the probability that an arbitrary item participates in less than $\nt$ tests, which can be made sufficiently small. Second, assuming that each item participates in at least $\nt$ tests, we compute the probability of misidentification, which again can be made sufficiently small. Formally, we have the following two lemmas, whose proofs are relegated to Appendices \ref{sec:num_tests_lb} and \ref{sec:num_tests_correctness}, respectively.
\begin{lemma} 
\label{lem:num_tests_lb}
With probability at least $1-\frac{\varepsilon}{2}$ over the test design, each item $i \in {\cal N}$ participates in at least $\nt$ tests.
\end{lemma}
\begin{lemma}
\label{lem:num_tests_correctness}
Conditioning on the event that each item participates in at least $\nt$ tests, with probability at least $1-\frac \varepsilon 2$ over the test design, all items are correctly identified using \eqref{eq:classification_small_q}.
\end{lemma}

Upon combining Lemmas \ref{lem:num_tests_lb} and \ref{lem:num_tests_correctness}, via the union bound, the probability that all items are correctly identified is bounded from below by  $1 - \varepsilon$,
which proves Proposition \ref{prop:error_small_q}.

\subsubsection{Performance of Decoding Rule 2 in (\ref{eq:classification_large_q})} \label{sec:prop2}
\begin{proposition} \label{prop:error_large_q}
Suppose that the decoding rule used is \eqref{eq:classification_large_q}. Then we show that the probability of error is at most $ \varepsilon$ 
if 
\begin{equation} 
\label{eq:T_large_q}
T \ge \fTt := \frac{13}{6(1-\cp)} \mo . 
\end{equation}
\end{proposition}

Similarly to Proposition \ref{prop:error_small_q}, this is established by proving the following two lemmas whose proofs can be found in Appendices~\ref{sec:num_tests_lb_l} and \ref{sec:num_tests_correctness_l}.
\begin{lemma} \label{lem:num_tests_lb_l}
With probability at least $1-\frac{\varepsilon}{2}$ over the test design, each item $i \in \cN$ participates in at most $T-\mo$ tests.
\end{lemma}
\begin{lemma}
\label{lem:num_tests_correctness_l}
Conditioning on the event that each item participates in at most $T-\mo$ tests, with probability at least $1-\frac{\varepsilon}{2}$ over the test design, all items are correctly identified using \eqref{eq:classification_large_q}.
\end{lemma}

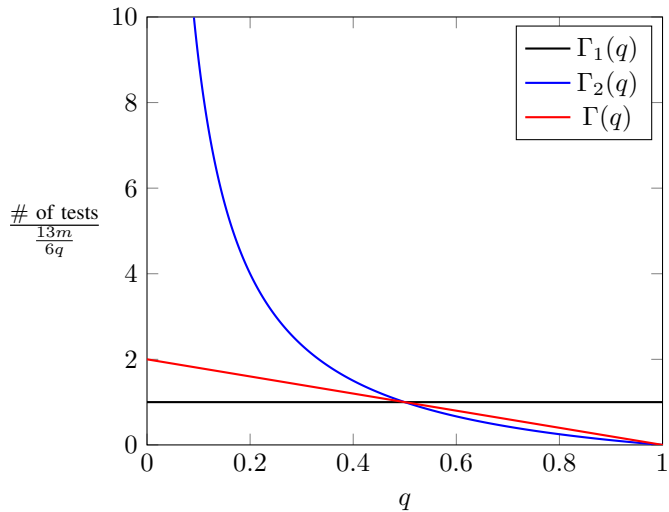
\begin{figure}
\centering
\begin{tikzpicture}
\begin{axis}[ylabel style={font=\large,rotate=-90},
	xmin=0,   xmax=1,
	ymin=0,   ymax=10,
	xlabel=$\cp$,
	ylabel=$\frac {\#\text{ of tests}} {\frac {13\nt}{6\cp}}$
]
\addplot[color=black, thick, samples=2000]{1};
\addlegendentry{$\fTo$}
\addplot[color=blue, domain=0.091:1, thick, samples=2000]{(1-x)/x};
\addlegendentry{$\fTt$}
\addplot[color=red, thick, samples=2000]{2-2*x};
\addlegendentry{$\fT$}
\end{axis}
\end{tikzpicture}
\caption{Comparisons of the number of tests defined in \eqref{eq:def:T}, \eqref{eq:T_small_q} and \eqref{eq:T_large_q}. The black line corresponds to the performance of decoding rule \eqref{eq:classification_small_q}, the blue line corresponds to that of decoding rule \eqref{eq:classification_large_q}, and the red line corresponds to that of our algorithm, which is a combination of the two decoding rules.}
\label{fig:T}
\end{figure}

Next, since both decoding rules are applicable to all $\cp \in (0,1)$, we shall compare the bound given in \eqref{eq:T_small_q} and \eqref{eq:T_large_q} and choose the more efficient one, i.e., the one that has a smaller lower bound on $T$.
Using \eqref{eq:delta:convert} along with \eqref{eq:def:m} and \eqref{eq:def:mo}, we have 
\begin{equation*} 
\label{eq:s_m}
\begin{aligned}
    \frac{13}{6(1-\cp)} \mo &= \frac{13}{6(1-\cp)} \times \frac{8.32\pmin}{(\pgapm)^2} \log \left( \frac{2n}{\varepsilon} \right) \\
    &= \frac{1-\cp}{\cp} \times \frac{13}{6\cp} \times \frac{8.32\pmin}{(\pgap)^2} \log \left( \frac{2n}{\varepsilon} \right) \\
    &= \frac{1-\cp}{\cp} \times \frac{13}{6\cp}\nt, 
\end{aligned}
\end{equation*}
which suggests that the decoding rule \eqref{eq:classification_small_q} is more efficient when $\cp \leq 1/2$, whereas the decoding rule \eqref{eq:classification_large_q} is more efficient when $\cp > 1/2$. 
By leveraging this observation, we argue that the value given in \eqref{eq:def:T} is always an upper bound on the minimum of  \eqref{eq:T_small_q} and \eqref{eq:T_large_q} for all $\cp \in (0,1)$, i.e.,
\begin{align*}
   \min\left\{ \fTo , \fTt \right\} \le \fT ,  \forall \cp \in (0,1).
\end{align*}
To see this, when $\cp \le 1/2$, we have 
\begin{equation*}
    \fTo = \frac{13}{6\cp} \nt \le \frac{13(1-\cp)}{3\cp} \nt = \fT;
\end{equation*}
on the other hand, when $\cp > 1/2$, we have
\begin{equation*}
    \fTt = \frac{13}{6(1-\cp)} \mo = \frac{13(1-\cp)}{6\cp^2}\nt < \frac{13(1-\cp)}{3\cp} \nt = \fT.
\end{equation*}
This is also illustrated in Fig.~\ref{fig:T}, where we plot the values of \eqref{eq:def:T}, \eqref{eq:T_small_q} and \eqref{eq:T_large_q} as a function of $\cp$.

Thus, for $\cp \leq 1/2$ (respectively $\cp > 1/2$), the number of tests $T$ given in \eqref{eq:def:T} can output the defective set $\cD$ with probability of error at most $\varepsilon$ using the decoding rule \eqref{eq:classification_small_q} (respectively \eqref{eq:classification_large_q}). This proves Theorem \ref{thm:main}-\ref{thm1:item1}.

\subsection{Proof of Theorem \ref{thm:main}-\ref{thm1:item3}} \label{sec:proof_thm_1_3}

We provide an explicit bound on the value of $\fTp$ defined in \eqref{eq:def:fTp} for some $\cp$, namely Proposition \ref{prop:existence_final} below, whose proof is presented in Appendix \ref{sec:existence_final}. 
\begin{proposition}
\label{prop:existence_final}
There exists some $\cpns\in(0,1)$ such that
\begin{equation} 
\label{eq:existence}
\fTpns \leq 376017 \fH d\log \left( \frac{2n}{\varepsilon } \right),
\end{equation} 
where $\fH$ is the ``sensitivity parameter" defined in \eqref{eq:def:sensi:para}.
\end{proposition}
Combined with \eqref{eq:1>=pmin>max:Delta}, \eqref{eq:def:fTp} and \eqref{eq:def:cpn}, this result yields
\begin{align}
  T=\left\lceil \fTn \right\rceil &= \left \lceil \pminn \fTpn \right \rceil \notag\\
  &\leq \left \lceil \pminn \fTpns \right \rceil \notag\\
  &\leq 376017  \pminn  \fH d\log \left( \frac{2n}{\varepsilon } \right)+1 \label{eq:T:ub:Pmin}\\
  &\leq 376017 \fH d\log \left( \frac{2n}{\varepsilon } \right)+1,
\end{align}
which proves Theorem \ref{thm:main}-\ref{thm1:item3}.
\subsection{Proof of Theorem \ref{thm:main}-\ref{thm1:item4}} \label{sec:proof_thm_1_4}

We first show in Lemma \ref{lem:q_range} below, whose proof is presented in Appendix \ref{app:lem_q_range}, that $\cpn$ can never be too small nor too large.
\begin{lemma} \label{lem:q_range}
We have
\begin{equation}
    \cpn \in \left( \frac{1}{376017d^3} , 1-\frac{1}{376017d^3}\right).
\end{equation}
\end{lemma}

Next, we show in Lemma \ref{lem:fT_error} below, whose proof is presented in Appendix \ref{app:lem_fT_error}, that for any $\cp$ that is within $\frac{1}{376017d^4}$ distance of $\cpn$, $\fT$ is bounded from above  by $64e^2 \fTn$. This means that our algorithm still performs well for an estimator of $\cpn$ with small error. 
\begin{lemma} \label{lem:fT_error}
For any $\cph \in \left[ \cpn-\frac{1}{376017d^4} , \cpn+\frac{1}{376017d^4} \right]$, 
\begin{equation}
    \fTh \le 64e^2 \fTn.
\end{equation}
\end{lemma}

Armed with Lemma \ref{lem:fT_error}, we are now ready to describe our algorithm to selecting $\cp$: First, calculate $\fT$ for all $\cp = \frac{j}{376017 d^4}$, $j = 1,2,\cdots, 376017d^4 - 1$. Then choose the one having the smallest $\fT$ value, denoted by $\cphglobal$. It follows that $\fThglobal \le \fTh \le 64e^2\fTn$. The computational complexity of computing binomial coefficients is $\cO(d^2 \log^2 d)$, and the computational complexity of multiplying the binomial coefficient (which comprises of $\cO(d \log d)$ bits) with $\cp^j (1-\cp)^{d-j}$ (which comprises  $\cO(d)$ bits), is $\cO(d^2 \log d)$. According to \eqref{eq:convert_delta}, the computational complexity of computing $\pgap$ is $\cO(d^3 \log^2 d)$. Hence the overall computational complexity of this algorithm is $\cO (d^7 \log^2d)$.

\subsection{Proof of Theorem \ref{thm:main}-\ref{thm1:item2}} \label{sec:proof_thm_1_2}
Given the tests and their outcomes, the computational complexity of counting all $\nti$ (respectively $\moi$) and $\ntip$ (respectively $\moip$) is $\cO (nT)$. Thus, the complexity of decoding is $\cO(nT)$. According to Theorem \ref{thm:main}-\ref{thm1:item3}, the complexity is at most $\cO \left(n \fH d\log \left( \frac{n}{\varepsilon} \right) \right)$.

\section{Proof of Theorem \ref{thm:converse}}
\label{sec:proof:converse}
In this section, for any given monotone test function $f(\cdot)$, we provide an information-theoretic lower bound on the number of tests required by \emph{any} non-adaptive group testing algorithm that is allowed to make an error with probability at most $\varepsilon$. 

Let us first introduce some notation which will be used in the proof. We use a binary vector $\boldsymbol{\cx}\in \{0,1\}^n$ to represent the set $\cN$, where $1$s indicate which items are defective.
To estimate $\boldsymbol{\cx}$, we perform $T$ suitable-designed tests, in which each test must be designed prior to observing any outcomes. Let $\boldsymbol{\cA}=(\cA_1,\dots,\cA_T)$ be a length $T$ vector, where $\cA_i$ denotes the number of defectives in the $i$-th test.
The test outcomes are represented by a binary vector $\boldsymbol{\cy}=(\cy_1,\dots,\cy_T)\in\{0,1\}^T$, where $\cy_i=1$ indicates the outcome of the $i$-th test is positive. We emphasize that $\cA_i$ is independent of $(\cy_1,\dots, \cy_{i-1},\cy_{i+1},\dots,\cy_{T})$. Given the tests and their outcomes, let $\boldsymbol{\cxh}$ be an estimate of $\boldsymbol{\cx}$.

 By standard information-theoretic definitions, we have
\begin{align} \label{eq:cv:etp}
    \etp(\boldsymbol{\cx}) &= \etp(\boldsymbol{\cx}|\boldsymbol{\cy}) + \mi(\boldsymbol{\cx};\boldsymbol{\cy})\notag\\
                           &=\etp(\boldsymbol{\cx}|\boldsymbol{\cy},\boldsymbol{\cxh}) + \mi(\boldsymbol{\cx};\boldsymbol{\cy})\notag\\
                           &\leq \etp(\boldsymbol{\cx}|\boldsymbol{\cxh}) + \mi(\boldsymbol{\cx};\boldsymbol{\cy})
\end{align}
where the second line follows since $\boldsymbol{\cxh}$ is a function of $\boldsymbol{\cy}$; the thrid line follows from the fact that conditioning reduces entropy.
Since the defective set $\cD$ is uniformly distributed over all length $n$ vector of Hamming weight $d$, we have
\begin{equation} \label{eq:cv:H(x)}
    \etp(\boldsymbol{\cx}) = \log \binom{n}{d}.
\end{equation}
By Fano's inequality,
\begin{equation} \label{eq:cv:H(x|xh)}
    \etp(\boldsymbol{\cx} | \boldsymbol{\cxh}) \le 1 + \varepsilon \log\binom{n}{d}.
\end{equation}

Let $\cy^{i-1}:=(\cy_1,\dots,\cy_{i-1})$ and $\cA^{i-1}:=(\cA_1,\dots,\cA_{i-1})$.\footnote{For $i=1$, we follow the convention that $\cy^{i-1}=\cA^{i-1}=\emptyset$.} Similar to channel coding (see for example \cite[Sec. 7.3]{yeung08}), it can be easily verified that \begin{align*}
(\boldsymbol{\cx},\cA^{i-1}, \cy^{i-1}) - \cA_i - \cy_i
\end{align*}
which implies
\begin{align}
\label{eq:markov:y_i:z_i}
(\boldsymbol{\cx}, \cy^{i-1}) - \cA_i - \cy_i.
\end{align}
Following a standard set of inequalities we have
\begin{align}
I(\boldsymbol{\cx};\boldsymbol{\cy})&=\sum\limits_{i=1}^{T} \left[\etp(\cy_i | \cy^{i-1}) - \etp(\cy_i |\boldsymbol{\cx},\cy^{i-1})\right] \notag\\
                                  &\leq \sum\limits_{i=1}^{T} \left[\etp(\cy_i  ) - \etp(\cy_i | \boldsymbol{\cx},\cy^{i-1})\right]  \notag\\
                                  &\leq\sum\limits_{i=1}^{T} \left[\etp(\cy_i  ) - \etp(\cy_i | \boldsymbol{\cx},\cy^{i-1},\cA_i)\right] \notag\\
                                  &= \sum\limits_{i=1}^{T} \left[\etp(\cy_i  ) - \etp(\cy_i | \cA_i)\right] \label{eq:cv:mi_3}
\end{align}
where the first line follows from chain rule; the second and third lines follow since conditioning reduces entropy;
the last line follows from \eqref{eq:markov:y_i:z_i}.


Let $\plsize_i$ denote the pool-size of the $i$-th test and $\Pr(\cA_i=\na)$ denote the probability that $\cA_i=\na$. We have
\begin{align}
    \Pr(\cA_i=\na) &= \frac{\binom{d}{\na}\binom{n-d}{\plsize_i-\na}}{\binom{n}{\plsize_i}}.
\end{align}
For simplicity, we define
\begin{equation}
\begin{aligned} \label{eq:cv:def:p_miu_v}
    \mean(\plsize_i) &\coloneqq \sum\limits_{\na=0}^{d} \Pr(\cA_i=\na) f(\na), \\
    \var(\plsize_i) &\coloneqq \sum\limits_{\na=0}^{d} \Pr(\cA_i=\na) \left(f(\na)-\mean(\plsize_i)\right)^2,
\end{aligned}
\end{equation}
where $\mean(\plsize_i)$ and $\var(\plsize_i)$ denote the mean and variance of $\Pr(\cy_i=1 | \cA_i)$.
It turns out we are able to bound the bracketed term $[\cdot]$ in \eqref{eq:cv:mi_3} as follows:
\begin{lemma}
\label{lem:cv_H}
\begin{equation}
\label{eq:bound:y-y|z}
\etp(\cy_i) - \etp(\cy_i | \cA_i) \leq \frac{\var(\plsize_i) \log e}{\mean(\plsize_i)(1-\mean(\plsize_i))}
\end{equation}
\end{lemma}
\begin{IEEEproof}
See Appendix \ref{app:lem_cv}.
\end{IEEEproof}
Now substituting \eqref{eq:bound:y-y|z} into \eqref{eq:cv:mi_3}, we see that
\begin{align}
     \mi(\boldsymbol{\cx};\boldsymbol{\cy}) 
    &\leq \sum\limits_{i=1}^{T} \frac{\var(\plsize_i)\log e}{\mean(\plsize_i)(1-\mean(\plsize_i))} \notag\\
    &\leq T \times \frac{\var(\plsize^*) \log e}{\mean(\plsize^*)(1-\mean(\plsize^*))},\label{eq:cv:maxH_3}
\end{align}
where the second line follows by defining
\begin{align} \label{eq:cv:def:X*}
\plsize^*:=\argmax_{\plsize\in \{1,\dots,n-1\}} \frac{\var(\plsize)}{\mean(\plsize)(1-\mean(\plsize))}.
\end{align}
The reason for restricting $\plsize\in \{1,\dots,n-1\}$ is that $\var(0)=\var(n)=0$.

Finally, combining \eqref{eq:cv:etp}, \eqref{eq:cv:H(x)}, \eqref{eq:cv:H(x|xh)} and \eqref{eq:cv:maxH_3}
gives us the desired result
\begin{equation}
\label{eq:cv:lower_bound}
    T \geq \frac{\mean(\plsize^*)\left(1-\mean(\plsize^*)\right)}{\var(\plsize^*) \log e} \left( (1-\varepsilon)\log\binom{n}{d} - 1 \right).
\end{equation}
This along with the definition of $\fh$ in \eqref{eq:def:concen:para} completes the proof of Theorem \ref{thm:converse}.

\section{Proof of Theorem \ref{thm:match}}
\label{sec:proof:match}
In this section, we argue that the upper bound in \eqref{eq:thm:T:ub} given by the proposed testing algorithm is at most a $\cO\left( \frac{\pminn}{\mean(\plsize^{*}) \left( 1-\mean(\plsize^{*}) \right)} \right)$ factor larger than the lower bound in \eqref{eq:T:lower:bound}.

Before presenting the proof, let us give some technical results that constitute the basic ingredients of the proof.
To proceed, recall the definitions of $\mean(\plsize), \var(\plsize)$ in \eqref{eq:def:u_v_x} and $\plsize^*$ in \eqref{eq:def:optimal-size}, which we repeat here for convenience: 
\begin{equation}
\label{eq:rep:def:mu:v:chi}
\begin{aligned} 
  \plsize^{*}&\coloneqq\argmin_{\plsize\in \{1,\dots,n-1\}} \frac{\mean(\plsize)(1-\mean(\plsize))}{\var(\plsize)},\\
   \mean(\plsize) &\coloneqq \sum\limits_{\idex=0}^{d} \frac{\binom{d}{\idex}\binom{n-d}{\plsize-\idex}}{\binom{n}{\plsize}} f(\idex), \\ 
    \var(\plsize) &\coloneqq \sum\limits_{\idex=0}^{d} \frac{\binom{d}{\idex}\binom{n-d}{\plsize-\idex}}{\binom{n}{\plsize}} \left(f(\idex)-\mean(\plsize)\right)^2.
\end{aligned}
\end{equation}
Define
\begin{align}
\mnp&:=\plsize^{*}\frac{d}{n}, \label{eq:def:vartheta}\\ 
\arb&:= 
    \left\{
    \begin{array}{ll}
    \lfloor \mnp \rfloor & \text{if} \; \mnp - \lfloor \mnp \rfloor < 0.5, \\
    \lceil \mnp \rceil & \text{if} \; \mnp - \lfloor \mnp \rfloor \ge 0.5.
    \end{array} 
    \right.  \label{eq:match:def:eta}
\end{align}
Since $0<\plsize^{*}<n$, we have $\mnp\in (0,d)$.
\begin{lemma}
\label{lem:match}
$\exists\,  \nan \in \{0,1,\cdots,d-1,d\} \backslash \{\arb\}$ such that
\begin{equation} 
\label{eq:mt:sttmt}
 \left(\apn(\nan) \frac{f(\nan) - f(\arb)}{ \nan-\mnp } \right) ^2  > \frac{\var(\plsize^{*})}{11}, 
\end{equation}
 where $\apn(\nan):=\min\left\{ |\nan-\mnp|, \sqrt{\nan+1}, \sqrt{d-\nan+1}, \sqrt{\mnp+1}, \sqrt{d-\mnp+1} \right\}$.
\end{lemma}
\begin{IEEEproof}
See Appendix \ref{app:lem_match}.
\end{IEEEproof}

The following proposition plays a key role in the proof of Theorem \ref{thm:match}.
\begin{proposition} \label{prop:match}
$\exists\, \lwm,\upm \in \{0,\dots,  d\}$ with $\lwm<\upm$ such that 
\begin{equation}
\label{eq:math:key-equation}
  \left(  \bet \frac{f(\upm)-f(\lwm)}{\upm-\lwm} \right)^2 \geq \frac{\var(\plsize^{*})}{176},
\end{equation}
where $\bet := \min \left\{ \upm-\lwm , \sqrt{\lwm+1} , \sqrt{d-\upm+1} \right\}$ and $\var(\plsize^{*})$ is defined in \eqref{eq:rep:def:mu:v:chi}.
\end{proposition}
\begin{IEEEproof}
Using $\nan$ in Lemma \ref{lem:match}, we construct a pair of $(\lwm,\upm)$ that satisfies \eqref{eq:math:key-equation}. Set
\begin{align}
   \lwm:= \min\{ \nan,\arb \}, \quad  \upm := \max\{ \nan,\arb \}.
\end{align}
It follows that 
\begin{align}
|\upm-\lwm|&=\left|\nan-\arb\right|\label{eq:u-l=k-the} ,\\
\sqrt{\lwm+1}&=\min\left\{ \sqrt{\nan+1}, \sqrt{\arb+1} \right\}   ,\label{eq:l:min:kthe}\\
\sqrt{d-\upm+1}&=\min\left\{\sqrt{d-\nan+1}, \sqrt{d-\arb+1}\right\} , \label{eq:u:max:kthe}\\
\left(f(\upm)-f(\lwm)\right)^2&=\left(f(\nan)-f\left(\arb\right)\right)^2. \label{eq:fu-fl=fk-fo}
\end{align}
From the definition of $\arb$ in \eqref{eq:match:def:eta}, we have $|\arb-\mnp |\leq 0.5$. From the assumption that $\nan \in \{0,1,\cdots,d-1,d\} \setminus \left\{ \arb \right\}$, we have $|\nan-\arb | \geq 1$. Using the triangle inequality, we have 
\begin{align*}
|\nan-\mnp|=|\nan-\arb+\arb-\mnp|\geq |\nan-\arb|-|\arb-\mnp|\geq 0.5\geq |\arb-\mnp |.
\end{align*}
Using this observation together with \eqref{eq:u-l=k-the}, it follows that
\begin{equation}
\label{eq:k-theta:u-l}
    4(\nan - \mnp)^2 \ge 2(\nan - \mnp)^2 + 2(\mnp - \arb)^2 \ge (\nan -\mnp+\mnp- \arb)^2 = (\upm - \lwm)^2.
\end{equation}
Recalling the definition of $\apn(\nan)$ in Lemma \ref{lem:match}, we have 
\begin{align}
\label{ineq:gama:beta}
    \apn(\nan) &= \min\left\{ |\nan-\mnp|, \sqrt{\nan+1}, \sqrt{d-\nan+1}, \sqrt{\mnp+1}, \sqrt{d-\mnp+1} \right\} \notag\\
    &\le \min\left\{ |\nan-\arb| + |\arb-\mnp|, \sqrt{\nan+1}, \sqrt{d-\nan+1}, \sqrt{\mnp+1}, \sqrt{d-\mnp+1} \right\} \notag\\
    &\le \min\left\{ |\nan-\arb| + 1, \sqrt{\nan+1}, \sqrt{d-\nan+1}, \sqrt{\arb+2}, \sqrt{d-\arb+2} \right\} \notag\\
    &\le \min\left\{ 2|\nan-\arb|, 2\sqrt{\nan+1}, 2\sqrt{d-\nan+1}, 2\sqrt{\arb+1}, 2\sqrt{d-\arb+1} \right\} \notag\\
    &= \min\left\{ 2|\upm-\lwm|, 2\sqrt{\lwm+1}, 2\sqrt{d-\upm+1} \right\} \notag\\
    &= 2 \bet,  
\end{align}
where the second line follows from the triangle inequality $|\nan-\mnp|\leq |\nan-\arb| + |\arb-\mnp|$; the third line is because $|\arb-\mnp |\leq 1$;
the fourth line follows from $|\nan-\arb|\geq 1$;
the fifth line follows from \eqref{eq:u-l=k-the}, \eqref{eq:l:min:kthe} and \eqref{eq:u:max:kthe}.
Combining \eqref{eq:fu-fl=fk-fo}, \eqref{eq:k-theta:u-l} \eqref{ineq:gama:beta}, along with \eqref{eq:mt:sttmt} in Lemma \ref{lem:match}, we obtain
\begin{equation}
  \left(  \bet \frac{f(\upm)-f(\lwm)}{\upm-\lwm} \right)^2 \geq \frac {\apn(\nan)^2}{4} \times\frac{(f(\nan)-f(\arb))^2}{4(\nan-\mnp)^2} > \frac{\var(\plsize^{*})}{176}.
\end{equation}
This completes the proof of Proposition \ref{prop:match}.
\end{IEEEproof}

Now we set out to prove Theorem \ref{thm:match}. Recalling the upper bound on $T$ in \eqref{eq:thm:T:ub}, we have
\begin{align*}
     T &\leq 376017  \pminn  \fH d\log \left( \frac{2n}{\varepsilon } \right)+1 \\
       &\leq  376017  \pminn \left(\frac{1}{\min\left\{\upm-\lwm , \sqrt{\lwm+1} , \sqrt{d-\upm+1}\right\}}\times \frac{\upm-\lwm}{f(\upm)-f(\lwm)} \right)^2 d\log \left( \frac{2n}{\varepsilon } \right)+1\\
       &\leq  376017  \pminn \frac{176}{\var(\plsize^*)} d\log \left( \frac{2n}{\varepsilon } \right)+1
\end{align*}
where the second inequality follows from the definition of $\fH$ in \eqref{eq:def:sensi:para} by letting $(\lw,\up)$ therein to be the pair $(\lwm,\upm)$ in Proposition \ref{prop:match}; the last inequality follows from Proposition \ref{prop:match}. This implies that the upper bound in \eqref{eq:thm:T:ub} scales as 
$\cO \left( \frac{\pminn}{\var(\plsize^*)} d \log \left( \frac{2n}{\varepsilon} \right) \right)$.
On the other hand, by the definition of $\fh$ in \eqref{eq:def:concen:para}, the lower bound in \eqref{eq:T:lower:bound} scales as $\Omega\left(\frac{\mean(\plsize^*)\left(1-\mean(\plsize^*)\right)}{\var(\plsize^*)}\log \binom{n}{d}\right)$. By standard arguments via Stirling’s approximation, $\log \binom{n}{d}$ is at least $d \log \frac n d$. Thus, under the assumptions that $d=n^\theta, 0\leq\theta< 1$, the number of tests $T$ required for $(1-\varepsilon)$-reliable recovery in Theorem \ref{thm:main} is up to a $\cO\left( \frac{\pminn}{\mean(\plsize^{*}) \left( 1-\mean(\plsize^{*}) \right)} \right)$ factor larger than the lower bound presented in Theorem \ref{thm:converse}. 

From Remarks \ref{remark:ub} and \ref{remark:lb}, we know that our upper and lower bounds scale as ${\cal O}\left(d^{2+o(1)}\log n\right)$ and $\Omega\left(d\log n\right)$, respectively. Therefore, the number of tests required in Theorem \ref{thm:main} is never more than a ${\cal O}(d^{1+o(1)})$ factor larger than the information-theoretic lower bound in Theorem \ref{thm:converse}.

\section{Simulation}
\label{sec:simulation}
In this section we report the results of our computer simulations to evaluate the performance of our proposed schemes.\footnote{The computing resource we use is an Intel(R) Xeon(R) CPU E5-2699 v4 @ 2.20GHz CPU.}
Our algorithm takes as input
\begin{itemize}
    \item number of items $n$;
    \item number of defectives $d$;
    \item test function $f$;
    \item number of tests $T$.
\end{itemize}
We then run the proposed test design and decoding rule multiple times to evaluate the probability of successful reconstruction.

\textbf{Testing:}
For given $(n,d,f,T)$, we randomly generate an array $\ary$ of length $n$ with $(n-d)$ 0s and $d$ 1s, where 0 represents non-defective and 1 represents defective. Then we choose the parameter $\cp$ accordingly, and randomly generate a $T\times n$ matrix $\mtx$ where each entry is i.i.d. Bernoulli($\cp$). Each row of $\mtx$ corresponds to a distinct test, and each column corresponds to a distinct item. Finally, we compute $\rs =(\rse_1,\dots,\rse_T)=  \mtx  \ary $ and generate the test outcomes according to $f(\rse_i), i=1,\dots, T$.

\textbf{Decoding:}
Depending on the value of $\cp$, we then use the decoding rules \eqref{eq:classification_small_q} or \eqref{eq:classification_large_q}. Let $\simdoc$ be the estimation of the decoder. The test succeeds if $\simdoc = \ary$, and fails otherwise.

\subsection{ Simulation result for threshold test function in Corollary \ref{coro:special:cases}-\ref{coro_threshold}}
\begin{figure*}
    \centering
      \subcaptionbox{\scriptsize{\it Example test function defined in~\eqref{eq:threshold_L=5}: The test outcome is positive if and only if at least $6$ items in a pool are defective.}\label{fig:threshold}}
      {\begin{tikzpicture} [scale=2]
\draw [<->] (0,2) node [left] {$f(x)$} -- (0,0) node [below left] {} -- (3.5,0) node [below] {$x$};
\draw (0,.8) node [left] {$0.5$};
\draw [fill] (0,.8) circle [radius=.025];
\draw (0,1.6) node [left] {$1$};
\draw [fill] (0,1.6) circle [radius=.025];
\draw (0.5,0) node [below] {\small $5$}  (0.8,0) node [below] {\small $6$} (2.8,0) node [below] {$d$};
\draw[thick, red] (0, 0)--(0.5,0);
\draw[thick, red] (0.8,0)-- (0.8,1.6)--(2.8, 1.6);
\draw [fill] (0.5,0) circle [radius=.025] (0.8,0) circle [radius=.025] (2.8,0) circle [radius=.025];
\end{tikzpicture}}
\hfil
 \subcaptionbox{\scriptsize{\it  The $x$-axis plots the number of tests $T$ as a multiple of $d\log n $, and the $y$-axis plots the probability of successful reconstruction, for fixed $n=2000, d=20$. When $T \gtrapprox 19.2d\log n $, the probability of successful reconstruction exceeds $0.99$. }\label{fig:case1:waterfall}}
   { \includegraphics[width=8cm]{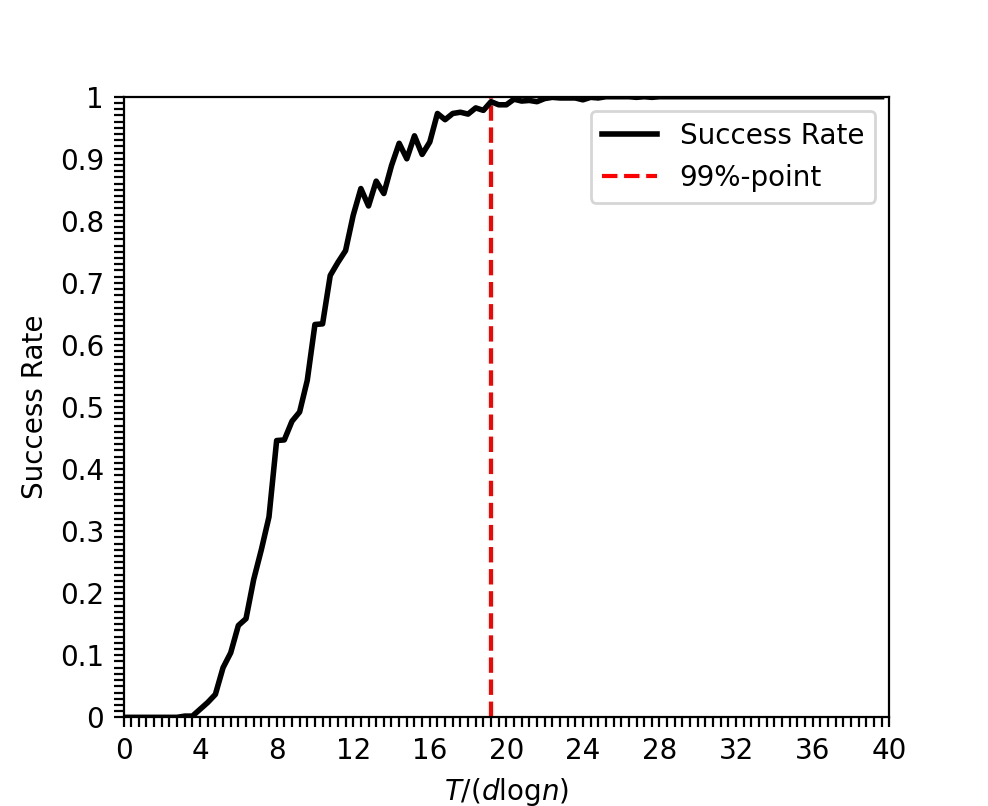}}
   \hfil
    \subcaptionbox{\scriptsize{\it   The $x$-axis corresponds to the number of defectives $d$ ranging from $20$ to $120$, and the $y$-axis corresponds to the number of tests $T$ as a multiple of $d\log n $--the number of items $n$ is fixed to be $2000$. Each pixel is coloured according to the probability of successful reconstruction-- the lighter the colour, the higher the probability of reconstruction success. For each value of $d$, the corresponding red dot in that column represents the number of tests for which this probability first equals $0.99$. The horizontal blue dashed line indicates that when $T \gtrapprox 17.2d\log n $, the probability of successful reconstruction generally exceeds $0.99$.} \label{fig:case1:heatmap:d}}
       { \includegraphics[width=8cm]{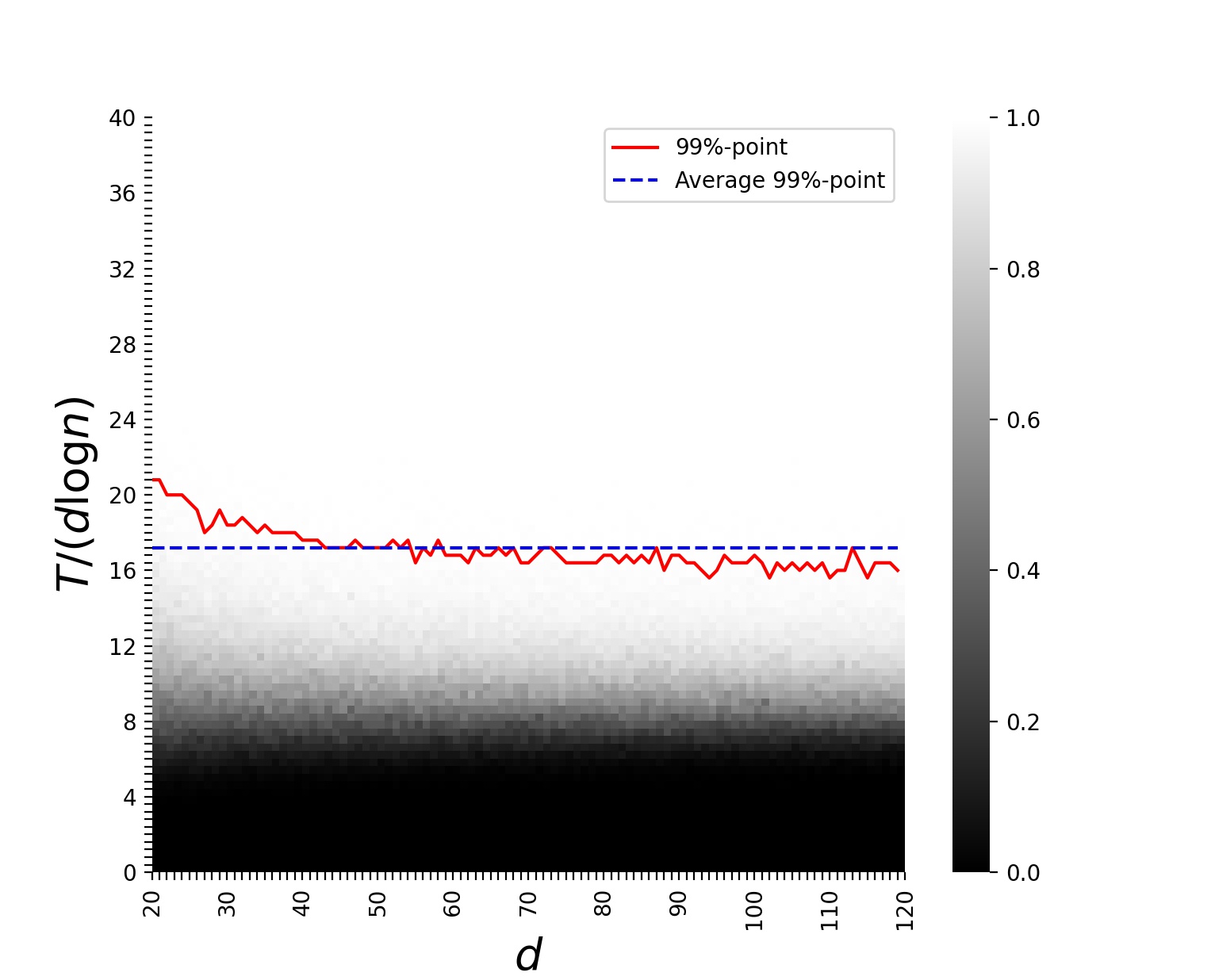}}
  \hfil
    \subcaptionbox{\scriptsize{\it The $x$-axis denotes the number of items $n$ ranging from $2000$ to $6000$, and the $y$-axis denotes the number of tests $T$ as a multiple of $d\log n $--the number of defectives $d$ equals $20$. Each pixel is coloured according to the probability of successful reconstruction -- the lighter the colour, the higher the probability of reconstruction success. For each value of $n$, the corresponding red dot in that column represents the number of tests for which this probability first equals $0.99$. The horizontal blue dashed line indicates that when $T \gtrapprox 20.4 d\log n $, the probability of successful reconstruction generally exceeds $0.99$.} \label{fig:case1:heatmap:n}}
       { \includegraphics[width=8cm]{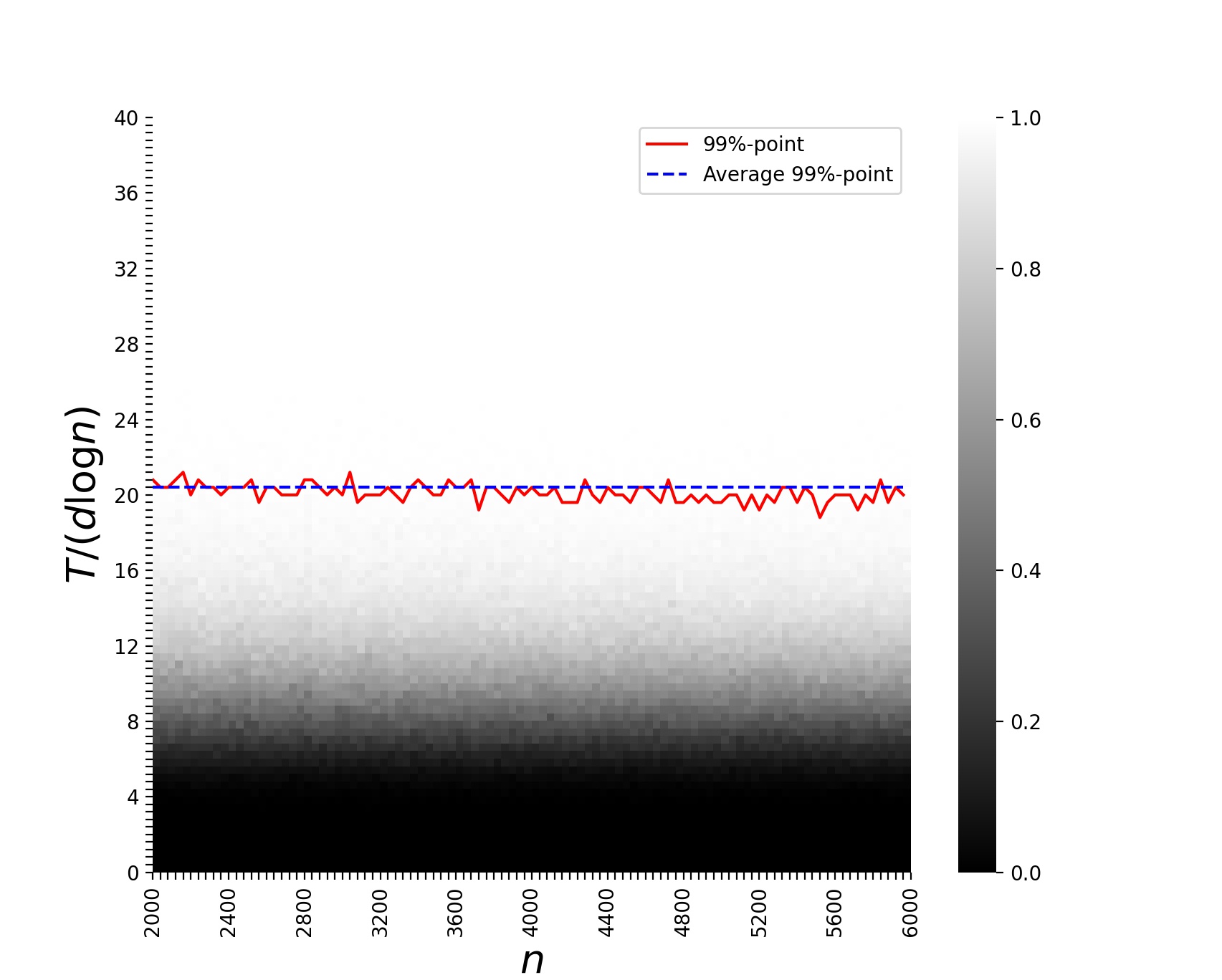}}       
       \caption{{\it Simulation result for threshold test function in Corollary \ref{coro:special:cases}-\ref{coro_threshold}.}}
    \label{sim:fig:threshold}
\end{figure*}

Consider the function $f(\cdot)$ defined in \eqref{eq:threshold:function} for $\ts = 5$ i.e., 
\begin{equation} \label{eq:threshold_L=5}
    f(x) = 
    \left\{
    \begin{array}{ll}
    0 & \text{if} \; x \le 5, \\
    1 & \text{if} \; x > 5,
    \end{array} 
    \right.
\end{equation}
which is also illustrated in Figure~\ref{fig:threshold}. In Corollary \ref{coro:special:cases}-\ref{coro_threshold}, we have shown that our algorithm is order-wise optimal and the number of tests $T$ scales as $\Theta (d \log n)$. For ease of implementation, we assign $\cp = \frac{5}{d}$. In the waterfall plot in Figure \ref{fig:case1:waterfall}, $n=2000$, $d=20$, the $x$-axis plots the number of tests $T$ ranging from $\tmin = 0$ to $\tmax = \left\lfloor 40d \log n  \right\rfloor$ with step size $\tgap = \left\lfloor \frac{\tmax - \tmin}{100} \right\rfloor$, and the $y$-axis plots the probability of successful reconstruction calculated by $1000$ trials for each test $ T = \tmin + \vari \times \tgap , \vari \in \{0,1,\dots,100\}$. When $T \gtrapprox 19.2d\log n $, the probability of successful reconstruction generally exceeds $0.99$.\footnote{Here and below, ``generally exceeds'' means that the average of itself and two tests prior to it is larger than $0.99$.} In the heat-map in Figure \ref{fig:case1:heatmap:d}, $n=2000$, the $x$-axis denotes the number of defectives $d$ ranging from $20$ to $120$, and the $y$-axis denotes the number of tests $T$ as a multiple of $d\log n $. In the heat-map in Figure \ref{fig:case1:heatmap:n}, $d=20$, the $x$-axis corresponds to the number of items $n$ ranging from $2000$ to $6000$, and the $y$-axis corresponds to the number of tests $T$ as a multiple of $d\log n $. In both Figures \ref{fig:case1:heatmap:d} and \ref{fig:case1:heatmap:n}, each pixel is coloured according to the probability of successful reconstruction calculated by $1000$ trials for each test $T$-- the lighter the colour, the higher the probability of reconstruction success. For each value of $d$ (respectively $n$) in Figure \ref{fig:case1:heatmap:d} (respectively Figure \ref{fig:case1:heatmap:n}), the corresponding red dot in that column represents the number of tests for which this probability first equals $0.99$. The horizontal blue dashed line indicates that when $T \gtrapprox 17.2d\log n $ (respectively $20.4d\log n $), the probability of successful reconstruction generally exceeds $0.99$.

\begin{figure*}
    \centering
      \subcaptionbox{\scriptsize{\it Example test function $f$ defined in~\eqref{eq:linear:function}: The probability that test outcome is positive increases linearly.}\label{fig:f:case2}}
      {\begin{tikzpicture}[scale=2] 
\draw [<->] (0,2) node [left] {$f(x)$} -- (0,0) node [below left] {} -- (3.5,0) node [below] {$x$};
\draw (0,0) node [left] {$0$};
\draw (0,0) node [below] {$0$};
\draw [fill] (0,0) circle [radius=.025];
\draw (0,.8) node [left] {$0.5$};
\draw [fill] (0,.8) circle [radius=.025];
\draw (0,1.6) node [left] {$1$};
\draw [fill] (0,1.6) circle [radius=.025];
\draw[thick, red] (0, 0)--(2.8, 1.6);
\draw (1.4,0) node [below] {$\frac{d}{2}$};
\draw [fill] (1.4,0) circle [radius=.025];
\draw (2.8,0) node [below] {$d$};
\draw [fill] (2.8,0) circle [radius=.025];
\end{tikzpicture}}
\hfil
 \subcaptionbox{\scriptsize{\it  The $x$-axis plots the number of tests $T$ as a multiple of $d^2\log n $, and the $y$-axis plots the probability of successful reconstruction, for fixed $n=2000$ and $ d=20$. When $T \gtrapprox 21.6d^2\log n $, the probability of successful reconstruction exceeds $0.99$. }\label{sim:fig:waterfall_linear}}
   { \includegraphics[width=8cm]{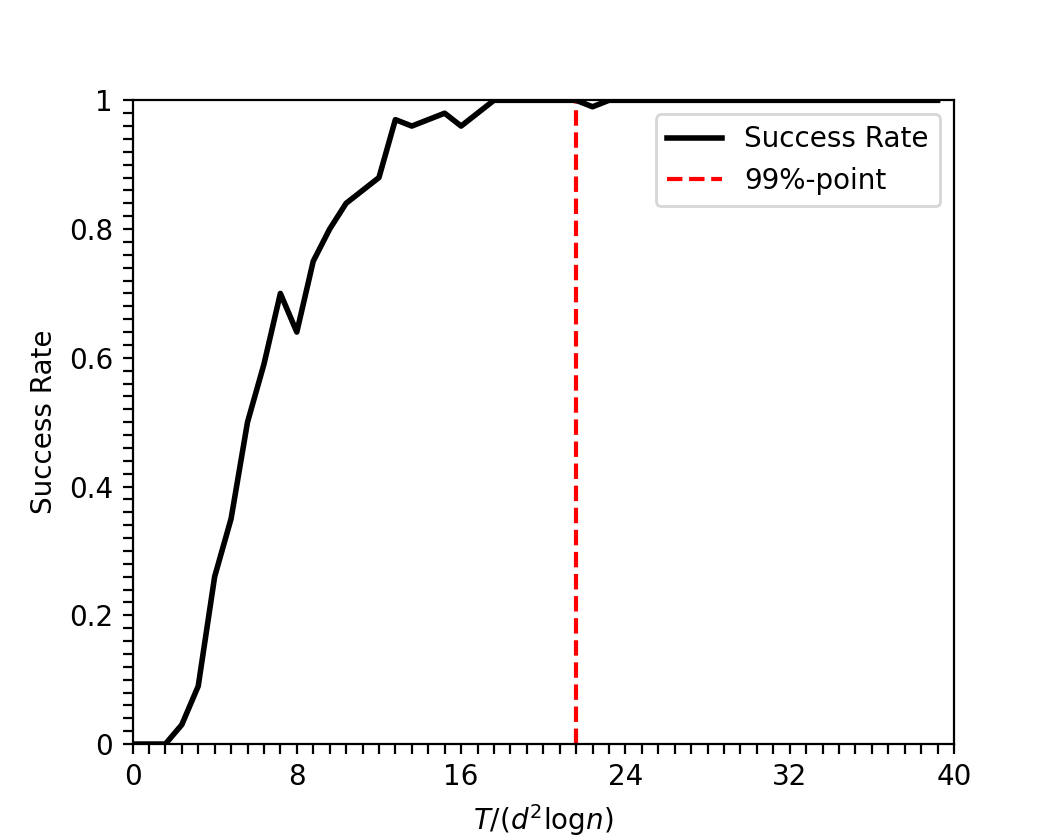}}
   \hfil
    \subcaptionbox{\scriptsize{\it  The $x$-axis corresponds to the number of defectives $d$ ranging from $20$ to $70$, and the $y$-axis corresponds to the number of tests $T$ as a multiple of $d^2\log n $, for fixed number of items $n=2000$. Each pixel is coloured according to the probability of successful reconstruction-- the lighter the colour, the higher the probability of reconstruction success. For each value of $d$, the corresponding red dot in that column represents the number of tests for which this probability first equals $0.99$. The horizontal blue dashed line indicates that when $T \gtrapprox 20.0d^2\log n $, the probability of successful reconstruction generally exceeds $0.99$.} \label{sim:fig:heatmap_linear_d}}
       { \includegraphics[width=8cm]{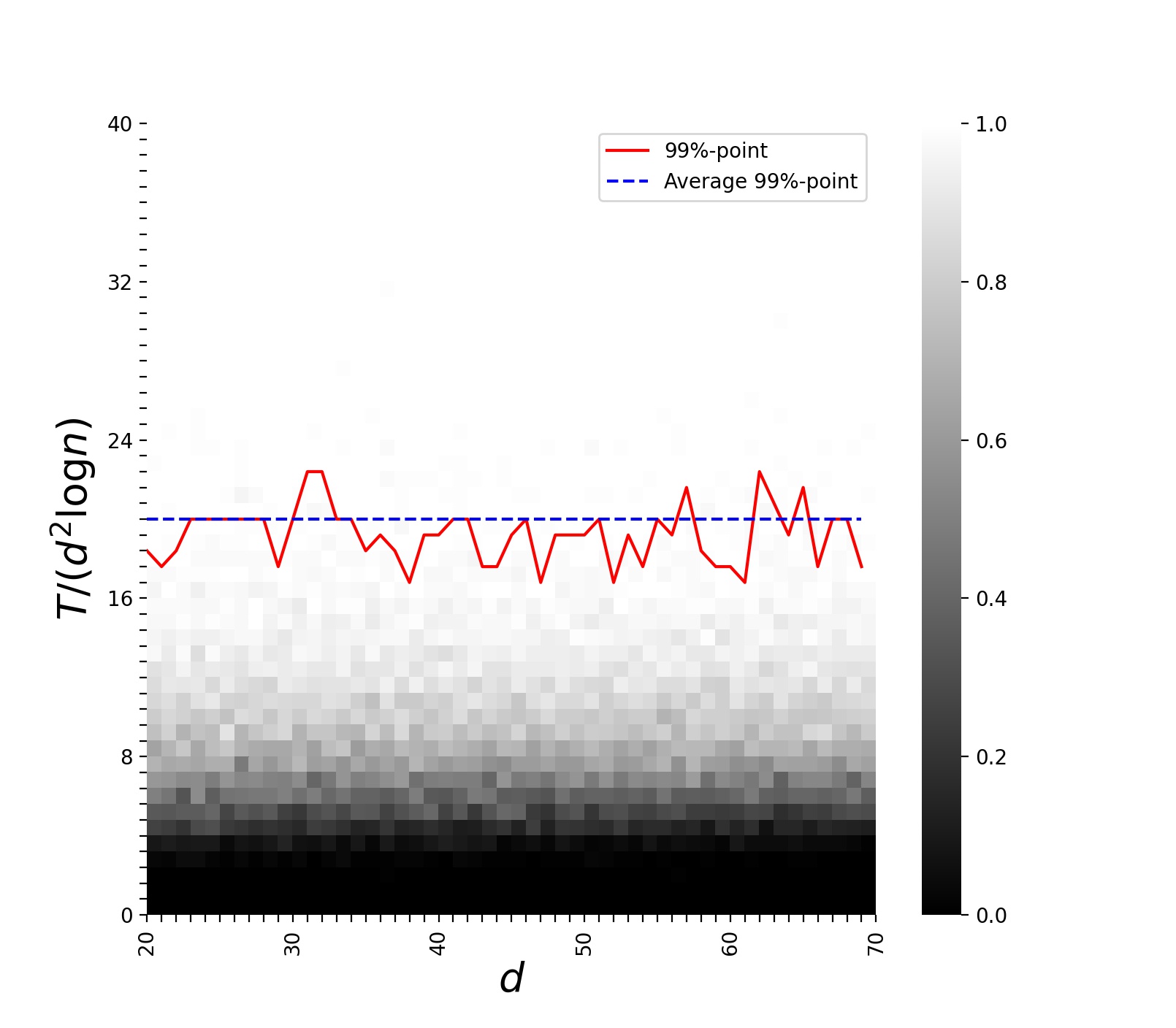}}
  \hfil
    \subcaptionbox{\scriptsize{\it The $x$-axis denotes the number of items $n$ ranging from $2000$ to $4000$, and the $y$-axis denotes the number of tests $T$ as a multiple of $d^2\log n $, for fixed number of defectives $d=20$. Each pixel is coloured according to the probability of successful reconstruction -- the lighter the colour, the higher the probability of reconstruction success. For each value of $n$, the corresponding red dot in that column represents the number of tests for which this probability first equals $0.99$. The horizontal blue dashed line indicates that when $T \gtrapprox 19.6 d^2\log n $, the probability of successful reconstruction generally exceeds $0.99$.} \label{sim:fig:heatmap_linear_n}}
       { \includegraphics[width=8cm]{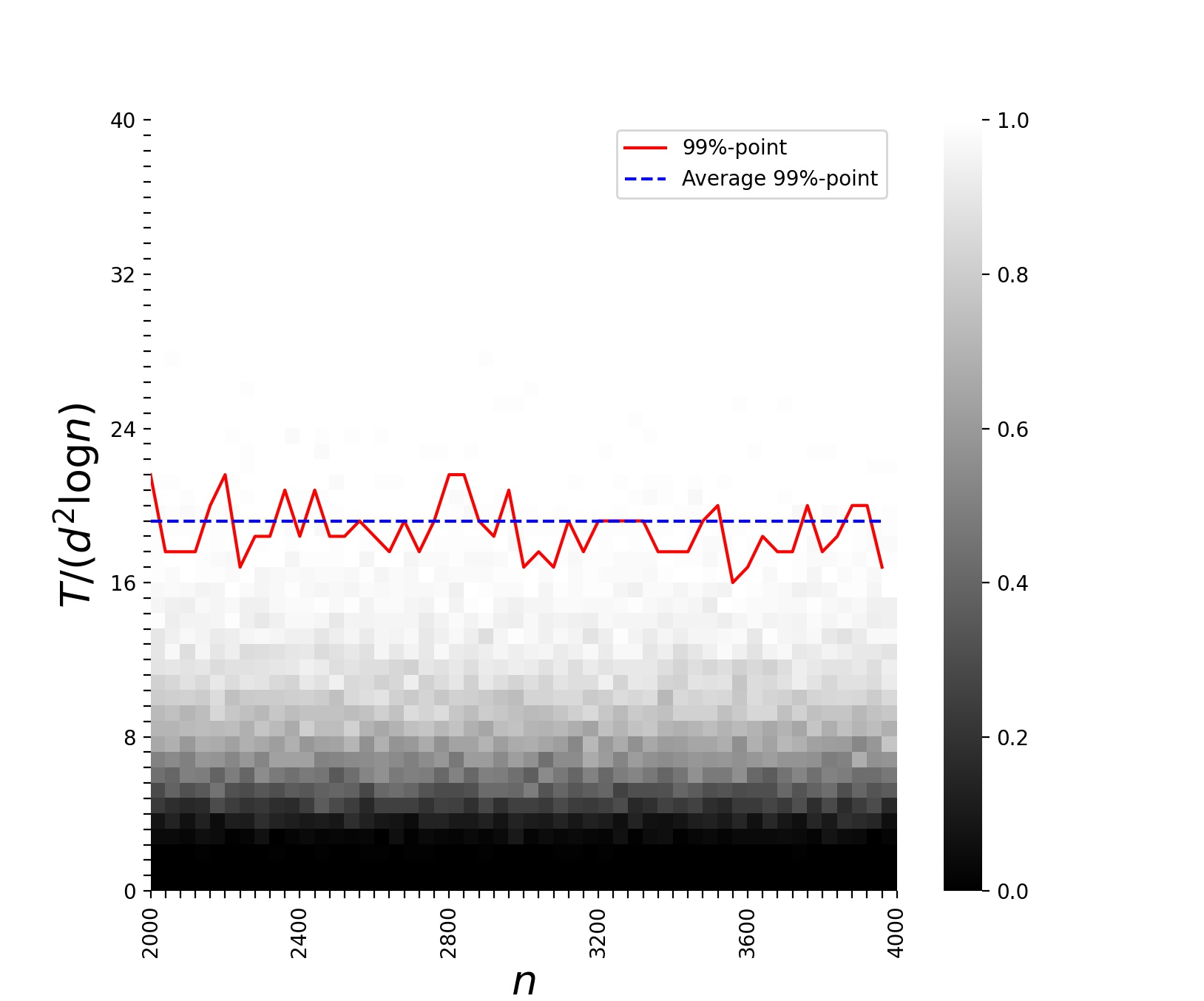}}       
       \caption{{\it Simulation result for linear test function in Corollary \ref{coro:special:cases}-\ref{coro_linear}.}}
    \label{sim:fig:linear}
\end{figure*}

\subsection{ Simulation result for linear test function in Corollary \ref{coro:special:cases}-\ref{coro_linear}}
We now consider the function $f(\cdot)$ defined in \eqref{eq:linear:function} and illustrated in Figure~\ref{fig:f:case2}. In Corollary \ref{coro:special:cases}-\ref{coro_linear}, we have shown that our algorithm is order-wise optimal and the number of tests $T$ scales as $\Theta (d^2 \log n)$. For ease of implementation, we assign $\cp = \frac{1}{2}$. In the waterfall plot in Figure \ref{sim:fig:waterfall_linear}, $n=2000$, $d=20$, the $x$-axis plots the number of tests $T$ ranging from $  \tmin '= 0$ to $\tmax '= \left\lfloor 40d^2 \log n  \right\rfloor$ with step size $\tgap' = \left\lfloor \frac{\tmax - \tmin}{100} \right\rfloor$, and the $y$-axis plots the probability of successful reconstruction calculated by $100$ trials for each test $ T = \tmin' + \vari \times \tgap' , \vari \in \{0,1,\dots,100\}$. When $T \gtrapprox  21.6  d^2 \log n$, the probability of successful reconstruction exceeds $0.99$. In the heat-map figure \ref{sim:fig:heatmap_linear_d}, $n=2000$, the $x$-axis denotes the number of defectives $d$ ranging from $20$ to $70$, and the $y$-axis denotes the number of tests $T$ as a multiple of $d^2\log n $. In the heat-map figure \ref{sim:fig:heatmap_linear_n}, $d=20$, the $x$-axis corresponds to the number of items $n$ ranging from $2000$ to $4000$, and the $y$-axis corresponds to the number of tests $T$ as a multiple of $d^2\log n $. In both figures \ref{sim:fig:heatmap_linear_d} and \ref{sim:fig:heatmap_linear_n}, each pixel is coloured according to the probability of successful reconstruction calculated by $100$ trials for each test $T$-- the lighter the colour, the higher the probability of reconstruction success. For each value of $d$ (respectively $n$) in figure \ref{sim:fig:heatmap_linear_d} (respectively figure \ref{sim:fig:heatmap_linear_n}), the corresponding red dot in that column represents the number of tests for which this probability first equals $0.99$. The horizontal blue dashed line indicates that when $T \gtrapprox 20.0d^2\log n $ (respectively $19.6d^2\log n $), the probability of successful reconstruction generally exceeds $0.99$.

\newpage
\subsection{Simulation result for Conjecture \ref{conj:optimal}}

\begin{figure*}
    \centering
      \subcaptionbox{\scriptsize{\it Example test function $f$ defined in~\eqref{eq:simu:conj:dw}.}\label{fig:f:d^w}}
      {\begin{tikzpicture}[scale=2] 
\draw [<->] (0,2) node [left] {$f(x)$} -- (0,0) node [below left] {} -- (3.5,0) node [below] {$x$};
\draw (0,0) node [left] {$0$};
\draw (0,0) node [below] {$0$};
\draw [fill] (0,0) circle [radius=.025];
\draw (0,.8) node [left] {$0.5$};
\draw [fill] (0,.8) circle [radius=.025];
\draw (0,1.6) node [left] {$1$};
\draw [fill] (0,1.6) circle [radius=.025];
\draw[thick, red] (0, 0)--(1, 1.6);
\draw[thick, red] (1, 1.6)--(2.8, 1.6);
\draw (1,0) node [below] {$d^{\frac{2}{3}}$};
\draw [fill] (1,0) circle [radius=.025];
\draw (2.8,0) node [below] {$d$};
\draw [fill] (2.8,0) circle [radius=.025];
\end{tikzpicture}}
\hfil
 \subcaptionbox{\scriptsize{\it Performance of our algorithm for the ``partial linear'' function. }\label{sim:fig:d^w}}
    { \includegraphics[width=8cm]{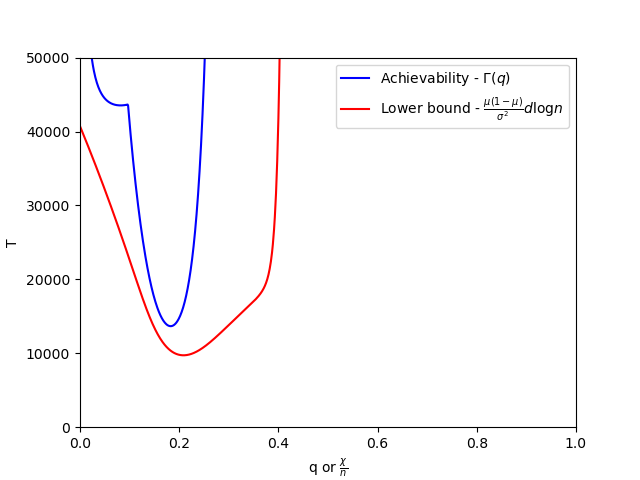}}
  \hfil
    \subcaptionbox{\scriptsize{\it Example test function $f$ defined in~\eqref{eq:simu:conj:sigmoid}. } \label{fig:f:sigmoid}}
      {\begin{tikzpicture}[scale=2] 
\draw [<->] (0,2) node [left] {$f(x)$} -- (0,0) node [below left] {} -- (3.5,0) node [below] {$x$};
\draw (0,0) node [left] {$0$};
\draw (0,0) node [below] {$0$};
\draw [fill] (0,0) circle [radius=.025];
\draw (0,.8) node [left] {$0.5$};
\draw [fill] (0,.8) circle [radius=.025];
\draw (0,1.6) node [left] {$1$};
\draw [fill] (0,1.6) circle [radius=.025];
\draw[smooth,thick,red] plot[domain=0:2.8] (\x,{exp(\x*10/2.8-5)/(1+exp(\x*10/2.8-5))*1.6});
\draw (1.4,0) node [below] {$\frac{d}{2}$};
\draw [fill] (1.4,0) circle [radius=.025];
\draw (2.8,0) node [below] {$d$};
\draw [fill] (2.8,0) circle [radius=.025];
\end{tikzpicture}}
    \hfil
    \subcaptionbox{\scriptsize{\it Performance of our algorithm for the sigmoid function.} \label{sim:fig:sigmoid}}
       { \includegraphics[width=8cm]{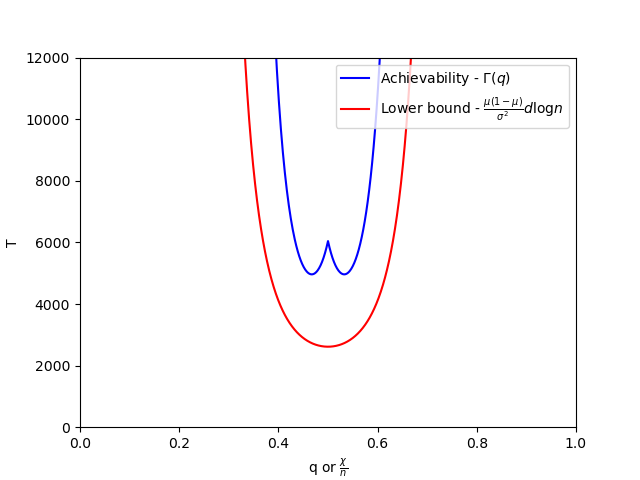}} 
       \caption{{\it Simulation result for Conjecture \ref{conj:optimal}.}}
    \label{sim:fig:conj}
\end{figure*}

We consider two test functions and show the corresponding performance of our algorithm. Consider one specific ``partial linear'' function defined in \eqref{eq:linear-alpha:function} in Example \ref{ex:dwf}:
\begin{equation} \label{eq:simu:conj:dw}
    f(x) = 
    \left\{
    \begin{array}{ll}
    \frac{x}{d^{2/3}} & x \in \left[0, d^{2/3}\right]\cap \mathbb{Z}^+, \\
    1 & \text{otherwise} ,
    \end{array} 
    \right.
\end{equation}
Letting $n=1250$ and $d=125$, the performance of our algorithm is presented in Fig. \ref{sim:fig:d^w}. Then consider the well-known sigmoid function:
\begin{equation} \label{eq:simu:conj:sigmoid}
    f(x) = \frac{e^{\frac{x}{2} - \frac{d}{4}}}{e^{\frac{x}{2} - \frac{d}{4}}+1}.
\end{equation}
Letting $n=2000$ and $d=100$, the performance of our algorithm is presented in Fig. \ref{sim:fig:sigmoid}. One can see from the figures that $\frac{\min_{\cp\in(0,1)} \fT}{\fh d \log n} \le 2$ in both two cases, which supports our conjecture.

 \section*{Acknowledgment}
The authors wish to acknowledge useful discussions with Profs. Oliver Johnson and Jonathan Scarlett. We would also like to thank the anonymous reviewers of the conference version for many helpful suggestions which greatly improved the quality of this work.

\newpage
\appendices
\makeatletter
\renewcommand{\thesubsection}{\thesectiondis-\arabic{subsection}}
\renewcommand{\thesubsectiondis}{\arabic{subsection}.}

\makeatother

\label{appendix}

\section{Proof of Lemma \ref{lem:sensi:para:bound}}
\label{app:sensi:para:bound}

Since $ \min\left\{\up-\lw , \sqrt{\lw+1} , \sqrt{d-\up+1}\right\} \leq \up-\lw $,
it follows from the definition of $\fH$ in \eqref{eq:def:sensi:para} that   
\begin{align*}
     \fH &\geq \min_{0\leq \lw<\up \leq d}\left(\frac{1}{\up-\lw}\times \frac{\up-\lw}{f(\up)-f(\lw)} \right)^2 \\
         &\geq \frac{1}{\left(f(d)-f(0)\right)^2}
\end{align*} 
where the second line follows from the assumption that $f(\cdot)$ is monotonically increasing.

The upper bound is proved by taking into account the ``shape'' of the monotone test function $f(\cdot)$. For any $\ups\in \left(0,1\right]$, define 
\begin{align}
\label{eq:def:k:logu}
    \logu := \left\lceil \log \frac{1}{\ups} +1 \right\rceil.
\end{align}
For notational convenience, let $$\fgap:=f(d)-f(0)>0.$$ There are two possible cases for $f(\cdot)$: 
\begin{enumerate}[label=\roman*)]
    \item $f\left(\left\lceil \frac {d} {2}\right \rceil\right)\geq f(0)+\frac {\fgap}{2}$; \label{f:item:type1}
    \item $f\left(\left\lceil \frac {d} {2}\right \rceil\right) < f(0)+\frac {\fgap}{2}$.   \label{f:item:type2}
\end{enumerate}

Case \ref{f:item:type1}: For $f\left(\left\lceil \frac {d} {2}\right \rceil\right)\geq f(0)+\frac {\fgap}{2}$, define a sequence $\{\seq_{\ind}\}$ such that 
\begin{equation} \label{eq:def:seq}
    \seq_\ind := 
    \left\{
    \begin{array}{ll}
    \left\lfloor \frac{1}{2} d^{1-2^{-\ind+1}} \right\rfloor & \ind = 1,2,\cdots,\logu, \\
    \left\lceil \frac{1}{2} d\right\rceil & \ind = \logu+1.
    \end{array} 
    \right.
\end{equation}
From \eqref{eq:def:seq} we have that for $\ind \in \{1,\cdots,\logu-1\}$, 
\begin{align}
    \frac{\seq_{\ind+1}^2}{\seq_{\ind}+1} \le \left( \frac{d^{1-2^{-\ind}}}{2} \right)^2 \frac{2}{d^{1-2^{-\ind+1}}} = \frac{d}{2}; \label{eq:seq}
\end{align}
for $\ind = \logu$,
\begin{align}
    \frac{\seq_{\ind+1}^2}{\seq_{\ind}+1} \le \left( \frac{d+1}{2} \right)^2 \frac{2}{d^{1-2^{-\logu+1}}} \le d^2 \frac{2}{d^{1-2^{-\logu+1}}} \leq 2d^{1+\ups}, \label{eq:seq_k}
\end{align}
where the second inequality follows the fact that $\frac{d+1}{2}\leq d, \forall d\geq 1 $; the last inequality follows from the definition of $\logu$ in \eqref{eq:def:k:logu}.
Combining \eqref{eq:seq} and \eqref{eq:seq_k}, we see that 
\begin{align}
\label{eq:seq:ratio}
    \frac{\seq_{\ind+1}^2}{\seq_{\ind}+1}  \leq 2d^{1+\ups}, \quad\forall \ind\in \{1,\dots,\logu\}. 
\end{align}
Note that the sequence $\{\seq_{\ind}\}$ is increasing in terms of $\ind$. It then follows that $\seq_{\ind}+\seq_{\ind+1}\leq  \left\lfloor  \frac{1}{2} d\right\rfloor + \left\lceil \frac{1}{2} d\right\rceil=d, \;\forall \ind\in \{1,\dots,\logu\}$. This implies
\begin{align}
    \seq_{\ind}+1 \leq d-\seq_{\ind+1}+1, \quad\forall \ind\in \{1,\dots,\logu\}. \label{eq:remove:d-u+1}
\end{align}
Then we argue that $\exists\exiind \in \{ 1,\cdots,\logu \}$ such that
\begin{equation} 
\label{eq:seq:exiind}
    f(\seq_{\exiind+1}) - f(\seq_{\exiind}) \ge \frac{\fgap}{2\logu}, 
\end{equation}
because, if to the contrary that such a $\exiind$ does not exist, then  
\begin{align*}
   f\left( \left\lceil \frac{d}{2} \right\rceil \right) - f(0)=  f(\seq_{\logu+1}) - f(\seq_1) = \sum_{i=1}^k\left( f(\seq_{\ind+1}) - f(\seq_{\ind})\right) < k \frac{\fgap}{2\logu}=\frac{\fgap}{2}
\end{align*}
contradicts our assumption that $f\left(\left\lceil \frac {d} {2}\right \rceil\right)\geq f(0)+\frac {\fgap}{2}$.

Next, letting $\lw = \seq_{\exiind}$ and $\up = \seq_{\exiind+1}$, we obtain from \eqref{eq:remove:d-u+1} that 
\begin{align*}
    \min\left\{\up-\lw , \sqrt{\lw+1} , \sqrt{d-\up+1}\right\}=\min\left\{ \seq_{\exiind+1}-\seq_{\exiind} , \sqrt{\seq_{\exiind}+1}\right\}
\end{align*}
It then follows from \eqref{eq:def:sensi:para} that 
\begin{align}
 \label{eq:gamma_seq_3}
    \fH&\leq \left(\frac{1}{\min\left\{\seq_{\exiind+1}-\seq_{\exiind} , \sqrt{\seq_{\exiind}+1 }\right \}}\times \frac{\seq_{\exiind+1}-\seq_{\exiind}}{f(\seq_{\exiind+1})-f(\seq_{\exiind})}\right)^2\notag\\ 
    &\leq \frac{4\logu^2}{\fgap^2} \times \max \left\{ 1 , \frac{(\seq_{\exiind+1}-\seq_{\exiind})^2}{\seq_{\exiind}+1} \right\}  \notag\\ 
    &\leq  \frac{4\logu^2}{\fgap^2} \times \max \left\{ 1 , \frac{\seq_{\exiind+1}^2}{\seq_{\exiind}+1} \right\}  \notag\\
    &\leq  \frac{8\logu^2}{\fgap^2} \times d^{1+\ups} \notag\\ 
    &\leq \frac{8}{\fgap^2}\left( \log \frac{1}{\ups}+2 \right)^2 d^{1+\ups} 
\end{align}
where the second line follows from \eqref{eq:seq:exiind}; the fourth line follows from \eqref{eq:seq:ratio}; the last line follows from the definition of $\logu$ in~\eqref{eq:def:k:logu}.

Case \ref{f:item:type2}: For $f\left(\left\lceil \frac {d} {2}\right \rceil\right) < f(0)+\frac {\fgap}{2}$, we define \begin{align}
\label{eq:def:fbar}
    \fbar(x) := 1-f(d-x),\quad x\in\{0,\dots, d\}.
\end{align} 
It follows that
\begin{equation}
    \fbar \left( \left\lceil \frac{d}{2} \right\rceil \right) - \fbar(0) \geq \fbar \left( \left\lfloor \frac{d}{2} \right\rfloor \right) - \fbar(0) 
    =  1 - f\left( d - \left\lfloor \frac{d}{2} \right\rfloor \right)  - \left( 1 - f(d) \right) 
    = f(d) - f\left(\left\lceil \frac {d} {2}\right \rceil\right)  
    > \frac{\fgap}{2}.
\end{equation}
Consider the sequence $\{\seq_{\ind}\}$ defined in \eqref{eq:def:seq}. 
Following the same argument as above, we have that $\exists \exiind\in \{ 1,\cdots,\logu \}$ such that 
\begin{equation} 
\label{eq:fbar:point}
    \fbar(\seq_{\exiind+1}) - \fbar(\seq_{\exiind}) \ge \frac{\fgap}{2\logu}.
\end{equation}
Using \eqref{eq:fbar:point} along with the definition of $\fbar$ in \eqref{eq:def:fbar} implies  
\begin{equation}
    f(d - \seq_{\exiind}) - f(d - \seq_{\exiind+1}) =  1 - \fbar(\seq_{\exiind})  - ( 1 - \fbar(\seq_{\exiind+1}) ) = \fbar(\seq_{\exiind+1}) - \fbar(\seq_{\exiind}) \ge \frac{\fgap}{2\logu}.
\end{equation}
Setting $\lw = d - \seq_{\exiind+1}$ and $\up = d - \seq_{\exiind}$, we have from \eqref{eq:remove:d-u+1} that 
\begin{align*}
     \min\left\{\up-\lw , \sqrt{\lw+1} , \sqrt{d-\up+1}\right\}=\min\left\{ \seq_{\exiind+1}-\seq_{\exiind} , \sqrt{\seq_{\exiind}+1}\right\}
\end{align*}
Similar to the derivation of \eqref{eq:gamma_seq_3}, we have from \eqref{eq:def:sensi:para} that   
\begin{align}
 \fH&\leq \left(\frac{1}{\min\left\{\seq_{\exiind+1}-\seq_{\exiind}, \sqrt{\seq_{\exiind}+1 }\right \}}\times \frac{d-\seq_{\exiind}-\left(d-\seq_{\exiind+1}\right)}{f(d-\seq_{\exiind})-f(d-\seq_{\exiind+1})}\right)^2\notag\\ 
  &\leq \frac{8}{\fgap^2} \left(\log \frac{1}{\ups}+2 \right)^2 d^{1+\ups}
\end{align}

Summarizing the two cases, we see that for any monotone test function $f(\cdot)$, 
\begin{equation} 
\label{eq:seq:final}
  \fH\leq \frac {8}{\left(f(d)-f(0)\right)^2}\left(\log \frac 1 \ups +2\right)^2 d^{1+\ups}.
\end{equation}
For $d\geq 2$, upon setting $\ups = \frac{1}{\log d}$, we obtain from \eqref{eq:seq:final} that 
\begin{equation}
    \fH \le \frac{16}{\left(f(d)-f(0)\right)^2} \left(\log\log d+2\right)^2 d,
\end{equation}
which completes the proof.

\section{Proof of Lemma \ref{lem:relation:quantities}}
\label{app:convert_delta_nabla}
We expand $\pn, \pp,\pnm$ and $\pp$ using elementary combinatorial and algebraic identities. 
Regarding $\pn$, we have that 
\begin{align}
\label{eq:convert_p-}
\pn &= \sum\limits_{j=0}^{d} \binom{d}{j} \cp^j (1-\cp)^{d-j} f(j) \notag\\ 
  &= (1-\cp)^d f(0)+ \sum\limits_{j=1}^{d-1} \binom{d}{j} \cp^j (1-\cp)^{d-j} f(j) +\cp^d f(d) \notag\\ 
 &=(1-\cp)^d f(0)+ \sum\limits_{j=1}^{d-1} \left( \binom{d-1}{j} + \binom{d-1}{j-1} \right) \cp^j (1-\cp)^{d-j} f(j)  +\cp^d f(d) \notag\\ 
&= \sum\limits_{j=0}^{d-1} \binom{d-1}{j} \cp^j (1-\cp)^{d-j} f(j) + \sum\limits_{j=1}^{d} \binom{d-1}{j-1} \cp^j (1-\cp)^{d-j} f(j).
\end{align}
Regarding $\pp$, by relabelling, we have that
\begin{align}
\pp &= \sum\limits_{j=0}^{d-1} \binom{d-1}{j} \cp^j (1-\cp)^{d-1-j} f(j+1) \label{eq:convert_p+_1} \\ 
&= \sum\limits_{j=1}^{d} \binom{d-1}{j-1} \cp^{j-1} (1-\cp)^{d-j} f(j). \label{eq:convert_p+_2}
\end{align}
From \eqref{eq:convert_p+_1} and \eqref{eq:convert_p+_2}, we can rewrite $\pp$ as 
\begin{align} 
\label{eq:convert_p+}
\pp &= (1-\cp) \sum\limits_{j=0}^{d-1} \binom{d-1}{j} \cp^j (1-\cp)^{d-1-j} f(j+1) + \cp \sum\limits_{j=1}^{d} \binom{d-1}{j-1} \cp^{j-1} (1-\cp)^{d-j} f(j) \notag\\
&= \sum\limits_{j=0}^{d-1} \binom{d-1}{j} \cp^j (1-\cp)^{d-j} f(j+1) + \sum\limits_{j=1}^{d} \binom{d-1}{j-1} \cp^j (1-\cp)^{d-j} f(j).
\end{align}
Combining \eqref{eq:convert_p-} and \eqref{eq:convert_p+}, we can write $\pgap$ as
\begin{equation}
\label{eq:convert_delta}
\begin{aligned}
\pgap &= \pp - \pn \\ 
&= \sum\limits_{j=0}^{d-1} \binom{d-1}{j} \cp^j (1-\cp)^{d-j} (f(j+1) - f(j)).
\end{aligned}
\end{equation}
Under the assumption that $f(\cdot)$ is monotonically increasing and $f(0) < f(d)$, we conclude that 
\begin{align}
\label{eq:P+>=P-}
 \pgap > 0 \; \text {  or equivalently  }\;   \pp >  \pn.
\end{align}
Similarly, regarding $\pnm$, we have
\begin{align*}
    \pnm &= \sum\limits_{j=0}^{d} \binom{d}{j} \cp^j (1-\cp)^{d-j} f(j) \\ 
    &= \sum\limits_{j=0}^{d-1} \binom{d-1}{j} \cp^j (1-\cp)^{d-j} f(j) + \sum\limits_{j=1}^{d} \binom{d-1}{j-1} \cp^j (1-\cp)^{d-j} f(j).
\end{align*}
And regarding $\ppm$, we have
\begin{align*}
    \ppm &= \sum\limits_{j=0}^{d-1} \binom{d-1}{j} \cp^j (1-\cp)^{d-1-j} f(j) \\ 
    &= \sum\limits_{j=1}^{d} \binom{d-1}{j-1} \cp^{j-1} (1-\cp)^{d-j} f(j-1) \\ 
    &= (1-\cp) \sum\limits_{j=0}^{d-1} \binom{d-1}{j} \cp^j (1-\cp)^{d-1-j} f(j) + \cp \sum\limits_{j=1}^{d} \binom{d-1}{j-1} \cp^{j-1} (1-\cp)^{d-j} f(j-1) \\
    &= \sum\limits_{j=0}^{d-1} \binom{d-1}{j} \cp^j (1-\cp)^{d-j} f(j) + \sum\limits_{j=1}^{d} \binom{d-1}{j-1} \cp^j (1-\cp)^{d-j} f(j-1).
\end{align*}
It then follows that 
\begin{align} 
\label{eq:Opposite-Delta>0}
    \pgapm &=\pnm-\ppm \\
    & =\sum\limits_{j=1}^{d} \binom{d-1}{j-1} \cp^j (1-\cp)^{d-j} (f(j)-f(j-1)) \notag \\ 
    &= \sum\limits_{j=0}^{d-1} \binom{d-1}{j} \cp^{j+1} (1-\cp)^{d-j-1} (f(j+1)-f(j)) \notag\\ 
    &= \frac{\cp}{1-\cp} \pgap,
\end{align}
where the last equality follows from \eqref{eq:convert_delta}. This together with \eqref{eq:P+>=P-} implies that 
\begin{align}
\label{eq:Q->=Q+}
 \pgapm > 0 \; \text {  or equivalently  }\;  \pnm >  \ppm.
\end{align}
Using the monotonicity of $f(\cdot)$, we see that
\begin{align}
\pp &\le  \sum\limits_{j=0}^{d-1} \binom{d-1}{j} \cp^j (1-\cp)^{d-1-j} f(d)=f(d)  \label{eq:P+<f(d)}\\ 
\ppm &\ge \sum\limits_{j=0}^{d-1} \binom{d-1}{j} \cp^j (1-\cp)^{d-1-j} f(0)=f(0)\label{eq:Q+<f(d)}
\end{align}
Combining \eqref{eq:P+>=P-}, \eqref{eq:Q->=Q+}, \eqref{eq:P+<f(d)} and \eqref{eq:Q+<f(d)}, along with the definitions of $\pn$ in \eqref{eq:def:p-p+} and $\pnm$ in \eqref{eq:def:p-pp+p}, we obtain
\begin{align}
\label{eqs:relation:P:Q}
    f(d)\ge  \pp > \pn = \pnm > \ppm  \ge f(0).
\end{align} 
Recalling the definition
\begin{align}
    \pmin =\min\left\{ \pp,1-\ppm \right\}, \label{eq:def:pmin:repeat}
\end{align} 
we have from \eqref{eqs:relation:P:Q} that
\begin{align}
   1 \geq \pmin &\geq \min\{\pp,1-\pn\} \notag\\
                 &\geq \min\{\pp-\pn,\pp-\pn\} \notag\\
                 &= \pgap, \label{eq:1>=pmin>=Delta}
\end{align}
and 
\begin{align}
\label{eq:pmin>=bigtriangledown}
        \pmin & \geq \min\{\pnm,1-\ppm\} \notag\\
              &\geq\min\{\pnm-\ppm,\pnm-\ppm\} \notag\\
              &=\pgapm,
\end{align}
Combining \eqref{eq:1>=pmin>=Delta} and \eqref{eq:pmin>=bigtriangledown}, we have that
\begin{align*}
     1  \geq  \pmin \geq  \max\left\{\pgap, \pgapm\right\}.
\end{align*}
This concludes the proof.


\section{Proofs of tail-bound lemmas}\label{app:tail-bound}
The proofs of the tail-bound lemmas that bound the probability of error of our decoding rules in Theorem~\ref{thm:main} are collected in this Appendix.
\subsection{Proof of Lemma \ref{lem:num_tests_lb}}
\label{sec:num_tests_lb}

We shall use the following well-known Chernoff bound \cite{chernoff}.
\begin{fact}[Chernoff bound \cite{chernoff}]\label{chernoff_bound}
Suppose that $X \sim \text{Binomial}(n,p)$. Then, for any $\delta' \in (0,1)$, we have
\begin{equation} 
\begin{aligned}
    \Pr\left(X \le (1-\delta')np\right) &\leq \exp\left(-\frac{\delta'^2}{2} np \right) \leq \exp\left(-\frac{\delta'^2}{3} np \right), \\
    \Pr\left(X \ge (1+\delta')np\right) &\leq \exp\left(-\frac{\delta'^2}{2+\delta'} np \right) \le \exp\left(-\frac{\delta'^2}{3} np \right).
\end{aligned}
\end{equation}
\end{fact}

We now proceed with the proof of Lemma \ref{lem:num_tests_lb}. Recall that in \eqref{eq:T_small_q}, $T$ was selected to satisfy
$T \ge \frac{13}{6\cp} m$.
Consider an arbitrary item $i\in \mathcal{N}$. 
Since each item participates in a test i.i.d. with probability $\cp$,  
the expected number of tests item $i$ involving is
\begin{align}
    \mathbb{E}(\nti) &= \cp T \notag \\ 
    &\ge \frac{13}{6} \nt \notag \\ 
    &= 2\nt + \frac{1}{6} \times \frac{12\pmin}{(\pgap)^2} \ln \left( \frac{2n}{\varepsilon} \right) \notag\\
    &\ge 2\nt + 2\ln \left( \frac{2n}{\varepsilon} \right), \label{eq:ETi_4}
\end{align}
where the third line follows from the definition of $\nt$ in \eqref{eq:def:m}; the last line follows from \eqref{eq:1>=pmin>max:Delta}. 

Let $\mathcal{E}_i$ be the event that item $i$ participates in less than $\nt$ tests. By Fact~\ref{chernoff_bound}, we have
\begin{equation*}
    \begin{aligned}
    \Pr (\mathcal{E}_i)&= \Pr \left(\nti < \nt = \left( 1 - \frac{ \mathbb{E}(\nti)-\nt}{ \mathbb{E}(\nti)} \right)  \mathbb{E}(\nti) \right) \\
    &\leq \exp \left( - \left( \frac{ \mathbb{E}(\nti)-\nt}{ \mathbb{E}(\nti)} \right)^2 \, \frac{ \mathbb{E}(\nti)}{2} \right) \\
    &= \exp \left( - \frac{ \mathbb{E}(\nti)}{2} + \nt - \frac{\nt^2}{2 \mathbb{E}(\nti)} \right) \\
    &\leq \exp \left( -\frac{ \mathbb{E}(\nti)}{2} + \nt \right) \\
    &\leq \exp \left( -\ln \left( \frac{2n}{\varepsilon} \right) \right)
    = \frac{\varepsilon}{2n},
    \end{aligned} 
\end{equation*}
where the last inequality follows from \eqref{eq:ETi_4}.
By the union bound, the probability that all items participate in at least $\nt$ tests can be bounded from below by
\begin{equation*}
    1 - \Pr\left( \bigcup\limits_{i\in\cN}\mathcal{E}_i \right) \ge 1 - \sum\limits_{i\in\cN} \Pr(\mathcal{E}_i) > 1 - \frac{\varepsilon}{2n} \times n = 1 - \frac{\varepsilon}{2}.
\end{equation*}

\subsection{Proof of Lemma \ref{lem:num_tests_correctness}}
\label{sec:num_tests_correctness}
To begin with, assume that each item participates in at least $\nt$ tests, i.e., $\nti \ge \nt$ for all items $i$. As discussed in Decoding Rule $1$, we identify item $i$ via (\ref{eq:classification_small_q}). Two types of error can happen:
\begin{enumerate}
\item Item $i$ is non-defective, but is identified as defective, i.e., false alarm;
\item Item $i$ is defective, but is identified as non-defective, i.e., missed detection.
\end{enumerate}
We will bound the probabilities of $1)$ and $2)$ occuring separately as follows. For notational simplicity, let $\spn = \pn$, $\spp = \pp$, $\spgap = \pgap$, $\spmin = \pmin$. Recall that by Definition \ref{item:def:Delta}, $\spgap=\spp-\spn$. 

\subsubsection{\underline{False alarm for item $i$}} 
In this scenario, each test outcome is positive with probability $\spn$, and
\begin{equation}
    \frac{\ntip}{\nti} > \frac{\spn+\spp}{2}.
\end{equation}
Let $\fd$ denote the probability of this false alarm. Since the test outcomes are independent due to the tests being constructed in an i.i.d. manner, $\ntip \sim \text{Binomial}(\nti,\spn)$. From Fact~\ref{chernoff_bound}, $\fd$ can be bounded as
\begin{align}
\label{eq:false_defective_1}
\fd &= \Pr \left( \frac{\ntip}{\nti} > \frac{\spn+\spp}{2} \right) \notag\\
&= \Pr \left( \ntip > \left( 1+\frac{\spgap}{2\spn} \right) \spn \nti \right) \notag\\ 
&\le \exp\left( - \frac{1}{3} \times \frac{\spgap^2}{4\spn^2} \times \spn \nti \right) \notag\\ 
&= \exp \left( -\frac{\spgap^2}{12\spn} \nti \right) \notag\\
&\le \exp \left( -\frac{\spgap^2}{12\spn} \nt \right).
\end{align}
On the other hand, we also have $\ntin \sim \text{Binomial} (\nti,1-\spn)$. From Fact~\ref{chernoff_bound}, $\fd$ can also be bounded as
\begin{align}
\label{eq:false_defective_2}
\fd &= \Pr \left( \frac{\ntip}{\nti} > \frac{\spn+\spp}{2} \right) \notag\\
&= \Pr \left( \frac{\ntin}{\nti} < 1 - \frac{\spn+\spp}{2} \right) \notag\\ 
&= \Pr \left( \ntin < \left( 1 - \frac{\spgap}{2(1-\spn)} \right) (1 - \spn) \nti \right) \notag\\ 
&\le \exp\left( - \frac{1}{3} \times \frac{\spgap^2}{4(1-\spn)^2} (1 - \spn) \nti \right) \notag\\ 
&= \exp \left( - \frac{\spgap^2}{12(1-\spn)} \nti \right) \notag\\
&\le \exp \left( - \frac{\spgap^2}{12(1-\spn)} \nt \right).
\end{align}
Combining \eqref{eq:false_defective_1} and \eqref{eq:false_defective_2}, we obtain
\begin{align} 
\label{eq:false_defective}
\fd &\leq \exp\left( -\frac{\spgap^2}{12\min\{\spn,1-\spn\}}\nt \right) \notag\\
   &\leq \exp\left( - \frac{\spgap^2}{12\spmin} \nt \right)  \notag\\
   &\leq \frac{\varepsilon}{2n},
\end{align}
where the second inequality follows from \eqref{eq:def:Pmin} and \eqref{ineq:relation:P:Q}; the last inequality follows by substituting the definition of $\nt$ in \eqref{eq:def:m}.  
\subsubsection{\underline{Missed detection for item $i$}} The calculations are similar to those above analyzing the probability of a false alarm for item $i$. In this case, each test outcome is positive with probability $\spp$, and
\begin{equation}
    \frac{\ntip}{\nti} \le \frac{\spn+\spp}{2}.
\end{equation}
Let $\fn$ denote the probability of this false non-defective. Again since the test outcomes are independent, we have $\ntip \sim \text{Binomial}(\nti,\spp)$. From Fact~\ref{chernoff_bound}, we can bound $\fn$ as
\begin{align}
\label{eq:false_non_defective_1}
\fn &= \Pr \left( \frac{\ntip}{\nti} \le \frac{\spn+\spp}{2} \right) \notag\\
&= \Pr \left( \ntip \le \left( 1-\frac{\spgap}{2\spp} \right) \spp \nti \right) \notag\\ 
&\le \exp\left( - \frac{1}{3} \times \frac{\spgap^2}{4\spp^2} \times \spp \nti \right) \notag\\ 
&= \exp \left( -\frac{\spgap^2}{12\spp} \nti \right) \notag\\
&\le \exp \left( -\frac{\spgap^2}{12\spp} \nt \right).
\end{align}
From another perspective, $\ntin \sim \text{Binomial} (\nti,1-\spp)$. From Fact~\ref{chernoff_bound}, we can also bound $\fn$ as
\begin{align}
\label{eq:false_non_defective_2}
\fn &= \Pr \left( \frac{\ntip}{\nti} \le \frac{\spn+\spp}{2} \right) \notag\\
&= \Pr \left( \frac{\ntin}{\nti} \ge 1 - \frac{\spn+\spp}{2} \right) \notag\\ 
&= \Pr \left( \ntin \ge \left( 1 + \frac{\spgap}{2(1-\spp)} \right) (1 - \spp) \nti \right) \notag\\ 
&\le \exp\left( - \frac{1}{3} \times \frac{\spgap^2}{4(1-\spp)^2} (1 - \spp) \nti \right) \notag\\ 
&= \exp \left( - \frac{\spgap^2}{12(1-\spp)} \nti \right) \notag\\
&\le \exp \left( - \frac{\spgap^2}{12(1-\spp)} \nt \right).
\end{align}
Combining \eqref{eq:false_non_defective_1} and \eqref{eq:false_non_defective_2}, we see that
\begin{align} 
\label{eq:false_non_defective}
\fn &\leq \exp\left( -\frac{\spgap^2}{12 \min\{\spp,1-\spp\}} \nt \right) \notag\\
    &\leq \exp\left( - \frac{\spgap^2}{12\spmin} \nt \right) \notag\\
    &\leq \frac{\varepsilon}{2n},
\end{align}
where the second inequality follows from \eqref{eq:def:Pmin} and \eqref{ineq:relation:P:Q}; the last inequality follows from the definition of $\nt$ in \eqref{eq:def:m}.

From \eqref{eq:false_defective} and \eqref{eq:false_non_defective} we conclude that for any item $i\in\mathcal{N}$, the probability of  misidentification (either false alarm or missed detection) is smaller than $\frac{\varepsilon}{2n}$. Therefore, when all items participate in at least $\nt$ tests, by the union bound the probability that all items are correctly identified is bounded from below by
\begin{equation}
    1 - \frac{\varepsilon}{2n} \times n = 1 - \frac{\varepsilon}{2},
\end{equation}
which concludes the proof of Lemma~\ref{lem:num_tests_correctness}.

\subsection{Proof of Lemma \ref{lem:num_tests_lb_l}} \label{sec:num_tests_lb_l}
Recall that in \eqref{eq:T_large_q}, $T$ was chosen to satisfy $T \ge \frac{13}{6(1-\cp)} \mo$. 
Consider an arbitrary item $i\in \cN$. 
Since each item does not participate in a test i.i.d. with probability $1-\cp$,  
the expected number of tests without item $i$ is
\begin{align}
    \mathbb{E}(\moi) &= (1-\cp) T \notag \\ 
    &\ge \frac{13}{6} \mo \notag \\ 
    &= 2\mo + \frac{1}{6} \times \frac{12\pmin}{(\pgapm)^2} \ln \left( \frac{2n}{\varepsilon} \right) \notag\\
    &\ge 2\mo + 2\ln \left( \frac{2n}{\varepsilon} \right), \label{eq:l:ETi_4}
\end{align}
where the third line follows from the definition of $\mo$ in \eqref{eq:def:mo}; the last line follows from \eqref{eq:1>=pmin>max:Delta}.

Let $\cEl_i$ be the event that item $i$ participates in more than $T-\mo$.  In other words, the number of tests without item $i$ is less than $\mo$. By Fact~\ref{chernoff_bound}, we have
    \begin{align*}
    \Pr (\cEl_i)&= \Pr \left(\moi < \mo = \left( 1 - \frac{ \mathbb{E}(\moi)-\mo}{ \mathbb{E}(\moi)} \right)  \mathbb{E}(\moi) \right) \\
    &\leq \exp \left( - \left( \frac{ \mathbb{E}(\moi)-\mo}{ \mathbb{E}(\moi)} \right)^2 \, \frac{ \mathbb{E}(\moi)}{2} \right) \\
    &= \exp \left( - \frac{ \mathbb{E}(\moi)}{2} + \mo - \frac{\mo^2}{2 \mathbb{E}(\moi)} \right) \\
    &\leq \exp \left( -\frac{ \mathbb{E}(\moi)}{2} + \mo \right) \\
    &\leq \exp \left( -\ln \left( \frac{2n}{\varepsilon} \right) \right)
    = \frac{\varepsilon}{2n},
    \end{align*} 
    where the last inequality follows from \eqref{eq:l:ETi_4}.
By the union bound, the probability that each item participates in at most $T-\mo$ tests can be bounded from below by
\begin{equation}
    1 - \Pr\left( \bigcup\limits_{i}\cEl_i \right) \ge 1 - \sum\limits_i \Pr(\cEl_i) > 1 - \frac{\varepsilon}{2n} \times n = 1 - \frac{\varepsilon}{2}.
\end{equation}

\subsection{Proof of Lemma \ref{lem:num_tests_correctness_l}}
\label{sec:num_tests_correctness_l}
To begin with, assume that each item participates in at most $T-\mo$ tests, i.e., $\moi \ge \mo$ for all items $i$. As discussed in Decoding Rule $2$, we identify item $i$ via (\ref{eq:classification_large_q}). Two types of error can happen:
\begin{enumerate}
\item Item $i$ is non-defective, but is identified as defective, i.e., false alarm;
\item Item $i$ is defective, but is identified as non-defective, i.e., missed detection.
\end{enumerate}
We will bound the probabilities of $1)$ and $2)$ occuring separately as follows. For notational simplicity, let $\spnl = \pnm$, $\sppl = \ppm$, $\spgapl = \pgapm$, $\spmin = \pmin$. Recall that by Definition \ref{item:def:Delta}, $\spgapl=\spnl-\sppl $. 

\subsubsection{\underline{False alarm for item $i$}} 
In this scenario, each test outcome is positive with probability $\spnl$, and
\begin{equation}
    \frac{\moip}{\moi} \le \frac{\spnl+\sppl}{2}.
\end{equation}
Let $\fdl$ denote the probability of this false alarm. Since the test outcomes are independent due to the tests being constructed in an i.i.d. manner, $\moip \sim \text{Binomial}(\moi,\spnl)$. From Fact~\ref{chernoff_bound}, $\fdl$ can be bounded as
\begin{align}
\label{eq:false_defective_1_l}
\fdl &= \Pr \left( \frac{\moip}{\moi} \le \frac{\spnl+\sppl}{2} \right) \notag\\
&= \Pr \left( \moip \le \left( 1-\frac{\spgapl}{2\spnl} \right) \spnl \moi \right) \notag\\ 
&\le \exp\left( - \frac{1}{3} \times \frac{\spgapl^2}{4\spnl^2} \times \spnl \moi \right) \notag\\ 
&= \exp \left( -\frac{\spgapl^2}{12\spnl} \moi \right) \notag\\
&\le \exp \left( -\frac{\spgapl^2}{12\spnl} \mo \right).
\end{align}
On the other hand, we also have $\moin \sim \text{Binomial} (\moi,1-\spnl)$. From Fact~\ref{chernoff_bound}, $\fdl$ can also be bounded as

\begin{align}
\label{eq:false_defective_2_l}
\fdl &= \Pr \left( \frac{\moip}{\moi} \le \frac{\spnl+\sppl}{2} \right) \notag\\
&= \Pr \left( \frac{\moin}{\moi} \ge 1 - \frac{\spnl+\sppl}{2} \right) \notag\\ 
&= \Pr \left( \moin \ge \left( 1 + \frac{\spgapl}{2(1-\spnl)} \right) (1 - \spnl) \moi \right) \notag\\ 
&\le \exp\left( - \frac{1}{3} \times \frac{\spgapl^2}{4(1-\spnl)^2} (1 - \spnl) \moi \right) \notag\\ 
&= \exp \left( - \frac{\spgapl^2}{12(1-\spnl)} \moi \right) \notag\\
&\le \exp \left( - \frac{\spgapl^2}{12(1-\spnl)} \mo \right).
\end{align}
Combining \eqref{eq:false_defective_1_l} and \eqref{eq:false_defective_2_l}, we obtain
\begin{equation} \label{eq:false_defective_l}
\fdl \le \exp\left( -\frac{\spgapl^2}{12 \min\{\spnl, 1-\spnl\}} \mo \right) \le \exp\left( - \frac{\spgapl^2}{12\spmin} \mo \right) \le \frac{\varepsilon}{2n},
\end{equation}
where the second inequality follows from \eqref{eq:def:Pmin} and \eqref{ineq:relation:P:Q}; the last inequality follows by substituting the definition of $\mo$ in \eqref{eq:def:mo}.  
\subsubsection{\underline{Missed detection for item $i$}} The calculations are similar to those above analyzing the probability of a false alarm for item $i$. In this case, each test outcome is positive with probability $\sppl$, and
\begin{equation}
    \frac{\moip}{\moi} > \frac{\spnl+\sppl}{2}.
\end{equation}
Let $\fnl$ denote the probability of this false non-defective. Again since the outcomes are independent, we have $\moip \sim \text{Binomial}(\moi,\sppl)$. From Fact~\ref{chernoff_bound}, we can bound $\fnl$ as
\begin{align}
\label{eq:false_non_defective_1_l}
\fnl &= \Pr \left( \frac{\moip}{\moi} > \frac{\spnl+\sppl}{2} \right) \notag\\
&= \Pr \left( \moip > \left( 1+\frac{\spgapl}{2\sppl} \right) \sppl \moi \right) \notag\\ 
&\le \exp\left( - \frac{1}{3} \times \frac{\spgapl^2}{4\sppl^2} \times \sppl \moi \right) \notag\\ 
&= \exp \left( -\frac{\spgapl^2}{12\sppl} \moi \right) \notag\\
&\le \exp \left( -\frac{\spgapl^2}{12\sppl} \mo \right).
\end{align}
From another perspective, $\moin \sim \text{Binomial} (\moi,1-\sppl)$. From Fact~\ref{chernoff_bound}, we can also bound $\fnl$ as
\begin{align}
\label{eq:false_non_defective_2_l}
\fnl &= \Pr \left( \frac{\moip}{\moi} > \frac{\spnl+\sppl}{2} \right) \notag\\
&= \Pr \left( \frac{\moin}{\moi} < 1 - \frac{\spnl+\sppl}{2} \right) \notag\\ 
&= \Pr \left( \moin < \left( 1 - \frac{\spgapl}{2(1-\sppl)} \right) (1 - \sppl) \moi \right) \notag\\ 
&\le \exp\left( - \frac{1}{3} \times \frac{\spgapl^2}{4(1-\sppl)^2} (1 - \sppl) \moi \right) \notag\\ 
&= \exp \left( - \frac{\spgapl^2}{12(1-\sppl)} \moi \right) \notag\\
&\le \exp \left( - \frac{\spgapl^2}{12(1-\sppl)} \mo \right).
\end{align}
Combining \eqref{eq:false_non_defective_1_l} and \eqref{eq:false_non_defective_2_l}, we see that
\begin{equation} \label{eq:false_non_defective_l}
\fnl \le \exp\left( -\frac{\spgapl^2}{12 \min\{\sppl,1-\sppl\}}\mo \right) \le \exp\left( - \frac{\spgapl^2}{12\spmin} \mo \right) \le \frac{\varepsilon}{2n},
\end{equation}
where the second inequality follows from \eqref{eq:def:Pmin} and \eqref{ineq:relation:P:Q}; the last inequality follows from the definition of $\mo$ in \eqref{eq:def:mo}.

From \eqref{eq:false_defective_l} and \eqref{eq:false_non_defective_l} we conclude that for any item $i\in \mathcal{N}$, the probability of misidentification (either false alarm or missed detection) is smaller than $\frac{\varepsilon}{2n}$. Therefore, when each item participates in at most $T-\mo$ tests, by the union bound the probability that all items are correctly identified is bounded from below by
\begin{equation}
    1 - \frac{\varepsilon}{2n} \times n = 1 - \frac{\varepsilon}{2},
\end{equation}
which concludes the proof of Lemma~\ref{lem:num_tests_correctness_l}.

\section{Proof of Proposition \ref{prop:existence_final}}
\label{sec:existence_final}
From \eqref{eq:def:fTp}, we know that the value of $\fTp$ depends on the choice of $\cp$. 
For $\cp\in \left(\frac 1 d , \frac {d-1}{d}\right)$ with $d\geq 3$, the following result asserts that by choosing $\cp$ properly, we can give an explicit bound on the value of $\fTp$.


\begin{proposition} 
\label{prop:existence}
Let $d\geq 3$. For any $\lw',\up' \in \{1 ,\dots, d-1\}$ with $\lw' < \up'$ and $f(\lw')<f(\up')$, there exists $\cpns \in \left(\frac {\lw'} {d}, \frac {\up'} {d}\right)$ such that $\fTpns$ in \eqref{eq:def:fTp} satisfies
\begin{equation}
\label{eq:existence_interval_d}
\fTpns \leq  31334.75 \times \frac{1}{\ap^2} \left( \frac{\up'-\lw'}{f(\up')-f(\lw')} \right)^2  d\log \left( \frac{2n}{\varepsilon } \right)
\end{equation} 
where $\ap := \frac{1}{2} \min\left\{\up'-\lw',\sqrt{\lw'},\sqrt{d-\up'}\right\}$.
\end{proposition}
\begin{IEEEproof}
The proof of Proposition \ref{prop:existence} is rather involved and is 
deferred to Appendix \ref{sec:existence}.
\end{IEEEproof}

Below, the boundary points are handled separately. 

\begin{proposition} 
\label{prop:existence_margin}
\begin{enumerate}
  \item  By choosing $\cpns = \frac {1} {d+1}$, which implicitly requires $f(1) - f(0)  > 0$,\footnote{\label{fn:boundary:cases}If $f(1) - f(0) =0 $ or $f(d) - f(d-1) =0$, we can  choose $\cpns$ as in Proposition \ref{prop:existence}.} $\fTpns$ in \eqref{eq:def:fTp} satisfies 
\begin{equation} 
\label{eq:existence_margin_z}
    \fTpns \le \frac{266.45}{(f(1) - f(0))^2} d \log \left( \frac{2n}{\varepsilon} \right).
\end{equation}
\item  By choosing $\cpns =\frac {d} {d+1}$, which implicitly requires $f(d) - f(d-1) > 0$,$^{\ref{fn:boundary:cases}}$ $\fTpns$ in \eqref{eq:def:fTp} satisfies  
\begin{equation} 
\label{eq:existence_margin_d}
    \fTpns \leq \frac{266.45}{(f(d) - f(d-1))^2} d \log \left( \frac{2n}{\varepsilon} \right).
\end{equation}
\end{enumerate}
\end{proposition}


\begin{IEEEproof}
We prove the first part of the lemma directly.
Upon choosing $\cpns = \frac {1} {d+1}$, we have from \eqref{eq:convert_delta} that 
\begin{align*}
    \spgap(\cpns) &= \sum\limits_{j=0}^{d-1} \binom{d-1}{j} \cpns^j (1-\cpns)^{d-j} (f(j+1) - f(j)) \\ 
    &\ge \cpns^0 (1-\cpns)^{d}  (f(1) - f(0)) \\
    &\ge \frac{f(1) - f(0)}{e}, 
\end{align*}
where the last inequality follows from the fact that $\left(1-\frac{1}{d+1}\right)^d$ is decreasing in $d$ and $\lim_{d\to \infty}\left(1-\frac{1}{d+1}\right)^d=\frac 1 e$.
Then, plugging this into \eqref{eq:def:fTp}, we obtain
\begin{align*}
    \fTpns = \frac{36.06(1-\cpns)}{\cpns (\spgap(\cpns))^2} \log \left( \frac{2n}{\varepsilon} \right) \le \frac{266.45}{(f(1) - f(0))^2} d \log \left( \frac{2n}{\varepsilon} \right) ,
\end{align*}
which yields \eqref{eq:existence_margin_z} as desired.

We now turn to prove the second part of the lemma. Under the choice $\cpns = \frac {d} {d+1}$, we have from \eqref{eq:convert_delta} that
\begin{align*}
    \spgap(\cpns) &= \sum\limits_{j=0}^{d-1} \binom{d-1}{j} \cpns^j (1-\cpns)^{d-j} (f(j+1) - f(j)) \\
    &\ge \cpns^{d-1} (1-\cpns)^{1} (f(d) - f(d-1)) \\ 
    &= \left( 1-\frac{1}{d+1} \right)^d \frac{f(d)-f(d-1)}{d} \\
    &\ge \frac{f(d) - f(d-1)}{ed}. 
\end{align*}
Then, plugging this into \eqref{eq:def:fTp}, we obtain
\begin{align*}
    \fTpns = \frac{36.06(1-\cpns)}{\cpns (\spgap(\cpns))^2} \log \left( \frac{2n}{\varepsilon} \right) \le \frac{266.45}{(f(d) - f(d-1))^2} d \log \left( \frac{2n}{\varepsilon} \right).
\end{align*}
This completes the proof of Proposition \ref{prop:existence_margin}. 
\end{IEEEproof}

Proposition \ref{prop:existence_final} is proved by unifying Propositions \ref{prop:existence} and  \ref{prop:existence_margin}. To begin with, consider any  $\lw,\up \in \{0,\dots, d\}$ such that $\lw < \up$ and $f(\lw)<f(\up)$.\footnote{If there is no such pair of $(\lw,\up)$, we have $f(x)$=constant for all $x$. In this case, the defective set $\mathcal{D}$ can never be recovered.}  
To simplify the notation, define
\begin{align}
  \label{eq:def:beta}
    \bet := \min \left\{ \up-\lw , \sqrt{\lw+1} , \sqrt{d-\up+1} \right\}.
\end{align}
Consider the following four cases:
\begin{enumerate}[label=\roman*)]
\item $1 \le \lw < \up \le d-1$; \label{unify:item1}
\item $0 = \lw < \up \le d-1$; \label{unify:item2}
\item $1 \le \lw < \up = d$; \label{unify:item3}
\item $0 = \lw < \up = d$. \label{unify:item4}
\end{enumerate}

Case \ref{unify:item1}: For $1 \le \lw < \up \le d-1$, we have 
\begin{align}
\label{eq:alph:beta:case1}
    \bet &= \min\left\{ \up-\lw , \sqrt{\lw+1} , \sqrt{d-\up+1} \right\} \le \min\left\{ \sqrt{2}(\up-\lw) , \sqrt{2\lw} , \sqrt{2(d-\up)} \right\}.
\end{align}
Then, applying Proposition \ref{prop:existence} with $\lw'=\lw$ and $\up'=\up$, we have that $\bet\leq  2\sqrt{2}\ap$ and  
\begin{align}
    \fTpns 
    &\le 250678 \times \frac{1}{\bet^2} \left( \frac{\up-\lw}{f(\up)-f(\lw)} \right)^2  d\log \left( \frac{2n}{\varepsilon } \right) \notag\\ 
    &\le \exifinal
\end{align}
for some $\cpns \in \left(\frac \lw d, \frac \up d\right)$. 

Case \ref{unify:item2}: For $0 = \lw < \up \le d-1$, we have $\bet=1$ by definition \eqref{eq:def:beta}. 
We proceed with two further sub-cases.
\begin{itemize}
\item \underline{$\up=1$}:  From Proposition \ref{prop:existence_margin} we see that
    \begin{align}
    \ftaup(\cpns) &\le \frac{266.45}{(f(1)-f(0))^2} d \log \left( \frac{2n}{\varepsilon} \right) \notag\\
    &\le \exifinal
    \end{align}
    for $\cpns=\frac{1}{d+1}\in \left(\frac{\lw}{d},\frac{\up}{d}\right)$.
    
\item \underline{$\up\geq 2$}: Both Propositions \ref{prop:existence} and  \ref{prop:existence_margin} are applicable, we prefer to choose the one with smaller upper bound. Applying Proposition \ref{prop:existence} with $\lw'=1$ and $\up'=\up$, we have that $\alp=\frac 1 2$ and 
 \begin{align}
    \ftaup(\cpns') \le 125339 \times \left( \frac{\up-1}{f(\up)-f(1)} \right)^2  d\log \left( \frac{2n}{\varepsilon } \right)
\end{align}
for some $\cpns' \in \left(\frac 1 d, \frac \up d\right)$.
On the other hand, Proposition \ref{prop:existence_margin} gives
\begin{align}
    \ftaup(\cpns'') \le \frac{266.45}{(f(1)-f(0))^2} d \log \left( \frac{2n}{\varepsilon} \right)
\end{align}
for $\cpns''=\frac 1 {d+1}\in \left(\frac{\lw}{d},\frac{1}{d}\right)$.
Since
\begin{align}
    \min \left\{ \frac{1}{(f(1)-f(0))^2} , \frac{(\up-1)^2}{(f(\up)-f(1))^2} \right\}
    &\le 
     \frac{1+(\up-1)^2}{(f(1)-f(0))^2 + (f(\up)-f(1))^2} \notag\\
    &\le \frac{2\up^2}{(f(\up)-f(0))^2},
\end{align}
it follows that
\begin{align}
    \fTpns &\le \min \left\{ 125339 \times \left( \frac{\up-1}{f(\up)-f(1)} \right)^2  d\log \left( \frac{2n}{\varepsilon } \right) , \frac{576.96}{(f(1)-f(0))^2} d \log \left( \frac{2n}{\varepsilon} \right) \right\} \notag\\
    &\le 125339 \times \min \left\{ \frac{1}{(f(1)-f(0))^2} , \frac{(\up-1)^2}{(f(\up)-f(1))^2} \right\} \times d\log \left( \frac{2n}{\varepsilon } \right) \notag\\
    &\le 250678 \times \frac{\up^2}{(f(\up)-f(0))^2}  d\log \left( \frac{2n}{\varepsilon } \right) \notag\\
    &= 250678 \times \frac{1}{\bet^2} \left( \frac{\up-\lw}{f(\up)-f(\lw)} \right)^2  d\log \left( \frac{2n}{\varepsilon } \right) \notag\\
    &\le \exifinal
\end{align}
for some $\cpns \in \left( \frac{\lw}{d} , \frac{\up}{d} \right)$.
\end{itemize}

Case \ref{unify:item3}: For $1 \leq \lw < \up = d$, we have $\bet=1$ by definition \eqref{eq:def:beta}. 
The proof is similar to Case \ref{unify:item2}. Consider the following two sub-cases.
\begin{itemize}
\item \underline{$\lw=d-1$}:  From Proposition \ref{prop:existence_margin} we see that
    \begin{align}
    \ftaup(\cpns) &\le \frac{266.45}{(f(d)-f(d-1))^2} d \log \left( \frac{2n}{\varepsilon} \right) \notag\\
    &\le \exifinal
    \end{align}
    for $\cpns=\frac{d}{d+1}\in \left(\frac{\lw}{d},\frac{\up}{d}\right)$.
    
\item \underline{$\lw\leq d-2$}: Applying Proposition \ref{prop:existence} with $\lw'=\lw$ and $\up'=d-1$, we have that $\alp=\frac 1 2$ and 
\begin{align}
    \ftaup(\cpns')  \le 125339 \times \left( \frac{d-1-\lw}{f(d-1)-f(\lw)} \right)^2  d\log \left( \frac{2n}{\varepsilon } \right)
\end{align}
for some $\cpns' \in \left(\frac \lw d, \frac {d-1} {d}\right)$.
On the other hand, Proposition \ref{prop:existence_margin} gives
\begin{align}
   \ftaup(\cpns'') \le \frac{266.45}{(f(d)-f(d-1))^2} d \log \left( \frac{2n}{\varepsilon} \right)
\end{align}
for $\cpns''=\frac 1 {d+1}\in \left(\frac{d-1}{d},\frac{\up}{d}\right)$. Since 
\begin{align}
    \min \left\{ \frac{1}{(f(d)-f(d-1))^2} , \frac{(d-1-\lw)^2}{(f(d-1)-f(\lw))^2} \right\}
    &\leq \frac{1+(d-1-\lw)^2}{(f(d)-f(d-1))^2 + (f(d-1)-f(\lw))^2} \notag\\ 
    &\le \frac{2(d-\lw)^2}{(f(d)-f(\lw))^2}, 
\end{align}
it follows that 
\begin{align}
    \fTpns &\le \min \left\{ 125339 \times \left( \frac{d-1-\lw}{f(d-1)-f(\lw)} \right)^2  d\log \left( \frac{2n}{\varepsilon } \right) , \frac{266.45}{(f(d)-f(d-1))^2} d \log \left( \frac{2n}{\varepsilon} \right) \right\} \notag\\
    &\le 125339 \times \min \left\{ \frac{1}{(f(d)-f(d-1))^2} , \frac{(d-1-\lw)^2}{(f(d-1)-f(\lw))^2} \right\} \times d\log \left( \frac{2n}{\varepsilon } \right) \notag\\
    &\le 250678 \times \frac{(d-\lw)^2}{(f(d)-f(\lw))^2}  d\log \left( \frac{2n}{\varepsilon } \right) \notag\\
    &= 250678 \times \frac{1}{\bet^2} \left( \frac{\up-\lw}{f(\up)-f(\lw)} \right)^2  d\log \left( \frac{2n}{\varepsilon } \right) \notag\\ 
    &\le \exifinal
\end{align}
for some $\cpns \in \left( \frac{\lw}{d} , \frac{\up}{d} \right)$. 
\end{itemize}

Case \ref{unify:item4}: For $0 = \lw < \up = d$, we have $\bet=1$ by definition \eqref{eq:def:beta}. The proof is similar to Case \ref{unify:item2}. Consider three sub-cases:
\begin{itemize}
\item \underline{$d= 1$}: In this sub-case, the two bounds in Proposition \ref{prop:existence_margin} are identical and give 
    \begin{align}
    \ftaup(\cpns) &\le \frac{266.45}{(f(1)-f(0))^2} d \log \left( \frac{2n}{\varepsilon} \right) \notag\\
    &\le \exifinal
    \end{align}
    for $\cpns=\frac{1}{d+1}\in \left(\frac{\lw}{d},\frac{\up}{d}\right)$.
    
\item \underline{$d= 2$}: From Proposition \ref{prop:existence_margin} we see that
   \begin{align}
    \ftaup(\cpns') \le \frac{266.45}{(f(1)-f(0))^2} d \log \left( \frac{2n}{\varepsilon} \right)
\end{align}
for $\cpns'=\frac 1 {d+1} \in \left(\frac {\lw} d, \frac {1} {d}\right)$, and 
\begin{align}
   \ftaup(\cpns'') \le \frac{266.45}{(f(2)-f(1))^2} d \log \left( \frac{2n}{\varepsilon} \right)
\end{align}
for $\cpns''=\frac {d} {d+1} \in \left(\frac {1} d, \frac {\up} {d}\right)$.
Since
\begin{align}
    \min \left\{ \frac{1}{(f(1)-f(0))^2} , \frac{1}{(f(2)-f(1))^2} \right\}
    &\le\frac{2}{(f(1)-f(0))^2 + (f(2)-f(1))^2} \notag\\
    &\le \frac{4}{(f(2)-f(0))^2},
\end{align}
it follows that 
\begin{align}
    \fTpns &\le \min \left\{ \frac{266.45}{(f(1)-f(0))^2} d \log \left( \frac{2n}{\varepsilon} \right),\frac{266.45}{(f(2)-f(1))^2} d \log \left( \frac{2n}{\varepsilon} \right)\right\} \notag\\
    &\le 266.45 \times \min \left\{ \frac{1}{(f(1)-f(0))^2} , \frac{1}{(f(2)-f(1))^2}\right\} \times d\log \left( \frac{2n}{\varepsilon } \right) \notag\\
    &\leq 266.45 \times \frac{4}{(f(2)-f(0))^2}  d\log \left( \frac{2n}{\varepsilon } \right) \notag\\
    &= 266.45 \times \frac{1}{\bet^2} \left( \frac{\up-\lw}{f(\up)-f(\lw)} \right)^2  d\log \left( \frac{2n}{\varepsilon } \right) \notag\\
    &\le \exifinal 
\end{align}
for some $\cpns \in \left( \frac{\lw}{d} , \frac{\up}{d} \right)$. 
    
\item \underline{$d\geq 3$}: Applying Proposition \ref{prop:existence} with $\lw'=1$ and $\up'=d-1$, we have that $\alp=\frac 1 2$
\begin{align}
    \ftaup(\cpns')  \le 125339 \times \left( \frac{d-2}{f(d-1)-f(1)} \right)^2  d\log \left( \frac{2n}{\varepsilon } \right)
\end{align}
for some $\cpns' \in \left(\frac 1 d, \frac {d-1} {d}\right)$. 
On the other hand, Proposition \ref{prop:existence_margin} gives
\begin{align}
    \ftaup(\cpns'') \le \frac{266.45}{(f(1)-f(0))^2} d \log \left( \frac{2n}{\varepsilon} \right)
\end{align}
for $\cpns''=\frac 1 {d+1} \in \left(\frac {\lw} d, \frac {1} {d}\right)$, and 
\begin{align}
   \ftaup(\cpns''') \le \frac{266.45}{(f(d)-f(d-1))^2} d \log \left( \frac{2n}{\varepsilon} \right)
\end{align}
for $\cpns'''=\frac {d} {d+1} \in \left(\frac {d-1} d, \frac {\up} {d}\right)$. Since 
\begin{align}
    &\min \left\{ \frac{1}{(f(1)-f(0))^2} , \frac{1}{(f(d)-f(d-1))^2} ,  \frac{(d-2)^2}{(f(d-1)-f(1))^2} \right\} \notag\\
    &\leq \frac{1+1+(d-2)^2}{(f(1)-f(0))^2 + (f(d)-f(d-1))^2 + (f(d-1)-f(1))^2} \notag\\ 
   &\le \frac{3d^2}{(f(d)-f(0))^2}, 
\end{align}
it follows that 
\begin{align}
    \fTpns &\le \min \Bigg\{ 125339\times \left( \frac{d-2}{f(d-1)-f(1)} \right)^2 d\log \left( \frac{2n}{\varepsilon } \right) , \frac{266.45}{(f(1)-f(0))^2} d \log \left( \frac{2n}{\varepsilon} \right) ,\notag \\
    &   \kern23em\frac{266.45}{(f(d)-f(d-1))^2} d \log \left( \frac{2n}{\varepsilon} \right) \Bigg\} \notag\\
    &\le 125339 \times \min \left\{\frac{(d-2)^2}{(f(d-1)-f(1))^2}, \frac{1}{(f(1)-f(0))^2} , \frac{1}{(f(d)-f(d-1))^2}  \right\} \times d\log \left( \frac{2n}{\varepsilon } \right) \notag\\
    &\leq 376017 \times \frac{d^2}{(f(d)-f(0))^2}  d\log \left( \frac{2n}{\varepsilon } \right) \notag\\
    &= \exifinal 
\end{align}
for some $\cpns \in \left( \frac{\lw}{d} , \frac{\up}{d} \right)$. 
\end{itemize}

Summarizing the above, we see that for any $0\leq \lw<\up\leq d$, there exists $\cpns \in \left( \frac{\lw}{d} , \frac{\up}{d} \right)$ such that
\begin{align*}
    \fTpns \leq \exifinal .
\end{align*}
This along with the definitions of $\fH$ and $\bet$ in \eqref{eq:def:sensi:para} and \eqref{eq:def:beta}, respectively, yields
\begin{align}
    \fTpns \leq 376017 \fH d\log \left( \frac{2n}{\varepsilon } \right) 
\end{align}
for some $\cpns\in (0,1)$.  Proposition \ref{prop:existence_final} is proved. 

\section{Proof of Proposition \ref{prop:existence}}
\label{sec:existence}
 The following technical result will serve as a stepping stone to establishing Proposition \ref{prop:existence}. 
\begin{lemma}
\label{lem:existence_pre}
Let $d\geq 3$. For any $\lw',\up' \in \{1 ,\dots, d-1\}$ with $\lw' < \up'$ and $f(\lw')<f(\up')$, there exists $\cpns \in \left(\frac {\lw'} {d}, \frac {\up'} {d}\right)$ such that $\fTpns $ defined in \eqref{eq:def:fTp} satisfies 
\begin{equation} 
\label{eq:existence_pre}
\fTpns  \leq 1253.39 \times \frac{ (d-\lw') \up' }{\ap^2 (d-\up')\lw'} \left( \frac{\up'-\lw'}{f(\up')-f(\lw')} \right)^2  d\log \left( \frac{2n}{\varepsilon } \right),
\end{equation} 
where $\ap := \frac{1}{2} \min\left\{\up'-\lw',\sqrt{\lw'},\sqrt{d-\up'}\right\}$.
\end{lemma}

\begin{IEEEproof}
See Appendix \ref{proof:lem:existence_pre}.
\end{IEEEproof}
Comparing Proposition \ref{prop:existence} and Lemma \ref{lem:existence_pre}, we see that the main difference is the $\frac{ (d-\lw') \up' }{(d-\up')\lw'}$ term. In the remainder of the proof, we manage to eliminate the $\frac{ (d-\lw') \up' }{(d-\up')\lw'}$ term from \eqref{eq:existence_pre}. Since $\ap = \frac{1}{2} \min\left\{\up'-\lw',\sqrt{\lw'},\sqrt{d-\up'}\right\}$, we consider the following two cases.

Case i): $\ap=\frac {\up'-\lw'} {2}$, i.e., 
\begin{align}
\up'-\lw' \leq \min\left\{ \sqrt{\lw'} , \sqrt{d-\up'} \right\}.
\end{align}
Then we immediately obtain
\begin{equation}
    \frac{(d-\lw')\up'}{(d-\up')\lw'} \le \frac{(d-\up' + \sqrt{d-\up'})(\lw' + \sqrt{\lw'})}{(d-\up')\lw'}  = \frac{ \sqrt{d-\up'} +1}{ \sqrt{d-\up'}}\cdot \frac{\sqrt{\lw'} + 1}{\sqrt{\lw'}}\le 2\times 2 < 25,
\end{equation}
where the second inequality follows from the assumption that $ 1 \le \lw' < \up' \le d-1$.
Substituting into \eqref{eq:existence_pre}, we have the desired result \eqref{eq:existence_interval_d}. 

Case ii): $\ap\neq \frac {\up'-\lw'} {2}$, i.e., 
\begin{equation}
\label{eq:case2:alpha}
    \ap = \frac{1}{2} \min\left\{ \sqrt{\lw'} , \sqrt{d-\up'}\right \} < \frac{\up'-\lw'}{2}. 
\end{equation}
Define 
\begin{align}
\ta &:= \left\lfloor \frac{\up'-\lw'}{\lceil 2\alp \rceil} \right\rfloor,\\
\sigi &:= \lw' + \exi\cdot\lceil 2\alp \rceil    \text { for } \exi = 0,\cdots, \ta-1,\label{eq:def:sigma}\\
\sigt &:= \up'.
\end{align}
 For the ease of notation, let 
 \begin{align}
 \label{eq:def:alpha:i}
 \alpi := \frac{1}{2} \min\left\{ \sigio-\sigi , \sqrt{\sigi} , \sqrt{d-\sigio}\right \}  \text { for }  \exi = 0,  \cdots , \ta-1. 
 \end{align}
The following lemma shows that for all $i=0,\dots, \ta-1$, $\alpi$ is bounded from below by $\ap$.
\begin{lemma} 
\label{lem:exi:alp}
$\alpi \ge \alp$ for all $\exi \in \{0,\cdots, \ta-1\}$. 
\end{lemma}
\begin{IEEEproof}
For $\exi = 0,\cdots, \ta-1$, we have from \eqref{eq:def:sigma} that 
\begin{align}
\label{eq:sigi:lu}
    \lw' \le \sigi < \sigio \le \up'.
\end{align} 
It follows that
\begin{equation}
\label{eq:aphi:second-thrid}
    \frac{1}{2} \min\left\{ \sqrt{\sigi} , \sqrt{d-\sigio} \right\} \ge \frac{1}{2} \min\left\{ \sqrt{\lw'} , \sqrt{d-\up'}\right \} = \alp.
\end{equation}
Next,  for $\exi = 0,\cdots, \ta-2$, 
\begin{equation}
\label{eq:aphi:first:tau-1}
    \frac{\sigio-\sigi}{2} = \frac{\lceil2\alp\rceil}{2} \ge \alp;
\end{equation}
for $\exi = \ta-1$, 
\begin{equation}
\label{eq:aphi:first:tau}
    \frac{\sigt-\sigto}{2} = \frac{\up'-\lw'-(\ta-1)\lceil2\alp\rceil}{2} \ge \frac{\up'-\lw' - \left( \frac{\up'-\lw'}{\lceil2\alp\rceil}-1 \right) \lceil2\alp\rceil}{2} = \frac{\lceil2\alp\rceil}{2} \ge \alp.
\end{equation}
Combining the above three inequalities, along with the definition of $\alpi$ in \eqref{eq:def:alpha:i}, yields the desired result.
\end{IEEEproof}

To complete the proof of Case ii), the following two lemmas will also be used.
\begin{lemma} 
\label{lem:exi:alp_dif}
$\frac{(d-\sigi)\sigio}{(d-\sigio)\sigi} \le 25$ for any $\exi = 0,\cdots, \ta-1$.  
\end{lemma}
\begin{IEEEproof}
Observe that for any $ \exi = 0,\cdots,\ta-2$, 
\begin{equation}
    \sigio - \sigi = \lceil 2\alp \rceil \leq 4\alp+2;
\end{equation}
and for $\exi = \ta-1$, 
\begin{align}
    \sigt-\sigto &= \up' - (\lw'+(\ta-1)\lceil2\alp\rceil) \notag\\ 
    &= \up' - \lw' - \left( \left\lfloor \frac{\up'-\lw'}{\lceil2\alp\rceil} \right\rfloor - 1 \right) \lceil 2\alp \rceil \notag\\ 
    &< \up' - \lw' - \left( \frac{\up'-\lw'}{\lceil2\alp\rceil} - 2 \right) \lceil 2\alp \rceil \notag\\ 
    &= 2 \lceil 2\alp \rceil \notag\\ 
    &\leq 4\alp + 2.
\end{align}
It then follows that for all $ \exi = 0,\cdots,\ta-1$,
\begin{align}
    \sigio - \sigi &\leq 4\alp + 2 \notag\\ 
    &= 2 \min\{\sqrt{\lw'} , \sqrt{d-\up'}\} + 2 \notag\\ 
    &\le 4 \min\{\sqrt{\lw'} , \sqrt{d-\up'}\} \notag\\ 
    &\le 4 \min\{\sqrt{\sigi} , \sqrt{d-\sigio}\},
\end{align}
where the equality follows from \eqref{eq:case2:alpha}; the second inequality follows from the assumption that $ 1 \le \lw' < \up' \le d-1$; the last inequality follows from \eqref{eq:sigi:lu}.
Using this observation along with $ 1 \leq \sigi\leq \sigio \leq d-1$, we obtain 
\begin{equation}
    \frac{(d-\sigi)\sigio}{(d-\sigio)\sigi} \leq \frac{(d-\sigio+4\sqrt{d-\sigio}) (\sigi+4\sqrt{\sigi})}{(d-\sigio)\sigi}=  \frac{(\sqrt{d-\sigio}+4)}{\sqrt{d-\sigio}}\cdot\frac {(\sqrt{\sigi}+4)}{\sqrt{\sigi}}\le 5\times 5=25,
\end{equation}
which finishes the proof.
\end{IEEEproof}

\begin{lemma} 
\label{lem:exi:slp}
$\exists \exind \in \{ 0,\cdots,\ta-1 \}$ such that $\frac{f(\sigindo)-f(\sigind)}{\sigindo-\sigind} \ge \frac{f(\up')-f(\lw')}{\up'-\lw'}$. 
\end{lemma}
\begin{IEEEproof}
Suppose to the contrary that $\frac{ f(\sigio)-f(\sigi)}{ (\sigio-\sigi) } < \frac{f(\up')-f(\lw')}{\up'-\lw'}$ for all $\exi = 0,\cdots,\ta-1 $. In other words,
\begin{equation} \label{eq:exi:contradiction_part}
    f(\sigio)-f(\sigi)  < \frac{f(\up')-f(\lw')}{\up'-\lw'}(\sigio-\sigi), \forall \exi = 0,\cdots,\ta-1 .
\end{equation}
Summing \eqref{eq:exi:contradiction_part} over all $\exi \in \{0,\cdots,\ta-1\} $, we obtain that
\begin{equation}
    f(\up') - f(\lw') = \sum\limits_{\exi=0}^{\ta-1} \left(f(\sigio)-f(\sigi)\right) < \sum\limits_{\exi=0}^{\ta-1} \frac{f(\up')-f(\lw')}{\up'-\lw'} (\sigio-\sigi) = f(\up') - f(\lw')
\end{equation}
yielding a contradiction. Lemma \ref{lem:exi:slp} is proved.
\end{IEEEproof}

We are now ready to finish the proof of Case ii) using the above results. Upon applying Lemma \ref{lem:existence_pre} with $\lw' = \sigind$ and $\up' = \sigindo$ as defined in Lemma~\ref{lem:exi:slp}, we have that
$\exists \cpns \in \left(\frac{\sigind}{d} , \frac{\sigindo}{d}\right) $    such that 
\begin{align} 
\fTpns  &\leq 1253.39 \times \frac{(d-\sigind ) \sigindo }{\alpind^2 (d-\sigindo)\sigind} \left( \frac{\sigindo-\sigind}{f(\sigindo)-f(\sigind)} \right)^2  d\log \left( \frac{2n}{\varepsilon } \right) \notag\\ 
&\le 31334.75 \times \frac{1}{\alpind^2} \left( \frac{\sigindo-\sigind}{f(\sigindo)-f(\sigind)} \right)^2  d\log \left( \frac{2n}{\varepsilon } \right) \notag\\ 
&\le 31334.75 \times \frac{1}{\alpind^2} \left( \frac{\up'-\lw'}{f(\up')-f(\lw')} \right)^2  d\log \left( \frac{2n}{\varepsilon } \right) \notag\\ 
&\le 31334.75 \times \frac{1}{\alp^2} \left( \frac{\up'-\lw'}{f(\up')-f(\lw')} \right)^2  d\log \left( \frac{2n}{\varepsilon } \right)  
\end{align} 
where the second inequality follows from Lemma  \ref{lem:exi:alp_dif}; the third inequality follows from Lemma \ref{lem:exi:slp}; the last inequality follows from Lemma \ref{lem:exi:alp}. 
This completes the proof of Proposition \ref{prop:existence}.


\section{Proof of Lemma \ref{lem:existence_pre}}
\label{proof:lem:existence_pre}

We now prove Lemma \ref{lem:existence_pre}, first giving some preliminary lemmas. 
\begin{lemma} \label{lem:stirling}
For any $j=1,\dots,d-1$, we have
\begin{equation*}
\frac{\sqrt{2\pi}}{e^2} \sqrt{\frac{d}{j(d-j)}} \leq \binom{d}{j} \left( \frac{j}{d} \right)^j \left( \frac{d-j}{d} \right)^{d-j} \leq \frac{e}{\pi} \sqrt{\frac{d}{j(d-j)}}.
\end{equation*}
\end{lemma}

\begin{IEEEproof}
We shall use the following well-known Stirling's approximation \cite{stirling} for the factorial function. 
\begin{fact}[Stirling's approximation \cite{stirling}] 
\begin{equation}
\label{StirlingApprox}
 \sqrt{2\pi} \, n^{n+\frac{1}{2}} \, e^{-n} \le n! \le e \, n^{n+\frac{1}{2}} \, e^{-n}.
\end{equation}
\end{fact}

Using the upper and lower bounds on $n!$ in \eqref{StirlingApprox}, we have
\begin{align*}
\binom{d}{j} \left( \frac{j}{d} \right)^j \left( \frac{d-j}{d} \right)^{d-j} &= \frac{d!}{j!(d-j)!} \left( \frac{j}{d} \right)^j \left( \frac{d-j}{d} \right)^{d-j} \\ 
&\ge \frac{\sqrt{2\pi} d^{d+1/2} e^{-d}}{e j^{j+1/2} e^{-j} e (d-j)^{d-j+1/2} e^{-d+j}} \left( \frac{j}{d} \right)^j \left( \frac{d-j}{d} \right)^{d-j} \\ 
&= \frac{\sqrt{2\pi}}{e^2} \sqrt{\frac{d}{j(d-j)}}.
\end{align*}
Similarly, we also have
\begin{align*}
\binom{d}{j} \left( \frac{j}{d} \right)^j \left( \frac{d-j}{d} \right)^{d-j} &= \frac{d!}{j!(d-j)!} \left( \frac{j}{d} \right)^j \left( \frac{d-j}{d} \right)^{d-j} \\ 
&\le \frac{e d^{d+1/2} e^{-d}}{\sqrt{2\pi} j^{j+1/2} e^{-j} \sqrt{2\pi} (d-j)^{d-j+1/2} e^{-d+j}} \left( \frac{j}{d} \right)^j \left( \frac{d-j}{d} \right)^{d-j} \\ 
&= \frac{e}{2\pi} \sqrt{\frac{d}{j(d-j)}}.
\end{align*}
Combining the two bounds gives the desired result. 
\end{IEEEproof}
\begin{lemma} \label{lem:int_j_alpha}
For any $j >  i  > 0$ and $d > j +  i  $, we have 
\begin{align} 
\int_{\frac {j}{d}}^{\frac {j+ i }{d}} \cp^j (1-\cp)^{d-j} \dif \cp &\geq \left( \frac{j}{d} \right)^j \left( \frac{d-j}{d} \right)^{d-j} \frac{j(d-j- i )}{ i  d^2} \left( 1 - \exp\left( \frac{-i ^2 d  }{j (d-j- i )} \right) \right), \label{eq:lem3_1} \\ 
\int_{\frac {j- i }{d}}^{\frac {j}{d}} \cp^j (1-\cp)^{d-j} \dif \cp &\geq \left( \frac{j}{d} \right)^j \left( \frac{d-j}{d} \right)^{d-j} \frac{(j- i )(d-j)}{i d^2 } \left( 1 - \exp\left( \frac{- i ^2 d }{(j- i )(d-j)} \right) \right). \label{eq:lem3_2}
\end{align}
\end{lemma}

\begin{IEEEproof}
We first prove \eqref{eq:lem3_1}. Define 
\begin{align}
\label{eq:def:varphi(x)}
 \gm(x) := j\ln(j+xd)+(d-j)\ln(d-j-xd), \quad x \in \left[-\frac { i }{d},\frac { i }{d}\right].   
\end{align}  
Taking the derivative of $\gm(x)$, we obtain that  
\begin{equation*}
\gm^\prime (x) = \frac{-x d^3 }{(j+xd)(d-j-xd)} \geq \frac{-x d^3 }{j(d-j- i )}\geq\frac{- i d^2  }{j(d-j- i )},\;\forall x \in \left[0,\frac { i }{d}\right].
\end{equation*}
This implies that 
\begin{equation}
\label{eq:lb:varphi(t)-varphi(0)}
\gm(\qp)-\gm(0) = \int_0^\qp \gm^\prime (x) \dif x \geq \int_0^\qp \frac{- i d^2 }{j(d-j- i )} \dif x = \frac{-  i d^2}{j(d-j- i )} \qp,\; \forall\qp \in \left[0,\frac  i  d\right]. 
\end{equation}
By the definition of $\gm$ in \eqref{eq:def:varphi(x)}, we have
\begin{align}
\label{eq:varphi(t)-varphi(0)}
\gm(\qp)-\gm(0)=\ln\left( \left( \frac{j+\qp d}{j} \right)^j \left( \frac{d-j-\qp d}{d-j} \right)^{d-j}\right),\;\forall  \qp \in \left[-\frac  i  d,\frac  i  d\right].
\end{align}
It follows that 
\begin{align*}
\int_{\frac {j}{d}}^{\frac {j+ i }{d}} \cp^j (1-\cp)^{d-j} \dif \cp &= \int_0^{\frac { i } {d}} \left( \frac{j+\qp d}{d} \right)^j \left( \frac{d-j-\qp d}{d} \right)^{d-j} \dif \qp \\ 
&= \int_0^{\frac { i } {d}} \left( \frac{j}{d} \right)^j \left( \frac{d-j}{d} \right)^{d-j}  \left( \frac{j+\qp d}{j} \right)^j \left( \frac{d-j-\qp d}{d-j} \right)^{d-j} \dif \qp \\ 
&= \left( \frac{j}{d} \right)^j \left( \frac{d-j}{d} \right)^{d-j}  \int_0^{\frac { i } {d}} \exp(\gm(\qp) - \gm(0)) \dif \qp \\ 
&\ge \left( \frac{j}{d} \right)^j \left( \frac{d-j}{d} \right)^{d-j}  \int_0^{\frac { i } {d}} \exp\left(\frac{- i d^2 }{j(d-j- i )} \qp\right) \dif \qp \\ 
&= \left( \frac{j}{d} \right)^j \left( \frac{d-j}{d} \right)^{d-j} \frac{j(d-j- i )}{ i d^2} \left( 1 - \exp\left( \frac{- i ^2 d }{j (d-j- i )} \right) \right),
\end{align*}
where the first line follows by setting $q$ to equal $t+\frac{j}{d}$ for some $t$; the third line follows from \eqref{eq:varphi(t)-varphi(0)}; the fourth line follows from \eqref{eq:lb:varphi(t)-varphi(0)}.
This proves the desired inequality~\eqref{eq:lem3_1}.

Next, we prove \eqref{eq:lem3_2} by a similar argument. We see that 
\begin{equation*}
\gm^\prime (x)  = \frac{-x d^3 }{(j+x d)(d-j-x d)} \le \frac{ i d^2 }{(j- i )(d-j)},\;\forall x \in \left[-\frac { i } {d},0\right]. 
\end{equation*}
This implies that 
\begin{equation*}
\gm(0)-\gm(\qp) = \int_\qp^0 \gm^\prime (x) \dif x \le \int_\qp^0 \frac{ i d^2 }{(j- i )(d-j)} \dif x = -\frac{ i d^2 }{(j- i )(d-j)} \qp,\;\forall \qp \in \left[-\frac  i  d,0\right].
\end{equation*}
Then we can deduce that 
\begin{align*}
\int_{\frac {j- i }{d} }^{\frac j d} \cp^j (1-\cp)^{d-j} \dif \cp &= \int_{-\frac { i } {d}}^0 \left( \frac{j+\qp d}{d} \right)^j \left( \frac{d-j-\qp d}{d} \right)^{d-j} \dif \qp \\ 
&= \int_{-\frac { i } {d}}^0 \left( \frac{j}{d} \right)^j \left( \frac{d-j}{d} \right)^{d-j}  \left( \frac{j+\qp d}{j} \right)^j \left( \frac{d-j-\qp d}{d-j} \right)^{d-j} \dif \qp \\ 
&= \left( \frac{j}{d} \right)^j \left( \frac{d-j}{d} \right)^{d-j}  \int_{-\frac { i } {d}}^0 \exp(\gm(\qp) - \gm(0)) \dif \qp \\ 
&\ge \left( \frac{j}{d} \right)^j \left( \frac{d-j}{d} \right)^{d-j}  \int_{-\frac { i } {d}}^0 \exp\left(\frac{ id^2  }{(j- i )(d-j)} \qp\right) \dif \qp \\ 
&= \left( \frac{j}{d} \right)^j \left( \frac{d-j}{d} \right)^{d-j} \frac{(j- i )(d-j)}{ i d^2 } \left( 1 - \exp\left( \frac{-  i ^2 d}{(j- i )(d-j)} \right) \right),
\end{align*}
which establishes the inequality~\eqref{eq:lem3_2}.
\end{IEEEproof}

\begin{lemma} \label{lem:int_U_L}
Let $\lw',\up' \in \{1, \dots, d-1\}$ with $d\geq 3$ and  $\lw' < \up'$, we have that for all $j \in [\lw',\up']$, 
\begin{equation*}
\int_{\frac {\lw'}{d}}^{\frac {\up'}{d}} \cp^j (1-\cp)^{d-j} \dif \cp \ge \left(\frac{j}{d}\right)^j \left(\frac{d-j}{d}\right)^{d-j} \frac{\ap}{2d},
\end{equation*}
where $\ap := \frac{1}{2} \min\left\{\up'-\lw' , \sqrt{\lw'} , \sqrt{d-\up'}\right\}$. 
\end{lemma}

\begin{IEEEproof}
We split the proof into the following two cases, depending on whether $j$ is in the range $j\in \left[\lw',\frac {\lw'+\up'} {2} \right]$, or in the range $j\in \left(\frac {\lw'+\up'} {2} ,\up'\right]$.

Case i): $\lw' \le j \le \frac{\up'+\lw'}{2}$. It follows that $j+\ap\leq \up'<d$ since $\ap\leq \frac{\up'-\lw'}{2}$ by definition. We also have $j>\ap>0$ since $\lw'>\ap$ by definition. Thus $j,d,i=\ap$ satisfy the premise of Lemma~\ref{lem:int_j_alpha}. Using the inequality~\eqref{eq:lem3_1} with $i=\ap$, we obtain 
\begin{align}
\label{eq:upint:1b1}
\int_{\frac {\lw'}{d}}^{\frac {\up'}{d}} \cp^j (1-\cp)^{d-j} \dif \cp &\ge \int_{\frac j d}^{\frac {j+\ap}{d}} \cp^j (1-\cp)^{d-j} \dif \cp \notag \\
&\ge \left( \frac{j}{d} \right)^j \left( \frac{d-j}{d} \right)^{d-j} \frac{j(d-j-\ap)}{\ap d^2} \left( 1 - \exp\left( \frac{-\ap^2 d }{j (d-j-\ap)} \right) \right).
\end{align}
Using the fact that  
\begin{align}
\label{eq:1-e-x:bound}
1-e^{-x} \geq \frac{1}{2} x(2-x),\;\forall x\ge 0,
\end{align} 
we have 
\begin{align}
\label{eq:upint:1b2}
\frac{j(d-j-\ap)}{\ap d^2} \left( 1 - \exp\left( \frac{- \ap^2 d }{j (d-j-\ap)} \right) \right) &\ge \frac{j(d-j-\ap)}{\ap d^2} \times \frac{1}{2} \frac{\ap^2 d}{j(d-j-\ap)} \left(2 - \frac{\ap^2 d}{j(d-j-\ap)}\right) \notag\\ 
&= \frac{\ap}{2d} \left(2 - \frac{ \ap^2 d}{j(d-j-\ap)}\right).
\end{align}
Next, we argue that $\frac{\ap^2 d}{j(d-j-\ap)}\leq 1$. This is done by dividing into the following two sub-cases: 
\begin{itemize}
    \item \underline{$j < \frac{d}{2}$}: 
\begin{equation*}
\frac{\ap^2  d}{j(d-j-\ap)} \leq \frac{\left(\frac{\sqrt{\lw'}}{2}\right)^2 d}{j\left(d-j-\frac{\sqrt{d-2}}{2}\right)} \leq \frac{\left(\frac{\sqrt{\lw'}}{2}\right)^2 d }{\lw'\left(\frac{d}{2} - \frac{\sqrt{d-2}}{2}\right)} = \frac{d}{2\left(d-\sqrt{d-2}\right)} \le 1,
\end{equation*}
where the first inequality follows from the  definition that $\ap \leq \frac{\sqrt{\lw'}}{2} \leq \frac{\sqrt{d-2}}{2}$; the second inequality follows from the fact that $\lw' \leq j < \frac{d}{2}$; the last inequality follows from the assumption that $d\geq 3$. 
\item \underline{$j \geq \frac{d}{2}$}:
\begin{equation*}
\frac{\ap^2  d}{j(d-j-\ap)} \leq \frac{\ap^2  d}{\frac d 2(d-\up')}\leq \frac{\left(\frac{\sqrt{d-\up'}}{2}\right)^2 d }{\frac d 2 \left(d-\up'\right)} = \frac{1}{2} \leq 1,
\end{equation*}
where the first inequality is because $j\geq  \frac{d}{2}$ and $j+\ap\leq \up'$ as argued above; the second inequality is because $\ap \leq \frac{\sqrt{d-\up'}}{2}$ by definition. 
\end{itemize}
Using this observation along with \eqref{eq:upint:1b1} and \eqref{eq:upint:1b2}, we conclude that  
\begin{align*}
\int_{\frac {\lw'}{d}}^{\frac {\up'}{d}} \cp^j (1-\cp)^{d-j} \dif \cp 
&\ge \left( \frac{j}{d} \right)^j \left( \frac{d-j}{d} \right)^{d-j} \frac{\ap}{2d}.
\end{align*}

Case ii): $\frac{\up'+\lw'}{2} < j \le \up'$. This case can be proved in a similar manner as the above one. Note that $j-\ap\geq \lw'>0$ since $\ap\leq \frac{\up'-\lw'}{2}$ by definition. We also have $j+\ap<d$ since $\ap<d-\up'$. Hence $j,d,i=\ap$ satisfy the premise of Lemma~\ref{lem:int_j_alpha}. Using the inequality \eqref{eq:lem3_2} with $i=\ap$, we have 
\begin{align*}
\int_{\frac {\lw'}{d}}^{\frac {\up'}{d}} \cp^j (1-\cp)^{d-j} \dif \cp &\geq \int_{\frac {j-\ap}{d}}^{\frac j d} \cp^j (1-\cp)^{d-j} \dif \cp \\ 
&\geq \left( \frac{j}{d} \right)^j \left( \frac{d-j}{d} \right)^{d-j} \frac{(j-\ap)(d-j)}{\ap   d^2} \left( 1 - \exp\left( \frac{-\ap^2  d}{(j-\ap)(d-j)} \right) \right).
\end{align*}
Applying the standard identity \eqref{eq:1-e-x:bound} again, we see that 
\begin{align*}
\frac{(j-\ap)(d-j)}{\ap d^2} \left( 1 - \exp\left( \frac{-\ap^2 d}{(j-\ap)(d-j)} \right) \right) &\ge \frac{(j-\ap)(d-j)}{\ap  d^2} \times \frac{1}{2} \frac{\ap^2 d } {(j-\ap)(d-j)} \left(2 - \frac{\ap^2 d}{(j-\ap)(d-j)}\right) \\ 
&= \frac{\ap}{2d} \left(2 - \frac{\ap^2  d}{(j-\ap)(d-j)}\right).
\end{align*}
Similar to the above case, we prove $\frac{\ap^2d}{(j-\ap)(d-j)}\leq 1$ by considering the following two sub-cases: 
\begin{itemize}
    \item \underline{$j < \frac{d}{2}$}: 
\begin{equation*}
\frac{\ap^2d}{(j-\ap)(d-j)} \leq \frac{\ap^2d}{\lw' \frac d 2}  \leq \frac{ \left(\frac{\sqrt{\lw'}}{2}\right)^2d}{\lw' \frac d 2} = \frac{1}{2} \leq  1,
\end{equation*}
where the first inequality follows by noting that $j < \frac{d}{2}$ and $j-\ap\geq \lw'$ as argued above; the second inequality follows from the fact that $\ap \leq \frac{\sqrt{\lw'}}{2}$. 
\item {\underline{$j \ge \frac{d}{2}$}:} {
\begin{equation*}
\frac{\ap^2d}{(j-\ap)(d-j)} \leq \frac{\ap^2d}{\left(\frac{d}{2} -\ap\right)(d-\up')} \leq \frac{ \left(\frac{\sqrt{d-\up'}}{2}\right)^2d}{\left(\frac{d}{2}-\frac {\sqrt{d-2}}{2}\right) (d-\up')} = \frac{d}{2\left(d-\sqrt{d-2}\right)} \leq  1,
\end{equation*}
where the first inequality is because $\frac{d}{2} \leq j\leq \up' $; the second inequality follows since $\ap \leq \frac{\sqrt{d-\up'}}{2}\leq \frac{\sqrt{d-2}}{2}$ by its definition; and the last inequality follows from $d\geq 3$. }
\end{itemize}
It follows that 
\begin{align*}
\int_{\frac {\lw'}{d}}^{\frac {\up'}{d}} \cp^j (1-\cp)^{d-j} \dif \cp 
&\ge \left( \frac{j}{d} \right)^j \left( \frac{d-j}{d} \right)^{d-j} \frac{\ap}{2d}.
\end{align*}
Summarizing the two cases, Lemma~\ref{lem:int_U_L} is proved. 
\end{IEEEproof}

Using the above results, we are now in a position to prove Lemma \ref{lem:existence_pre}.
Recalling from \eqref{eq:convert_delta} that
\begin{equation*}
\pgap = \sum\limits_{j=0}^{d-1} \binom{d-1}{j} \cp^j (1-\cp)^{d-j} (f(j+1) - f(j)).
\end{equation*}
Note that $\pgap$ is continuous w.r.t. $\cp$. Assuming $\lw', \up' \in\{1,\dots, d-1\}$ with $ \lw'<\up'$ and $f(\lw')<f(\up')$, we can calculate the integral of $\pgap$ for $\cp \in \left[ \frac{\lw'}{d} , \frac{\up'}{d} \right]$ as follows:
\begin{align}
\int_{\frac{\lw'}{d}}^{\frac{\up'}{d}} \pgap \dif q &= \int_{\frac{\lw'}{d}}^{\frac{\up'}{d}} \sum\limits_{j=0}^{d-1} \binom{d-1}{j} \cp^j (1-\cp)^{d-j} (f(j+1) - f(j)) \dif q \notag\\ 
&\ge \int_{\frac{\lw'}{d}}^{\frac{\up'}{d}} \sum\limits_{j=\lw'}^{\up'-1} \binom{d-1}{j} \cp^j (1-\cp)^{d-j} (f(j+1) - f(j)) \dif q \notag\\
&= \sum\limits_{j=\lw'}^{\up'-1} \frac{d-j}{d} \binom{d}{j} (f(j+1)-f(j)) \int_{\frac{\lw'}{d}}^{\frac{\up'}{d}} \cp^j (1-\cp)^{d-j} \dif q \notag\\ 
&\ge \sum\limits_{j=\lw'}^{\up'-1} \frac{d-j}{d} \binom{d}{j} (f(j+1)-f(j)) \left(\frac{j}{d}\right)^j \left(\frac{d-j}{d}\right)^{d-j} \frac{\ap}{2d} \label{eq:bound_int_4} \\ 
&\ge \sum\limits_{j=\lw'}^{\up'-1} \frac{\sqrt{2\pi}}{e^2} \sqrt{\frac{d}{j(d-j)}} \frac{d-j}{d} (f(j+1)-f(j)) \frac{\ap}{2d} \label{eq:bound_int_5}\\ 
&= \sum\limits_{j=\lw'}^{\up'-1} \sqrt{\frac{\pi \ap^2 (d-j)}{2e^4 j d^3}} (f(j+1)-f(j)) \notag\\ 
&\ge \sum\limits_{j=\lw'}^{\up'-1} \sqrt{\frac{\pi \ap^2 (d-\up')}{2e^4 \up' d^3}} (f(j+1)-f(j)) \notag\\ 
&= \sqrt{\frac{\pi \ap^2 (d-\up')}{2e^4 \up' d^3}} (f(\up')-f(\lw')), \label{eq:bound_int_8}
\end{align}
where \eqref{eq:bound_int_4} follows from Lemma~\ref{lem:int_U_L} and $\ap = \frac{1}{2} \min\left\{\up'-\lw' , \sqrt{\lw'} , \sqrt{d - \up'}\right\}$; \eqref{eq:bound_int_5} follows from Lemma~\ref{lem:stirling}.
By the mean value theorem, from \eqref{eq:bound_int_8}, we know there exists some $\cpns \in \left( \frac{\lw'}{d} , \frac{\up'}{d} \right)$ such that
\begin{equation*}
\pgapns \geq \frac{\sqrt{\frac{\pi \ap^2 (d-\up')}{2e^4 \up' d^3}} (f(\up')-f(\lw'))}{\frac {\up'} {d} - \frac {\lw'} {d}} = \sqrt{\frac{\pi \ap^2 (d-\up')}{2e^4 \up' d}} \frac{f(\up')-f(\lw')}{\up'-\lw'}.
\end{equation*}
Using this observation, we can bound $\fTpns$ in \eqref{eq:def:fTp} as
\begin{align*}
\fTpns = \frac{36.06(1-\cpns )}{\cpns  (\pgapns)^2} \log \left( \frac{2n}{\varepsilon} \right) 
&\leq  \frac{36.06(1-\cpns )}{\cpns}\times\frac{2e^4 \up' d}{\pi \ap^2 (d-\up')} \left(\frac{\up'-\lw'}{f(\up')-f(\lw')} \right)^2 \log \left( \frac{2n}{\varepsilon} \right)\\
&\leq 1253.39 \times \frac{ (d-\lw') \up' }{\ap^2 (d-\up')\lw'} \left( \frac{\up'-\lw'}{f(\up')-f(\lw')} \right)^2  d\log \left( \frac{2n}{\varepsilon } \right),
\end{align*}
which proves Lemma \ref{lem:existence_pre}.


\section{Proof of Lemma \ref{lem:q_range}}
\label{app:lem_q_range}
Suppose to the contrary that $\cpn \in \left(0,\frac{1}{376017d^3}\right]\cup \left[1-\frac{1}{376017d^3},1\right)$. From \eqref{eq:convert_delta} we can bound 
\begin{align*}
    \frac{\cpn }{1-\cpn} \left(\pgapn\right)^2&= \left( \sum\limits_{j=0}^{d-1} \binom{d-1}{j} \cpn^j (1-\cpn)^{d-j-1} ( f(j+1) - f(j) ) \right)^2 \cdot \cpn (1-\cpn) \notag \\
    &\leq\left( \sum\limits_{j=0}^{d-1} \binom{d-1}{j} \cpn^j (1-\cpn)^{d-j-1} \left( f(d) - f(0) \right) \right)^2 \cdot \cpn (1-\cpn) \notag \\
    & = (f(d) - f(0))^2 \cdot \cpn (1-\cpn)\notag\\
    &\leq  (f(d) - f(0))^2 \times \frac{1}{376017d^3}\left(1-\frac{1}{376017d^3}\right) \notag \\ 
    &\le (f(d) - f(0))^2 \times \frac{1}{376017d^3}. 
\end{align*}
This along with the definition of $\fTp$ in \eqref{eq:def:fTp} yields that 
\begin{equation}
\label{eq:q_range_con1}
    \fTpn  \geq 36.06 \times \frac{376017d^3}{ (f(d) - f(0))^2} \log \left( \frac{2n}{\varepsilon} \right).
\end{equation}
On the other hand, applying Proposition \ref{prop:existence_final} with $\lw = 0$ and $\up = d$, we have that $\exists \cpns \in (0,1)$ such that 
\begin{equation} \label{eq:q_range_con2}
    \fTpns \leq \frac{376017 d^3}{ (f(d) - f(0))^2} \log \left( \frac{2n}{\varepsilon} \right).
\end{equation}
From \eqref{eq:q_range_con1} and \eqref{eq:q_range_con2} we have that 
\begin{equation*}
    \fTpn > \fTpns,
\end{equation*}
which is a contradiction to the definition that $\cpn = \argmin_{\cpn \in (0,1)} \fTp$ in \eqref{eq:def:cpn}. Hence we prove Lemma \ref{lem:q_range}.

\section{Proof of Lemma \ref{lem:fT_error}}
\label{app:lem_fT_error}
For notational simplicity, let $\varsigma=\frac{1}{376017d^4}$. It follows from Lemma \ref{lem:q_range} that 
\begin{align}
    0<\frac {\varsigma}{\cpn} < \frac{1}{d} \text{ and  }    0< \frac {\varsigma}{1-\cpn} <  \frac{1}{d}.
\end{align}
For any $\cph \in \left[ \cpn-\varsigma , \cpn+\varsigma \right]$, we have from \eqref{eq:convert_delta} that 
\begin{align}
    &\frac{\cph }{1-\cph}(\pgaph)^2 \notag\\
    &= \left( \sum\limits_{j=0}^{d-1} \binom{d-1}{j} \cph^j (1-\cph)^{d-j-1} ( f(j+1) - f(j) ) \right)^2 \cdot \cph (1-\cph) \notag\\
    &\ge \left( \sum\limits_{j=0}^{d-1} \binom{d-1}{j} (\cpn-\varsigma)^j (1-\cpn-\varsigma)^{d-j-1} ( f(j+1) - f(j) ) \right)^2 \cdot (\cpn-\varsigma) (1-\cpn-\varsigma)  \notag\\
    &= \left( \sum\limits_{j=0}^{d-1} \binom{d-1}{j} \cpn^j\left(\frac{\cpn-\varsigma}{\cpn}\right)^j (1-\cpn)^{d-j-1}\left(\frac{1-\cpn-\varsigma}{1-\cpn}\right)^{d-j-1} ( f(j+1) - f(j) ) \right)^2 \cdot (\cpn-\varsigma) (1-\cph-\varsigma)  \notag\\
    &\geq \left( \sum\limits_{j=0}^{d-1} \binom{d-1}{j} \cpn^j\left(\frac{\cpn-\varsigma}{\cpn}\right)^{d-1} (1-\cpn)^{d-j-1}\left(\frac{1-\cpn-\varsigma}{1-\cpn}\right)^{d-1} ( f(j+1) - f(j) ) \right)^2 \cdot (\cpn-\varsigma) (1-\cph-\varsigma)  \notag\\
    &=\left( \sum\limits_{j=0}^{d-1} \binom{d-1}{j} \cpn^j (1-\cpn)^{d-j-1} ( f(j+1) - f(j) ) \right)^2 \cdot \cpn (1-\cpn) \cdot \left( \frac{\cpn-\varsigma}{\cpn} \right)^{2d-1} \left( \frac{1-\cpn-\varsigma}{1-\cpn} \right)^{2d-1} \notag\\
    &\ge \left( \sum\limits_{j=0}^{d-1} \binom{d-1}{j} \cpn^j (1-\cpn)^{d-j-1} ( f(j+1) - f(j) ) \right)^2 \cdot \cpn (1-\cpn) \cdot \left( 1 - \frac{1}{d} \right)^{2d-1} \left( 1 - \frac{1}{d} \right)^{2d-1} \notag\\ 
    &\ge \left( \sum\limits_{j=0}^{d-1} \binom{d-1}{j} \cpn^j (1-\cpn)^{d-j-1} ( f(j+1) - f(j) ) \right)^2 \times \cpn (1-\cpn) \times \frac{1}{8}\times \frac{1}{8} \notag\\ 
    &= \frac{1}{64} \times \frac{\cpn (\pgapn)^2}{1-\cpn}. \label{eq:pgaph}
\end{align}
Using \eqref{eq:pgaph} along with the definition of $\fTp$ in \eqref{eq:def:fTp}, we have
\begin{align}
\label{eq:Gammaprmie<=Gammap}
    \ftaup(\cph)\leq 64 \fTpn.
\end{align}
For any $\cph \in \left[ \cpn-\varsigma , \cpn+\varsigma \right]$, we also have 
\begin{align}
    \pph &= \sum\limits_{j=0}^{d-1} \binom{d-1}{j} \cph^j (1-\cph)^{d-1-j} f(j+1) \notag\\ 
        &\leq  \sum\limits_{j=0}^{d-1} \binom{d-1}{j} (\cpn+\varsigma)^j (1-\cpn+\varsigma)^{d-1-j} f(j+1) \notag\\ 
        &\le \sum\limits_{j=0}^{d-1} \binom{d-1}{j} \cpn^j (1-\cpn)^{d-1-j} f(j+1) \cdot \left( \frac{\cpn+\varsigma}{\cpn} \right)^{d-1} \left( \frac{1-\cpn+\varsigma}{1-\cpn} \right)^{d-1} \notag\\ 
        &\le \sum\limits_{j=0}^{d-1} \binom{d-1}{j} \cpn^j (1-\cpn)^{d-1-j} f(j+1) \cdot \left( 1+\frac{1}{d} \right)^{d-1} \left( 1+\frac{1}{d} \right)^{d-1} \notag\\ 
        &\le \sum\limits_{j=0}^{d-1} \binom{d-1}{j} \cpn^j (1-\cpn)^{d-1-j} f(j+1) \cdot e^2 \notag \\ 
        &= e^2 \ppn. \label{eq:pph}
\end{align}
And similarly,
\begin{align}
    1 - \ppmh &= \sum\limits_{j=0}^{d-1} \binom{d-1}{j} \cph^j (1-\cph)^{d-1-j} (1-f(j)) \notag\\ 
    &\le \sum\limits_{j=0}^{d-1} \binom{d-1}{j} \cpn^j (1-\cpn)^{d-1-j} (1-f(j)) \cdot \left( \frac{\cpn+\varsigma}{\cpn} \right)^{d-1} \left( \frac{1-\cpn+\varsigma}{1-\cpn} \right)^{d-1} \notag\\ 
    &\le \sum\limits_{j=0}^{d-1} \binom{d-1}{j} \cpn^j (1-\cpn)^{d-1-j} (1-f(j)) \cdot \left( 1+\frac{1}{d} \right)^{d-1} \left( 1+\frac{1}{d} \right)^{d-1} \notag\\ 
    &\le \sum\limits_{j=0}^{d-1} \binom{d-1}{j} \cpn^j (1-\cpn)^{d-1-j} (1-f(j)) \cdot e^2 \notag\\ 
    &= e^2 (1-\ppmn). \label{eq:ppmh}
\end{align}
Combining \eqref{eq:pph} and \eqref{eq:ppmh}, along with the definition of $\pmin$ in \eqref{eq:def:Pmin}, we have that 
\begin{equation} \label{eq:pminh}
    \pminh \le e^2 P_{min}(\cpn).
\end{equation}
Finally, using \eqref{eq:Gammaprmie<=Gammap} and \eqref{eq:pminh} along with \eqref{eq:def:fTp} implies that 
\begin{equation*}
    \fTh=\ftaup(\cph) \pminh  \le 64 \fTpn e^2 P_{min}(\cpn)=64e^2\fTn,
\end{equation*}
which completes the proof.

\section{Proof of Lemma \ref{lem:cv_H}}
\label{app:lem_cv}
The proof of Lemma \ref{lem:cv_H} will resort to the following technical lemma:
\begin{lemma} \label{lem:cv:series}
For any $ \vara \in (0,1)$ and $\varb \in (-\vara,1-\vara)$, it follows that 
\begin{equation*}
    -\vara \ln\vara + (\vara+\varb)\ln(\vara+\varb) \le \varb (1+ \ln \vara ) + \frac{\varb^2}{\vara}.
\end{equation*}
\end{lemma}
\begin{IEEEproof}
Since $\vara + \varb > 0$, $\frac{\varb}{\vara} > -1$, we have 
\begin{align*}
    0 &\le (\vara+\varb) \left( \frac{\varb}{\vara} - \ln\left( 1 + \frac{\varb}{\vara} \right) \right) \\ 
    &= \varb + \frac{\varb^2}{\vara} - (\vara+\varb)\ln(\vara+\varb) + (\vara+\varb)\ln\vara,
\end{align*}
which, via simple rearrangement, gives the promised inequality.
\end{IEEEproof}

We now set out to prove Lemma \ref{lem:cv_H}. First, we have
\begin{align}
\label{eq:cv:maxH}
   \etp(\cy_i) - \etp(\cy_i|\cA_i)&= \etp(\cy_i) - \sum\limits_{\na=0}^{d} \Pr(\cA_i=\na) \etp(\cy_i | \cA_i=\na) \notag\\
    &= \sum\limits_{\na=0}^{d} \Pr(\cA_i=\na) \left[\etp(\cy_i) - \etp(\cy_i | \cA_i=\na)\right]
\end{align}
For the sake of notational brevity, let $\mean=\mean(\plsize_i)$ and $\var=\var(\plsize_i)$. Since 
\begin{align*}
\Pr (\cy_i = 1|\cA_i=\na) =f(a),
\end{align*}
\begin{align*}
\Pr (\cy_i = 1)=\sum\limits_{\na=0}^{d} \Pr(\cA_i=\na)\Pr (\cy_i = 1|\cA_i=\na) =\sum\limits_{\na=0}^{d} \Pr(\cA_i=\na)f(a)=\mean,
\end{align*}
we have
\begin{align}
\label{eq:H(Y_i):H(Y_i|Z_i=a)}
\begin{aligned}
     \etp(\cy_i) &= -\mean \log \mean - (1-\mean) \log (1-\mean) \\
                 &=\left[-\mean \ln \mean - (1-\mean) \ln (1-\mean)\right]\log e, \\
     \etp(\cy_i | \cA_i=\na) &= - f(\na) \log f(\na) - (1-f(\na)) \log (1-f(\na))\\
     &= \left[- f(\na) \ln f(\na) - (1-f(\na)) \ln (1-f(\na))\right]\log e.
\end{aligned}
\end{align}
Next, we argue that 
\begin{align}
\label{upb:H(Y_i)-H(Y_i|Z_i=a)}
\kern-.5em     -\mean \ln \mean - (1-\mean) \ln (1-\mean) + f(\na) \ln f(\na) + (1-f(\na)) \ln (1-f(\na)) 
    &\leq (f(\na)-\mean) (\ln \mean - \ln(1-\mean)) + \frac{(f(\na)-\mean)^2}{\mean(1-\mean)}.
\end{align}
This is done for each of the following possible cases.
\begin{itemize}
\item When $f(\na)=0$, 
\begin{equation*}
    \LHS - \RHS = 1 - \frac{1}{1-\mean} - \ln(1-\mean) \le 0,\quad \forall\, 0 < \mean< 1.
\end{equation*}
\item When $f(\na)=1$, 
\begin{equation*}
    \LHS - \RHS = 1 - \frac{1}{\mean} - \ln\mean \le 0,\quad \forall\, 0< \mean< 1.
\end{equation*}
\item When $f(\na) \in (0,1)$, applying Lemma \ref{lem:cv:series} with $\vara=\mean$ and $\varb = f(\na)-\mean$ we obtain  
\begin{equation*}
    -\mean \ln \mean + f(\na) \ln f(\na) \le (f(\na)-\mean) (1+\ln\mean ) + \frac{(f(\na)-\mean)^2}{\mean}.
\end{equation*}
Applying Lemma \ref{lem:cv:series} again with $\vara=1-\mean$ and $\varb=\mean-f(\na)$, we obtain  
\begin{equation*}
    -(1-\mean) \ln (1-\mean) + (1-f(\na)) \ln (1-f(\na)) \leq (\mean-f(\na)) (1+\ln(1-\mean)) + \frac{(f(\na)-\mean)^2}{1-\mean}. 
\end{equation*}
Upon combining the above two inequalities, we have $ \LHS \leq \RHS$ as desired.
\end{itemize}
Substituting \eqref{eq:H(Y_i):H(Y_i|Z_i=a)} and \eqref{upb:H(Y_i)-H(Y_i|Z_i=a)} into \eqref{eq:cv:maxH}, we arrive at
\begin{align*}
      \etp(\cy_i) - \etp(\cy_i|\cA_i)
    &\leq \sum\limits_{\na=0}^{d} \Pr(\cA_i=\na) \left[  \left(f(\na)-\mean\right) \left(\ln \mean - \ln(1-\mean)\right) + \frac{(f(\na)-\mean)^2}{\mean(1-\mean)}\right] \log e  \\
    &=\frac{\var \log e}{\mean(1-\mean)}
\end{align*}
where the equality follows from \eqref{eq:cv:def:p_miu_v}. This proves Lemma \ref{lem:cv_H}.
\section{Proof of Lemma \ref{lem:match}}
\label{app:lem_match}
We prove the claim by contradiction. To begin with, assume the contrary is true, i.e., 
\begin{equation} \label{eq:mt:cv_aspt_previous}
    \left(\apn(\idex)(f(\idex) - f(\arb))\right)^2 \leq \frac{\var(\plsize^{*})} {11} (\idex-\mnp)^2 , \; \forall \idex \in \left\{0,1, \dots, d-1,d\right\} \setminus \{\arb\}.
\end{equation}
Noting that \eqref{eq:mt:cv_aspt_previous} always holds for $\idex = \arb$. It then follows that 
\begin{equation} \label{eq:mt:cv_aspt}
    \left(\apn(\idex)(f(\idex) - f(\arb))\right)^2 \leq \frac{\var(\plsize^{*})} {11} (\idex-\mnp)^2 , \quad \forall \idex \in \{0,1, \dots, d-1,d\}.
\end{equation}
In the sequel, we adopt the convention that $\frac{\idex-\mnp}{ \apn(\idex)}=\frac {0}{0}=1$ for $\idex=\mnp$.
Then equation \eqref{eq:mt:cv_aspt} can be equivalently written as
\begin{equation} 
\label{eq:match:le}
    \left(f(\idex) - f(\arb)\right)^2 \leq  \frac{\var(\plsize^{*})}{11} \cdot \left(\frac{\idex-\mnp}{\apn(\idex)}\right)^2 ,  \;\forall \idex \in \left\{0,1, \dots, d-1,d\right\}.
\end{equation}
For notational convenience, let
\begin{equation}
\label{eq:hypergeometric:pdf}
 p(\idex):= \frac{\binom{d}{\idex}\binom{n-d}{\plsize^{*}-\idex}}{\binom{n}{\plsize^{*}}} ,  \;\forall \idex \in \left\{0,1, \dots, d-1,d\right\}.
\end{equation}
By definition, $\left(p(0),\dots, p(d)\right)$ is the hypergeometric distribution with parameters $n,d$ and $\plsize^{*}$. The mean and variance formulae for hypergeometric distributions $(n,d, \plsize^{*})$ are, respectively, 
\begin{align}
\label{eq:mean:var:formulae}
   \plsize^{*} \frac d n  \;
   \text{  and }  \;
   \plsize^{*} \frac {d} {n}\cdot\frac {n-d} {n}\cdot\frac {n-\plsize^{*}} {n-1}.
\end{align}
Taking expectations on both sides of \eqref{eq:match:le} w.r.t. the distribution \eqref{eq:hypergeometric:pdf}, we get
\begin{equation}
    \cE \left(\left( f(\idex) - f(\arb)\right )^2\right) \le \frac{\var(\plsize^{*})}{11} \cE \left( \left(\frac{\idex-\mnp}{\apn(\idex)}\right)^2 \right) . \label{eq:mt:E_up}
\end{equation}
On the other hand, 
\begin{align}
    \cE \left(\left( f(\idex) - f(\arb) \right)^2\right)
    &=\cE\left(\left(f(\idex) - \mean(\plsize^{*})\right)^2\right) + \cE\left(\left(\mean(\plsize^{*}) - f(\arb)\right)^2\right) + \cE\left(2 \left(f(\idex) - \mean(\plsize^{*})\right) \left(\mean(\plsize^{*}) - f(\arb)\right)\right) \notag\\
    &= \var(\plsize^{*}) + (\mean(\plsize^{*}) - f(\arb))^2 + 0 \notag\\
    &\ge \var(\plsize^{*}), \label{eq:mt:E_low_3}
\end{align}
where the second line follows from \eqref{eq:rep:def:mu:v:chi} and \eqref{eq:hypergeometric:pdf}.
Combining \eqref{eq:mt:E_up} and \eqref{eq:mt:E_low_3}, we deduce that
\begin{equation} 
\label{eq:mt:1st_conclusion}
    \cE \left( \left(\frac{\idex-\mnp}{\apn(\idex)}\right)^2 \right) \ge 11.
\end{equation}
However, we will argue that $\cE \left( \left(\frac{\idex-\mnp}{\apn(\idex)}\right)^2 \right) < 11$, which is a contradiction to \eqref{eq:mt:1st_conclusion}. This is proved for each of the two possible cases:
\begin{enumerate}[label=\roman*)]
    \item $\mnp \in (0,1)\cup (d-1,d)$; \label{eq:bound:exp:case1}
    \item $\mnp \in[1, d-1]$.\label{eq:bound:exp:case2}
\end{enumerate}

Case \ref{eq:bound:exp:case1}: For $\mnp \in (0,1)\cup (d-1,d)$, we have for any $\idex \in \{0,\dots, d\}$ that 
\begin{equation}
    \min\left\{ \idex+1 , d-\idex+1 \right\} \ge 1 > \min\{ \mnp , d-\mnp \}.
\end{equation}
It then follows that 
\begin{align}
\label{eq:mt:value}
   \left(\frac{\idex-\mnp}{\apn(\idex)}\right)^2 &= \frac{(\idex-\mnp)^2}{\min \{ (\idex - \mnp)^2, \idex+1, d-\idex+1, \mnp+1, d-\mnp+1 \}} \notag\\
    &\le \frac{(\idex-\mnp)^2}{\min \{ (\idex - \mnp)^2, \idex+1, d-\idex+1, \mnp, d-\mnp \}} \notag\\
    &= \frac{(\idex-\mnp)^2}{\min \{ (\idex - \mnp)^2, \mnp, d-\mnp \}} \notag\\
    &\le \frac{(\idex-\mnp)^2}{(\idex - \mnp)^2} + \frac{(\idex-\mnp)^2}{\min\{\mnp, d-\mnp\}} \notag\\
    &= 1 + \frac{(\idex-\mnp)^2 \max\{\mnp, d-\mnp\}}{\mnp(d-\mnp)} \notag\\
    &\le 1 + \frac{(\idex-\mnp)^2 d}{\mnp(d-\mnp)}.
\end{align}
Taking expectations on both sides of \eqref{eq:mt:value} w.r.t. the hypergeometric distribution \eqref{eq:hypergeometric:pdf}, we get
\begin{align}
\label{eq:boundexp:case1}
    \cE \left(  \left(\frac{\idex-\mnp}{\apn(\idex)}\right)^2 \right) &\le 1+ \frac{ d}{\mnp(d-\mnp)}\cE \left((\idex-\mnp)^2  \right).
\end{align}
From the mean formula in \eqref{eq:mean:var:formulae}, we have that 
$\cE(\idex)= \plsize^{*} \frac d n =\mnp$. 
Then using the variance formula in \eqref{eq:mean:var:formulae}, we have
\begin{align*}
    \cE \left( (\idex-\mnp)^2\right) =  \plsize^{*} \frac{d}{n}\cdot \frac{n-d}{n}\cdot \frac{n-\plsize^{*}}{n-1},
\end{align*}
which implies 
 \begin{align}
 \label{eq:boundexp:comm}
  \frac{ d}{\mnp(d-\mnp)}  \cE \left( (\idex-\mnp)^2\right) =  \frac{n-d}{n-1} \leq 1.
\end{align}
Note that \eqref{eq:boundexp:comm} holds for all $\mnp\in(0,d)$.
Substituting \eqref{eq:boundexp:comm} into \eqref{eq:boundexp:case1}, we have  $\cE \left( \left(\frac{\idex-\mnp}{\apn(\idex)}\right)^2 \right) \leq 2$. 

Case \ref{eq:bound:exp:case2}: For $\mnp \in[1, d-1]$, we must have $d \ge 2$. It follows that for any $i\in \{0, \dots, d\}$,    
\begin{align}
    \left(\frac{\idex-\mnp}{\apn(\idex)}\right)^2  &= \frac{(\idex-\mnp)^2}{\min \{ (\idex - \mnp)^2, \idex+1, d-\idex+1, \mnp+1, d-\mnp+1 \}} \notag\\
    &\le \frac{(\idex-\mnp)^2}{\min \{ (\idex - \mnp)^2, \idex+1, d-\idex+1, \mnp, d-\mnp \}} \notag\\
    &\le \frac{(\idex-\mnp)^2}{(\idex - \mnp)^2} + \frac{(\idex-\mnp)^2}{\min \{ \idex+1, d-\idex+1 \}} + \frac{(\idex-\mnp)^2}{\min\{\mnp, d-\mnp\}} \notag\\
    &= 1 + \frac{(\idex-\mnp)^2 \max\{ \idex+1, d-\idex+1 \}}{(\idex+1)(d-\idex+1)}+ \frac{(\idex-\mnp)^2 \max\{\mnp, d-\mnp\}}{\mnp(d-\mnp)}  \notag\\
    &\le 1  + \frac{(\idex-\mnp)^2\left( d+1 \right)}{ (\idex+1) (d-\idex+1) } + \frac{(\idex-\mnp)^2 d}{\mnp(d-\mnp)}. \label{eq:mt:le_5}
\end{align}
Taking the expectation of \eqref{eq:mt:le_5} w.r.t. the hypergeometric distribution \eqref{eq:hypergeometric:pdf}, we have
\begin{align}
\label{eq:mt:le:exp}
    \cE \left(\left(\frac{\idex-\mnp}{\apn(\idex)}\right)^2\right) &\le 1 + \left( d+1 \right) \cE \left( \frac{(\idex-\mnp)^2}{(\idex+1)(d-\idex+1)} \right) + \frac{ d}{\mnp(d-\mnp)}\cE \left((\idex-\mnp)^2  \right) \notag\\
    &\leq 2 + \left( d+1 \right) \cE \left( \frac{(\idex-\mnp)^2}{(\idex+1)(d-\idex+1)} \right),
\end{align}
where the second line follows from \eqref{eq:boundexp:comm} since it continues to hold for this case.

Next, we proceed to bound the term on the right hand side of \eqref{eq:mt:le:exp}.
We can expand
\begin{align}
    \cE \left( \frac{(\idex-\mnp)^2}{(\idex+1)(d-\idex+1)} \right) 
    &= \sum \limits_{\idex=0}^{d} \left( \frac{\binom{d}{\idex}\binom{n-d}{\plsize^{*}-\idex}}{\binom{n}{\plsize^{*}}} \cdot \frac{(\idex-\mnp)^2}{(\idex+1)(d-\idex+1)} \right) \notag\\
    &= \sum \limits_{\idex=0}^{d} \left( \frac{\binom{d+2}{\idex+1}\binom{n-d}{\plsize^{*}-\idex}}{\binom{n+2}{\plsize^{*}+1}} \cdot \frac{(n+1)(n+2)}{(\plsize^{*}+1)(n-\plsize^{*}+1)(d+1)(d+2)} \cdot (\idex-\mnp)^2 \right) \notag\\
    &= \frac{ (n+1)(n+2)}{(\plsize^{*}+1)(n-\plsize^{*}+1)(d+1)(d+2)} \cdot \sum \limits_{\idex=0}^{d} \left( \frac{\binom{d+2}{\idex+1}\binom{n-d}{\plsize^{*}-\idex}}{\binom{n+2}{\plsize^{*}+1}} \cdot (\idex-\mnp)^2 \right) \notag\\
    &\leq \frac{ (n+1)(n+2)}{(\plsize^{*}+1)(n-\plsize^{*}+1)(d+1)(d+2)} \cdot \sum \limits_{\idex=-1}^{d+1} \left( \frac{\binom{d+2}{\idex+1}\binom{n-d}{\plsize^{*}-\idex}}{\binom{n+2}{\plsize^{*}+1}} \cdot (\idex-\mnp)^2 \right) \notag\\
    &= \frac{ (n+1)(n+2)}{(\plsize^{*}+1)(n-\plsize^{*}+1)(d+1)(d+2)} \cdot \sum \limits_{\idex=0}^{d+2} \left( \frac{\binom{d+2}{\idex}\binom{n-d}{\plsize^{*}+1-\idex}}{\binom{n+2}{\plsize^{*}+1}} \cdot (\idex-\mnp-1)^2 \right). \label{eq:exp-expand-case2}
\end{align}
Using the formula for the mean of hypergeometric distributions $(n+2,d+2, \plsize^{*}+1)$, we have 
$$\sum \limits_{\idex=0}^{d+2} \left( \frac{\binom{d+2}{\idex}\binom{n-d}{\plsize^{*}+1-\idex}}{\binom{n+2}{\plsize^{*}+1}} \cdot \idex \right)=(\plsize^{*}+1)\frac{d+2}{n+2}.$$
Then, using the formula for the variance of hypergeometric distributions $(n+2,d+2, \plsize^{*}+1)$, we have
\begin{align}  
\kern-2em \sum \limits_{\idex=0}^{d+2} \left( \frac{\binom{d+2}{\idex}\binom{n-d}{\plsize^{*}+1-\idex}}{\binom{n+2}{\plsize^{*}+1}} \cdot (\idex-\mnp-1)^2 \right) 
    &= \sum \limits_{\idex=0}^{d+1} \left( \frac{\binom{d+2}{\idex}\binom{n-d}{\plsize^{*}+1-\idex}}{\binom{n+2}{\plsize^{*}+1}} \cdot \left( \idex-(\plsize^{*}+1)\frac{d+2}{n+2} + (\plsize^{*}+1)\frac{d+2}{n+2}-\mnp-1 \right)^2 \right) \notag\\
    &= \sum \limits_{\idex=0}^{d+1} \left( \frac{\binom{d+2}{\idex}\binom{n-d}{\plsize^{*}+1-\idex}}{\binom{n+2}{\plsize^{*}+1}} \cdot \left( \idex-(\plsize^{*}+1)\frac{d+2}{n+2}   \right)^2\right) + \left((\plsize^{*}+1)\frac{d+2}{n+2}-\mnp-1\right)^2\notag\\    
    &=  (\plsize^{*}+1)\frac{d+2}{n+2}\cdot \frac{n-d}{n+2}\cdot \frac{n-\plsize^{*}+1}{n+1} + \left((\plsize^{*}+1)\frac{d+2}{n+2}-\mnp-1\right)^2 \notag\\
    &\le (\plsize^{*}+1)\frac{d+2}{n+2}\cdot \frac{n-d}{n+2}\cdot \frac{n-\plsize^{*}+1}{n+1} + 4, \label{eq:bound:var-case2}
\end{align}
where the last line follows from 
\begin{align*}
   0=  \plsize^{*}\frac{d}{n}-\mnp \leq  (\plsize^{*}+1)\frac{d+2}{n+2}-\mnp \leq (\plsize^{*}+1)\frac{d+2}{n}-\mnp =\frac{d+2(\plsize^{*}+1)}{n}\leq 3.
\end{align*}
Substituting \eqref{eq:bound:var-case2} into \eqref{eq:exp-expand-case2}, we conclude that
\begin{align}
   \cE \left( \frac{(\idex-\mnp)^2}{(\idex+1)(d-\idex+1)} \right)    
   &= \frac{1 }{d+1}\cdot \frac{n-d}{n+2}  + \frac{4 (n+1)(n+2)}{(\plsize^{*}+1)(n-\plsize^{*}+1)(d+1)(d+2)}\notag\\
   &\le \frac{1 }{d+1} +  \frac{4(n+1)(n+2)}{(\plsize^{*}+1)(n-\plsize^{*}+1)(d+1)(d+2)}\notag\\
    &\leq \frac{1 }{d+1} + \frac{4 }{d+1}\cdot \frac{n+2}{n+d}\cdot \frac{n+1}{n-\frac{n}{d}+1}\cdot \frac{d}{d+2} \notag\\
    &\le \frac{1 }{d+1} + \frac{4 }{d+1}\cdot  \frac{n+1}{n-\frac{n}{2}+1}\cdot \frac{nd+2d}{nd+2d+2n+d^2} \notag\\
    &< \frac{1 }{d+1} + \frac{4 }{d+1} \times 2 \times 1 \notag\\
    &= \frac{9}{d+1}, \label{eq:mt:le_3_9}
\end{align}
where the third line follows from 
\begin{align*}
    (\plsize^{*}+1)(n-\plsize^{*}+1)&=-\left(\plsize^{*}-\frac{n}{2}\right)^2+\frac{n^2}{4}+n+1\\ &\geq-\left(\frac{n}{d}-\frac{n}{2}\right)^2+\frac{n^2}{4}+n+1 \\
    &=\left(\frac{n}{d}+1\right)\left(n-\frac{n}{d}+1\right)
\end{align*}
since we have from \eqref{eq:def:vartheta} that $\frac{n}{d}\leq \plsize^{*}\leq n-\frac{n}{d}$ for $ 1\leq \mnp \leq d-1$.
Upon combining \eqref{eq:mt:le:exp} and \eqref{eq:mt:le_3_9}, we arrive at
\begin{align*}
    \cE \left(\frac{(\idex-\mnp)^2}{ \apn(\idex)^2}\right)
    < 2 + \left( d+1 \right) \cdot \frac{9}{d+1} = 11.
\end{align*}

Summarizing the above two cases, we see that $\cE \left( \left(\frac{\idex-\mnp}{\apn(\idex)}\right)^2 \right) < 11$,
which contradicts \eqref{eq:mt:1st_conclusion}. Therefore, the assumption in \eqref{eq:mt:cv_aspt_previous} is false and Lemma~\ref{lem:match} is proved.

\section{Proof of Corollary \ref{coro:special:cases}}
\label{app:coro:cases}
\subsection{Proof of Corollary \ref{coro:special:cases}-\ref{coro_classical}} 
\label{app:coro_classical}
\begin{IEEEproof}
For test function \eqref{f(x)_classical}, letting $\lw=0$ and $\up=1$, we have from definition \eqref{eq:def:sensi:para} that $\fH\leq 1$.\footnote{\label{label:ft:H(f)=1}Indeed, we have $\fH=1$ for this test function. The reverse inequality follows from \eqref{eq:sensi-para:lb:up}.} It then follows that the upper bound in \eqref{eq:thm:T:ub} scales as $\cO \left(d\log n\right)$.

On the other hand, recall from Remark \ref{remark:lb} that the lower bound in \eqref{eq:T:lower:bound} scales as $\Omega\left(\log \binom{n}{d}\right)$. Indeed, we can show that the lower bound is precisely $\log \binom{n}{d}$, i.e., $\fh=1$ for this test function. To see this, noting that $f(0)=0$ and $f(a)=1, \forall a \geq 1$, we can compute that
\begin{align}
\label{eq:mean:classical}
    \mean(\plsize) = \sum\limits_{\na=0}^{d} \frac{\binom{d}{\na}\binom{n-d}{\plsize-\na}}{\binom{n}{\plsize}} f(\na) = 1 - \frac{\binom{n-d}{\plsize}}{\binom{n}{\plsize}},\\
     \sum\limits_{\na=0}^{d} \frac{\binom{d}{\na}\binom{n-d}{\plsize-\na}}{\binom{n}{\plsize}} f^2(\na) = 1 - \frac{\binom{n-d}{\plsize}}{\binom{n}{\plsize}}.\label{eq:f2:classical}
\end{align}
Since $(f(\na) - \mean(\plsize))^2=f^2(\na) +\mean^2(\plsize)-2f(\na)\mean(\plsize)$, we have from \eqref{eq:mean:classical} and \eqref{eq:f2:classical} that
\begin{align*}
    \var(\plsize) &= \sum\limits_{\na=0}^{d} \frac{\binom{d}{\na}\binom{n-d}{\plsize-\na}}{\binom{n}{\plsize}} (f(\na) - \mean(\plsize))^2\\
    &= \sum\limits_{\na=0}^{d} \frac{\binom{d}{\na}\binom{n-d}{\plsize-\na}}{\binom{n}{\plsize}} f^2(\na) - \mean^2(\plsize) \\ 
    &= 1 - \frac{\binom{n-d}{\plsize}}{\binom{n}{\plsize}} - \left( 1 - \frac{\binom{n-d}{\plsize}}{\binom{n}{\plsize}} \right)^2 \\ 
    &= \frac{\binom{n-d}{\plsize}}{\binom{n}{\plsize}} \left( 1 - \frac{\binom{n-d}{\plsize}}{\binom{n}{\plsize}} \right).
\end{align*}
It follows that 
\begin{align*}
     \frac{\mean(\plsize)(1-\mean(\plsize))}{\var(\plsize)}=1, \; \forall \plsize\in \{1,\dots,n-1\}.
\end{align*}
Thus we have $\fh=1$ for this test function.
Now the expression \eqref{eq:T:lower:bound} reduces to the classical Fano's inequality based information theoretic lower bound \cite{sid2014} on the number of tests required for $(1-\varepsilon)$-reliable recovery
\begin{equation*}
    T \geq \frac{1}{\log e}\left((1-\varepsilon)\log\binom{n}{d} - 1 \right).
\end{equation*}
By standard arguments via Stirling’s approximation, this quantity scales as $\Omega\left(d\log\frac{n}{d}\right)$.

Finally, the assumption that $ d= n^{\theta},\theta\in(0,1)$ implies that our upper and lower bounds are order-wise tight, both scaling as $\Theta \left(d\log n\right)$.
\end{IEEEproof}

\subsection{Proof of Corollary \ref{coro:special:cases}-\ref{coro_threshold}}
\label{app:coro_threshold}
\begin{IEEEproof}
The proof is very similar to the proof of Corollary \ref{coro:special:cases}-\ref{coro_classical} and appears for completeness.
For test function \eqref{eq:threshold:function}, letting $\lw=\ell$ and $\up=\ell+1$, we have from \eqref{eq:def:sensi:para} that $\fH\leq 1. ^{\ref{label:ft:H(f)=1}}$ Substituting into \eqref{eq:thm:T:ub}, the upper bound scales as $\cO \left(d\log n\right)$.

On the other hand,  recall from Remark \ref{remark:lb} that the lower bound in \eqref{eq:T:lower:bound} scales as $\Omega\left(\log \binom{n}{d}\right)$.\footnote{By a similar argument to the one above, one can show the lower bound is precisely $\log \binom{n}{d}$, i.e., we also have $\fh=1$ for this test function.} 
By standard arguments via Stirling’s approximation, $\log \binom{n}{d}$ is at least $d\log\frac{n}{d}$. Using the assumption that $ d= n^{\theta},\theta\in(0,1)$, we see that our upper and lower bounds are order-wise tight, both scaling as $\Theta \left(d\log n\right)$.
\end{IEEEproof}

\subsection{Proof of Corollary \ref{coro:special:cases}-\ref{coro_linear}}
\label{app:coro_linear}
\begin{IEEEproof}
For linear test function \eqref{eq:linear:function}, letting $\lw = \left \lfloor \frac{d}{3} \right \rfloor$ and $\up = \left \lceil \frac{2d}{3} \right \rceil$, we have that 
\begin{align*}
\min\left\{\up-\lw , \sqrt{\lw+1} , \sqrt{d-\up+1}\right\} &= \min \left\{ \left \lceil \frac{2d}{3} \right \rceil- \left \lfloor \frac{d}{3} \right \rfloor , \sqrt{ \left \lfloor \frac{d}{3} \right \rfloor+1} , \sqrt{d-\left \lceil \frac{2d}{3} \right \rceil+1} \right\}\\
 &\geq \min \left\{ \frac{2d}{3}-  \frac{d}{3} , \sqrt{ \frac{d}{3} -1+1} , \sqrt{d-\left(\frac{2d}{3} +1\right)+1} \right\}\\
 &=\sqrt{\frac d 3}.
\end{align*}
It follows from \eqref{eq:def:sensi:para} that 
\begin{align*}
    \fH &\leq \left( \frac{1}{\sqrt{\frac d 3}} \times\frac{\left \lceil \frac{2d}{3} \right \rceil- \left \lfloor \frac{d}{3} \right \rfloor}{f\left(\left \lceil \frac{2d}{3} \right \rceil\right)-f\left( \left \lfloor \frac{d}{3} \right \rfloor\right)} \right)^2\\
    &=3d
\end{align*}
Plugging this into \eqref{eq:thm:T:ub}, the upper bound scales as $\cO \left( d^2 \log n\right)$. 

The mean and variance formulae for hypergeometric distributions with parameters $n, d$ and $\plsize$ are, respectively, $\plsize \frac d n$  and $\plsize \frac {d} {n}\frac {n-d} {n}\frac {n-\plsize} {n-1}$.
For this test function we can therefore compute that 
\begin{equation*}
    \mean(\plsize) = \sum\limits_{\na=0}^{d} \frac{\binom{d}{\na}\binom{n-d}{\plsize-\na}}{\binom{n}{\plsize}} f(\na) = \frac{1}{d} \sum\limits_{\na=0}^{d} \frac{\binom{d}{\na}\binom{n-d}{\plsize-\na}}{\binom{n}{\plsize}} \na = \frac{\plsize}{n},
\end{equation*}
and 
\begin{align*}
    \var(\plsize) &= \sum\limits_{\na=0}^{d} \frac{\binom{d}{\na}\binom{n-d}{\plsize-\na}}{\binom{n}{\plsize}} (f(\na) - \mean(\plsize))^2 \\ 
    &= \frac{1}{d^2}\sum\limits_{\na=0}^{d} \frac{\binom{d}{\na}\binom{n-d}{\plsize-\na}}{\binom{n}{\plsize}} \left( \na-d\mean(\plsize) \right)^2 \\ 
    &= \frac{1}{d^2}\sum\limits_{\na=0}^{d} \frac{\binom{d}{\na}\binom{n-d}{\plsize-\na}}{\binom{n}{\plsize}} \left( \na - \plsize\frac d n \right)^2 \\     
    &= \frac{1}{d^2} \cdot\plsize \frac {d} {n}\frac {n-d} {n}\frac {n-\plsize} {n-1} \\
    &=\frac{\plsize(n-\plsize)}{n^2}\cdot \frac{n-d}{d(n-1)}.
\end{align*}
It follows that 
\begin{align*}
    \frac{\mean(\plsize)(1-\mean(\plsize))}{\var(\plsize)}= \frac{d(n-1)}{n-d} 
    \ge d, \quad \forall\plsize\in\{1,\dots,n-1\}.
\end{align*}
This together with the definition of $\fh$ in \eqref{eq:def:concen:para} implies $\fh\geq d$. 
Plugging into \eqref{eq:T:lower:bound}, we have
\begin{equation*}
    T \geq \frac{1}{\log e} d \left( (1-\varepsilon)\log\binom{n}{d} - 1 \right) 
\end{equation*}
which, by standard arguments via Stirling’s approximation, scales as $ \Omega \left( d^2 \log \frac{n}{d}  \right)$.

Finally, under the assumption that $ d= n^{\theta},\theta\in(0,1)$, we see that both the upper and lower bounds scale as  $ \Theta \left( d^2 \log n  \right)$.

\end{IEEEproof}

\section{Proof of Corollary \ref{coro:noisy:function}}
\label{app:coro:noisy}
\begin{IEEEproof}
Applying the inequalities in \eqref{ineq:relation:P:Q} to the definition of $\pmin$ in \eqref{eq:def:Pmin}, we have
\begin{align}
\label{example:pmin:bounds}
   \min\{f(0) ,1-f(d)\}\leq \pmin \leq \min\{f(d) ,1-f(0)\}, \quad \forall \cp\in(0,1).
\end{align}
From Remark \ref{eq:mu:lb:ub:mono} we have
\begin{align}
    \label{example:mu:bounds}
 f(0)(1-f(d))\leq \mean(\plsize)(1- \mean(\plsize))\leq  f(d)(1-f(0)), \quad \forall \plsize\in\{1,\dots,n-1\}.
\end{align}
Combining \eqref{example:pmin:bounds} and \eqref{example:mu:bounds}, we see that 
\begin{align}
\label{eq:factor:Pmin_mu:lb:up}
 \frac{ \min\{f(0) ,1-f(d)\}}{f(d)(1-f(0))}  \leq  \frac{\pminn}{\mean(\plsize^*) \left( 1-\mean(\plsize^*) \right)} \leq \frac{ \min\{f(d) ,1-f(0)\}}{f(0)(1-f(d))}.
\end{align}
Recalling the definition of noisy test functions, we have $f(0),1-f(d)\in \Theta(1)$. It then follows from \eqref{eq:factor:Pmin_mu:lb:up} that $$\frac{\pminn}{\mean(\plsize^*) \left( 1-\mean(\plsize^*) \right)}\in \Theta(1).$$ This along with Theorem \ref{thm:match} yields that our bounds are order-wise tight. 
\end{IEEEproof}

\section{Proof of Lemma \ref{lem:estimation:d} (Estimating the Exact Number of Defectives)}
\label{app:esti_d}

Let us first analyze a useful subroutine, and then present the full algorithm.

\subsection{A useful subroutine}
Let $\ud \geq 2$ be a putative number of defective items, and consider the goal of deciding whether $d \le \ud-1$ or $d \ge \ud$. Towards this end, we use a Bernoulli test design in which each item is independently placed into each test with probability $\lomcp$. Let $\lompp$ denote the probability of having a positive test outcome conditioned on $d=\ud$. It follows that
\begin{equation} \label{eq:lompp}
    \lompp = \sum\limits_{j=0}^{\ud} \binom{\ud}{j} \lomcp^j (1-\lomcp)^{\ud-j} f(j).
\end{equation}
Similar to \eqref{eq:convert_delta}, define
\begin{equation} \label{eq:def:lompgap}
    \lompgap := \sum\limits_{j=0}^{\ud-1} \binom{\ud-1}{j} \lomcp^j (1-\lomcp)^{\ud-j} \left(f(j+1) - f(j)\right).
\end{equation}
The subroutine $\lom(\ud,\lomcp,\lomerr)$ for deciding whether $d \le \ud-1$ or $d \ge \ud$ is described in Algorithm \ref{algo:sub}.

\begin{algorithm}
    \caption{$\lom(\ud,\lomcp,\lomerr)$}\label{algo:sub}
    \begin{algorithmic}[1]
    \State Take $\lomrs$ tests of the Bernoulli test design with parameter $\lomcp$, where
     \begin{equation} \label{eq:def:lomrs}
    \lomrs := \left\lceil 8.32 \left( \frac{1-\lomcp}{\lomcp \lompgap} \right)^2 \log \frac{1}{\lomerr} \right\rceil. 
     \end{equation}
    \State Let $\lompo$ denote the number of tests with positive outcome within these $\lomrs$ tests. If
    \begin{equation} \label{eq:lom_l}
        \frac{\lompo}{\lomrs} \le \lompp - \frac{\lomcp}{2(1-\lomcp)} \lompgap,
    \end{equation}
    return  $d \le \ud-1$; otherwise, return $d \ge \ud$.
    \end{algorithmic}
\end{algorithm}
\begin{lemma}\label{lem:sub:error}
The error probability of $\lom(\ud,\lomcp,\lomerr)$ is at most $\lomerr$.
\end{lemma}
\begin{IEEEproof}
Let $\lomppo$ denote the probability of having a positive test outcome conditioned on $d=\ud-1$. It follows that
\begin{equation*} 
    \lomppo = \sum\limits_{j=0}^{\ud-1} \binom{\ud-1}{j} \lomcp^j (1-\lomcp)^{\ud-1-j} f(j).
\end{equation*}
Then we have
    \begin{align}
        \lompp - \lomppo &= \sum\limits_{j=0}^{\ud} \binom{\ud}{j} \lomcp^j (1-\lomcp)^{\ud-j} f(j) - \sum\limits_{j=0}^{\ud-1} \binom{\ud-1}{j} \lomcp^j (1-\lomcp)^{\ud-1-j} f(j) \notag\\
        &= (1-\lomcp) \sum\limits_{j=0}^{\ud} \left( \binom{\ud-1}{j} + \binom{\ud-1}{j-1} \right) \lomcp^j (1-\lomcp)^{\ud-1-j} f(j) - \sum\limits_{j=0}^{\ud-1} \binom{\ud-1}{j} \lomcp^j (1-\lomcp)^{\ud-1-j} f(j) \notag\\
        &= (1-\lomcp) \sum\limits_{j=1}^{\ud} \binom{\ud-1}{j-1} \lomcp^j (1-\lomcp)^{\ud-1-j} f(j) - \lomcp \sum\limits_{j=0}^{\ud-1} \binom{\ud-1}{j} \lomcp^j (1-\lomcp)^{\ud-1-j} f(j) \notag\\
        &= \lomcp \sum\limits_{j=1}^{\ud} \binom{\ud-1}{j-1} \lomcp^{j-1} (1-\lomcp)^{\ud-j} f(j) - \lomcp \sum\limits_{j=0}^{\ud-1} \binom{\ud-1}{j} \lomcp^j (1-\lomcp)^{\ud-1-j} f(j) \notag\\
        &= \lomcp \sum\limits_{j=0}^{\ud-1} \binom{\ud-1}{j} \lomcp^j (1-\lomcp)^{\ud-1-j} (f(j+1) - f(j)) \notag\\
        &= \frac{\lomcp}{1-\lomcp} \lompgap. \label{eq:lompgap_dif}
    \end{align}
    Using this observation, the threshold equation \eqref{eq:lom_l} is equivalent to
    \begin{equation}\label{eq:sub:threshold}
        \frac{\lompo}{\lomrs} \le \frac{\lompp + \lomppo}{2}.
    \end{equation}
    For $\lom(\ud,\lomcp,\lomerr)$ two types of error can happen:
    \begin{enumerate}[label=\roman*)]
    \item We have $d \le \ud-1$, but is claimed to be $d \ge \ud$;\label{sub:error:TypeI}
    \item We have $d \ge \ud$, but is claimed to be $d \le \ud-1$.\label{sub:error:TypeII}
    \end{enumerate}

    Case \ref{sub:error:TypeI}: It is worth noting that $\lompo \sim \text{Binomial} \left(\lomrs , \lomppd\right)$. From \eqref{eq:sub:threshold} we know the probability of error is
    \begin{align*}
        \Pr \left( \frac{\lompo}{\lomrs} > \frac{\lompp + \lomppo}{2} \right) &= \Pr \left( \lompo > \frac{\lompp + \lomppo}{2}\cdot\lomrs \right)\\
        &\le \Pr \left( \text{Binomial} \left(\lomrs , \lomppo\right) > \frac{\lompp + \lomppo}{2} \cdot \lomrs \right) \\
        &\le \exp \left( -\frac{1}{3} \left( \frac{\lompp-\lomppo}{2\lomppo} \right)^2 \cdot\lomrs \lomppo\right) \\
        &\le \exp \left( - \frac{\left(\lompp-\lomppo\right)^2}{12} \lomrs \right) \\
        &\leq \lomerr.
    \end{align*}
       where the first inequality follows from the fact that $\lomppd$ is monotonically increasing with respect to $d$ and $d \leq \ud-1$; the second inequality follows from Chernoff bound in Fact \ref{chernoff_bound}; the third inequality follows from the fact that $\lomppo\leq 1$; the last inequality follows by substituting \eqref{eq:def:lomrs} and \eqref{eq:lompgap_dif}.

    Case \ref{sub:error:TypeII}: The calculations are similar to Case \ref{sub:error:TypeI}. Once again,  $\lompo \sim \text{Binomial} \left(\lomrs , \lomppd\right)$. We know from \eqref{eq:sub:threshold} that the probability of error is
    \begin{align*}
        \Pr \left( \frac{\lompo}{\lomrs} \le \frac{\lompp + \lomppo}{2} \right) &= \Pr \left( \lompo \le \frac{\lompp + \lomppo}{2}\cdot\lomrs \right)\\
        &\le \Pr \left( \text{Binomial} \left(\lomrs , \lompp\right) \le \frac{\lompp + \lomppo}{2} \cdot \lomrs \right) \\
        &\le \exp \left( -\frac{1}{3} \left( \frac{\lompp-\lomppo}{2\lompp} \right)^2 \cdot\lomrs \lompp\right) \\
        &\le \exp \left( - \frac{\left(\lompp-\lomppo\right)^2}{12} \lomrs \right) \\
        &\leq \lomerr.
    \end{align*}
     where the first inequality follows from the fact that $\lomppd$ is monotonically increasing with respect to $d$ and $d \ge \ud$; the second inequality follows from Chernoff bound in Fact \ref{chernoff_bound}; the third inequality follow from the fact that $\lompp\leq 1$; the last inequality follows by substituting \eqref{eq:def:lomrs} and \eqref{eq:lompgap_dif}.

    Combining the two cases we conclude that the error probability of $\lom(\ud,\lomcp,\lomerr)$  is at most $\lomerr$.
\end{IEEEproof}

\subsection{Algorithm for exactly estimating $d$}
Armed with the above subroutine $\lom(\ud,\lomcp,\lomerr)$, the algorithm for exactly estimating $d$ is now described in Algorithm \ref{algo:est}. 
\begin{algorithm}
    \caption{Exact estimation of $d$}\label{algo:est}
    \begin{algorithmic}[1]
    \State $\textbf{Initialize } \lomerr \gets \frac{\esterr}{2\log n + 2}$, $\dup \gets 2$
    \While{true}
        \State $\textbf{set } \lomcps = \argmin_{\lomcp\in(0,1)} \frac{1-\lomcp}{\lomcp (\lompgapup)^2}$
        \State $\textbf{run } \lom(\dup,\lomcps,\lomerr)$
        \State $\textbf{if } d \le \dup-1 \textbf{ then halt}$
        \State $\textbf{else set } \dup \gets 2\dup$
    \EndWhile
    \State $\textbf{set } \dlw \gets \frac{\dup}{2}$

    \While{$\dup - \dlw \ge 2$}
        \State $\textbf{set } \dmd = \left\lfloor \frac{\dlw+\dup}{2} \right\rfloor \text{, } \lomcps = \argmin_{\lomcp\in(0,1)} \frac{1-\lomcp}{\lomcp (\lompgapmd)^2}$
        \State $\textbf{run } \lom(\dmd,\lomcps,\lomerr)$
        \State $\textbf{if } d \le \dmd-1 \textbf{ then set } \dup \gets \dmd$
        \State $\textbf{else set } \dlw \gets \dmd$
    \EndWhile
    \State $\textbf{output } \dlw$
    \end{algorithmic}
\end{algorithm}


We start by noting that both while loops in Algorithm \ref{algo:est} invoke at most $\log (2d)$ calls to the subroutine $\lom(\ud,\lomcp,\lomerr)$. By Lemma \ref{lem:sub:error} and the union bound, we know that the error probability of Algorithm \ref{algo:est} is bounded from above by $$2 \log (2d) \cdot \lomerr =2 \log (2d)\frac{\esterr}{2\log n + 2}\le \esterr.$$

From \eqref{eq:def:lomrs} we have
\begin{align*}
    \lomrss &= \left\lceil 8.32 \left( \frac{1-\lomcps}{\lomcps \lompgaps} \right)^2 \log \frac{1}{\lomerr}  \right\rceil\\
    &\le \left\lceil 8.32 \left( \frac{1-\lomcps}{\lomcps (\lompgaps)^2} \right)^2 \log \frac{1}{\lomerr} \right\rceil\\
    &\le 8.32 \left(376017 H(f,\ud) \, \ud\right)^2\log \frac{1}{\lomerr}+1  \\
    &= 8.32 \left(376017 H(f,\ud) \, \ud \right)^2 \log\left(\frac{2\log n+2} { \esterr }\right)+1.
\end{align*}
The first inequality follows by noting that $\lompgaps\leq 1$ since $\lompgap$ in \eqref{eq:def:lompgap} is the same as $\pgap$ in \eqref{eq:convert_delta} (with $(\ud,\lomcps)$ in place of $(d,\cp)$), and $\pgap\leq 1 $ for all $\cp$ by Lemma~\ref{lem:relation:quantities}. The second inequality can be justified as follows: The expression in $(\cdot)^2$ is similar to $\fTp$ in \eqref{eq:def:fTp}. By the same argument as in Proposition \ref{prop:existence_final} and the discussions that follow, we can bound the expression in $(\cdot)^2$ by $376017 H(f,\ud) \, \ud$, where $H(f,\ud)$ is the same as $\fH$ in \eqref{eq:def:sensi:para} but with $\ud $ in place of $d$.

Lastly, since $\lom(\ud,\lomcp,\lomerr)$ is called $\cO(\log d)$ times and $\ud\in\cO(d) $, the above inequality implies that the total number of tests in Algorithm \ref{algo:est} scales as $\cO \left(\left(\fH d\right)^2 \log d \log\left(\frac{\log n}{\esterr }\right)\right)$. This completes the proof of Lemma \ref{lem:estimation:d}.

\newpage
\bibliographystyle{IEEEtran}
\bibliography{IEEEabrv,references}

\end{document}